\documentclass[a4paper,11pt]{article}
\usepackage{jcappub}
\usepackage{lineno}
\usepackage{multirow}
\usepackage{graphicx}
\usepackage{amsmath}
\usepackage{amsfonts}
\usepackage{amssymb}
\usepackage{slashed} 
\usepackage{color}
\usepackage{hyperref}
\usepackage{units}
\usepackage{braket}
\usepackage{enumitem}
\usepackage{booktabs}
\usepackage[table,x11names]{xcolor}
\usepackage{empheq}
\usepackage{amsmath}
\usepackage{soul}
\nolinenumbers

\title{Angular momentum of vacuum bubbles in a first-order phase transition}
 \author[a,d]{Jan Tristram Acu\~{n}a,}
\author[b]{Danny Marfatia,}
\author[a,c]{Po-Yan Tseng}
\affiliation[a]{Department of Physics, National Tsinghua University, 101 Kuang-Fu Rd., Hsinchu 300044,\\ Taiwan ROC}
\affiliation[b]{Department of Physics and Astronomy, University of Hawai’i, Honolulu, HI 96822, USA}
\affiliation[c]{Physics Division, National Center for Theoretical Sciences,
Taipei 106319, Taiwan ROC}
\affiliation[d]{National Institute of Physics, University of the Philippines Diliman, Quezon City 1101, Philippines}
\emailAdd{jtacuna@gapp.nthu.edu.tw}
\emailAdd{dmarf8@hawaii.edu}
\emailAdd{pytseng@phys.nthu.edu.tw}
\date{\today}

\abstract{
The formation of primordial black holes (PBHs) during a first-order phase transition (FOPT) in a dark sector has been of recent interest. A quantity that characterizes a black hole is its spin. We carry out the first step towards determining the spin of such PBHs, by calculating the spin of spherical false vacuum bubbles induced by cosmological perturbations. The angular momentum is given by the product of density and velocity perturbations. We carefully track the evolution of background quantities and calculate the transfer functions during the FOPT.  We find that the dimensionless spin parameter $s = J/(G_{\rm N} M^2)$ of false vacuum bubbles of mass $M$ and angular momentum $J$, take a wide range of values from ${\cal{O}}(10^{-5})$ to ${\cal{O}}(10)$ for FOPTs between \unit[10]{keV} and \unit[100]{GeV} and a dark sector that is 0.1 to 0.4 times cooler than the visible sector. We also find a scaling relation between the root-mean-square value of the spin, the FOPT time scale, the bubble wall velocity, and the dark sector-to-visible sector temperature ratio.
}
\makeatletter
\gdef\@fpheader{}
\makeatother

\begin{document}
\maketitle
\flushbottom

\section{Introduction}
A well-known result in general relativity (GR) is that black holes can be characterized by their mass and spin. As discussed extensively in the literature, the angular momentum of primordial black holes (PBHs) is strongly related to the black hole production mechanism. A common scenario is that enhancements in the primordial curvature power spectrum increase the probability of obtaining rare density peaks above the critical threshold for collapse~\cite{Press:1973iz,Green:2004wb,Young:2020xmk,Yoo:2020dkz}. Computations of the spin of PBHs have been performed adopting various hypotheses, where they  predict spin values in the range $s \simeq 10^{-3} \sim 10^{-2}$~\cite{Mirbabayi:2019uph,DeLuca:2019buf,Harada:2020pzb,Banerjee:2023qya}. Alternatively, we focus on the scenario that PBHs are formed from a cosmological dark first-order phase transition (FOPT) without any enhancement in the primordial power spectrum.

The PBH may be formed from the direct collapse of false vacuum (FV) bubbles~\cite{Baker:2021nyl}. Another mechanism involves the formation of intermediate compact objects called Fermi balls (FBs), which are composed of degenerate dark fermions~\cite{Gross:2021qgx,Kawana:2021tde}. The FBs cool down by emitting light particles that may eventually trigger the collapse into PBHs. Both mechanisms efficiently produce subsolar mass FBs/PBHs for FOPT energy scales from $\mathcal{O}({\rm keV})$ to $\mathcal{O}(100\,{\rm GeV})$. In this work, the fundamental hypothesis in generating PBH spin is the idea that cosmological perturbations induce nonzero angular momentum. The cosmological perturbations are Gaussian random variables whose initial values are ultimately determined by the primordial curvature power spectrum. We show that the angular momentum and spin are Gaussian random variables and thus their distributions are determined by their root-mean-square (RMS) values. We restrict our calculation of the spin to the case of false vacuum (FV) bubbles for a number of reasons. In the case of direct collapse to PBHs, the process is nonlinear and may require techniques beyond perturbation theory. Similarly, in the case where FBs are formed, the emission of light species carry away angular momentum from the FV bubble. Tracking the subsequent evolution of the angular momentum is beyond the scope of our work.

To realize the FOPT scenario, which is an essential ingredient in the novel PBH formation mechanism that we feature in this work, it is necessary to introduce dark sector dynamics, since phase transitions in the Standard Model are known to be smooth crossovers. In particular, we consider a model framework with a dark scalar $\phi$ that triggers the FOPT. For the case in which  Fermi balls form in an intermediate step, we also require dark Dirac fermion $\chi$, which serve as the fundamental building block for Fermi balls. The size distribution of FV bubbles during the FOPT has been discussed in Ref.~\cite{Lu:2022paj},   
from which the mass and abundance of FB/PBH can be determined. In our work, we take the size of the FV sphere to be the critical radius at the time of percolation, within which no true vacuum bubble can nucleate.

Earlier studies on PBH formation from FOPTs focused on the PBH mass and have not considered the generation of angular momentum.  In this setup, it is possible to generate angular momentum from spatial inhomogenities in the wall velocity, or from perturbations in the density and velocity fields of the dark sector plasma, treated as a single fluid consisting of two phases. In this work, we adopt the latter approach, where we track the density, velocity, and metric perturbations from the time at which all modes are superhorizon, up to the time of percolation during the FOPT. We assume that the initial conditions are set by Gaussian primordial curvature perturbations, that are nearly scale-invariant. We adopt the uniform Hubble gauge to track the evolution of the perturbations. Taking into account some subtleties in defining angular momentum in GR, we express the angular momentum as a volume integral of a product of density and velocity perturbations, where we choose the volume of the proto-object to be spherical. As mentioned, the angular momentum and spin are random variables since the initial curvature perturbations are themselves Gaussian random variables. Thus, our goal is to calculate the RMS spin of FV bubbles during a dark FOPT at the time of percolation. 

We point out that the determination of the PBH spin, as well as its distribution, has been the subject of active investigation in the literature. Based on the seminal work~\cite{Choptuik:1992jv} on the critical behavior of BH masses, arising from the evolution of a family of ingoing packets of a scalar field,   
Ref.~\cite{Baumgarte:2016xjw} considered a radiation fluid characterized by the size of the density fluctuation and angular momentum. The main result of Ref. \cite{Baumgarte:2016xjw} is the presence of a power-law relation not only for the BH mass but also for the angular momentum. Building upon this work, Ref.~\cite{Chiba:2017rvs} obtained the mass and angular momentum distribution by assuming that the initial density fluctuation is Gaussian, while the parameter describing the initial rotation was assumed to be flat. The resulting spin $s 
\lesssim 0.4$; on the other hand, accretion effects on the mass and spin of the BHs were shown to be negligible. In the case where the PBH spin is induced by cosmological perturbations Ref.~\cite{Mirbabayi:2019uph} calculated the RMS value of the PBH spin, assuming that the PBHs are formed from the collapse of rare spherically symmetric density peaks. The spin of the PBH was shown to be a product of density and velocity perturbations, making it a \textit{second-order} quantity in perturbations. Assuming a narrow enhancement in the primordial curvature power spectrum, they obtained $s \simeq 0.01$, for PBHs comprising a sizable fraction of dark matter. In contrast, Ref.~\cite{DeLuca:2019buf} assumed that the proto-object that will eventually collapse into PBHs is ellipsoidal, and the distribution of the ellipsoidal configurations is determined by peak theory \cite{heavens1988tidal}. They showed that the PBH spin is a \textit{first-order} quantity in perturbations and obtained spin parameters at the percent level, given some assumptions in the enhancement in the curvature power spectrum. Ref.~\cite{Harada:2020pzb} followed the formalism laid down in Ref.~\cite{DeLuca:2019buf}, and provided some improvements in the analysis of the latter. Both approaches assumed that the initial PBH spin corresponds to the value at turn around time, \textit{i.e.}, the moment when the overdensity begins to collapse; in Ref.~\cite{Harada:2013epa}, they pointed out that the turn around time does not necessarily occur at the moment when the (enhanced) perturbation mode enters the horizon. As a result, Ref.~\cite{Harada:2013epa} predicted an RMS spin value of $\simeq 4.0 \times 10^{-3} (M/M_H)^{-1/3}$, where $M_H$ is the mass within the Hubble horizon at horizon entry of the enhanced perturbation. In addition, Ref.~\cite{Harada:2013epa} pointed out that the second-order approach in Ref.~\cite{Mirbabayi:2019uph} may give rise to spin estimates that are comparable with the first-order approach. In Ref.~\cite{Eroshenko:2021sez}, PBHs are formed from the collapse of domain walls and the angular momentum arises from the tidal forces from the surrounding radiation fluid acting on the walls, obtaining $s \simeq 10^{-4} (M/30 M_\odot)^{-1}$. Finally, recent work in Ref.~\cite{Banerjee:2023qya} used the result from Ref.~\cite{DeLuca:2019buf}, and combined it with the curvature perturbation obtained in Ref.~\cite{Liu:2022lvz,Elor:2023xbz}, where they obtained a spin value of around $s \simeq \mathcal{O}(10^{-3})$. The curvature perturbation is generated from the difference in nucleation times of different points in space, thereby generating regions that contain more vacuum energy than others. We point out that the applicability of the result from Ref.~\cite{DeLuca:2019buf} assumes that any modification in the curvature power spectrum should be made at the time when all perturbations are superhorizon.

The rest of the paper is devoted to the task of calculating the spin of FV bubbles at the percolation time, during an FOPT in the dark sector, induced by cosmological perturbations. In Section~\ref{sec:DarkSectorModel} we first introduce the dark sector model and the quartic scalar potential, before proceeding with the discussion of angular momentum in GR and in the specific case of a perturbed Friedmann–Robertson–Walker (FRW) spacetime in Section~\ref{sec:AngularMomCosmo}. Following Ref.~\cite{Schmid:1998mx}, we present the evolution equations for cosmological perturbations in the uniform Hubble gauge in Section~\ref{sec:CosmoPertUHG}. We separately consider the period from the time when all modes are superhorizon up to the critical point, and the period of FOPT between the critical point and the percolation time. Determining the evolution of the background quantities and the perturbations during the FOPT are crucial to our task, and we provide a detailed discussion of this part of the calculation in Section \ref{subsec:EvolPertCritToPerc}. We present our main results, estimates for the FV spin, and benchmark points in Section \ref{sec:MainResultsDisc}. Finally, we conclude in Section \ref{sec:Conclusion}.

\section{Dark first-order phase transitions}
\label{sec:DarkSectorModel}
In the case where PBHs are formed from the direct collapse of FV bubbles, it is sufficient to consider only a real scalar field $\phi$ in the dark sector. The scalar field triggers the FOPT by acquiring a nonzero expectation value at some critical point. On the other hand, to consider the formation of Fermi balls we include almost massless dark Dirac fermions 
$\chi$ in the particle spectrum. These fermions serve as the fundamental component of Fermi balls. Then the dark sector effective Lagrangian density can be written as
\begin{eqnarray}
   \label{DSLagrangian} \mathcal{L} = \frac{1}{2}\left(\partial \phi\right)^2-V_{\rm eff}(\phi,T)+\bar{\chi}\left(i\slashed{\partial} - m_\chi\right)\chi - g_\chi \phi \bar{\chi}\chi\,,
\end{eqnarray}
where $T$ is the temperature of the dark sector, and the effective potential is given by
\begin{eqnarray}
  \label{VeffDefn}  V_{\rm eff}(\phi, T) &=& V_4(\phi, T) - \frac{\pi^2}{90}g_\rho(\phi)T^4 - \Delta b(\phi)\,T^2 \phi^2\,,
\end{eqnarray}
where the quartic potential is
\begin{eqnarray}
  \label{DefV4} V_4(\phi,T) &=& \frac{\lambda}{4} \phi^4 -(AT+C) \phi^3 +D(T^2 - T_0^2) \phi^2\,.
\end{eqnarray}
The form of the effective potential may arise from a more fundamental theory, possibly with more particle species, where the usual procedure of accounting for the zero-temperature and finite-temperature contributions to the effective potential at one-loop order, and taking the high-temperature limit are carried out, as in Refs.~\cite{Dolan:1973qd,Anderson:1991zb}. The quantity $g_\rho(\phi)$ counts the number of effective relativistic degrees of freedom in the dark sector for a given classical field configuration $\phi$, where bosons and fermions are respectively weighted by $1$ and $7/8$. $\Delta b$ counts the number of degrees of freedom that are relativistic in the false vacuum but not in the true vacuum (TV). For simplicity, we take
\begin{eqnarray}
    \Delta b = 0\,,\quad g_\rho = 4.5\,.
\end{eqnarray}
We take $\lambda$, $A$, $C$, $D$, and $T_0^2$ to be positive real parameters and require that $T>T_0$ so that the potential has two minima with one at $\phi = 0$. At the critical temperature $T = T_{\rm c}$, the two minima are degenerate in energy. When the temperature of the dark sector further cools to $T < T_{\rm c}$, the $\phi = 0$ configuration will be regarded as the false vacuum, which has a higher energy than the other minimum, the true vacuum, given by
\begin{eqnarray}
    \phi_+(T) = \frac{3(AT+C)+\sqrt{9(AT+C)^2 - 8\lambda D (T^2 - T_0^2)}}{2\lambda}.
\end{eqnarray}
During this period, regions of spacetime will begin to nucleate TV bubbles, at a rate per unit volume given by the nucleation rate $\Gamma$, which depends on the underlying particle physics model. The bubbles of TV expand, and the TV bubbles subsequently form an infinite connected network of bubbles, which defines the time of \textit{percolation}. A useful criterion for calculating the percolation time is to identify the FV fraction $F$, defined as the spatial volume fraction occupied by FV bubbles, with $1/e$. 

Note that the Yukawa term in Eq.~(\ref{DSLagrangian}) is mainly relevant for the Fermi ball scenario of PBH formation, but it can also be present in the scenario where PBHs are directly formed from FV bubbles. Firstly, the Yukawa term allows the filtered DM scenario to be realized, where the $\chi$ species are trapped inside FV bubbles. A necessary condition for this to occur is that the $\chi$ mass difference $\Delta m_\chi = g_\chi \phi_+$ between the TV and FV is much larger than the typical kinetic energy of $\chi$, roughly given by $T$, so that the flux of $\chi$ particles through the bubble walls is exponentially suppressed. Secondly, the Yukawa term provides an attractive force between the $\chi$ particles to trigger an instability that leads to the formation of PBH from FBs. Finally, under certain conditions, the Yukawa interaction ensures that the $\chi$ and $\phi$ species are tightly coupled with each other. The necessary condition for this to occur can be derived by requiring that interactions involving $\chi$ and $\phi$, \textit{e.g.},~$\chi \bar{\chi} \rightarrow \phi \phi$, occur at a much faster rate than the Hubble expansion, so that we have
\begin{eqnarray}
    \frac{n_\chi \langle \sigma v\rangle_{\chi \bar{\chi} \rightarrow \phi \phi}}{H} \sim g_{\rho,{\rm SM}}^{-1/2} g_\chi^4 r_T\frac{M_{\rm Pl}}{T_{\rm SM}} \gg 1\,,
\end{eqnarray}
where $r_T \equiv T/T_{\rm SM}$ is the dark-to-SM-sector temperature ratio, and $g_{\rho,{\rm SM}}$ is the effective number of relativistic degrees of freedom in the SM sector at temperature $T_{\rm SM}$. 

The dynamics of the FOPT is determined by tracking the cosmological evolution of the plasma, in the presence of both SM and dark sector degrees of freedom, treating each component as a fluid with a known equation of state. For the dark sector we can calculate the thermodynamic quantities---free energy, pressure, entropy, and energy---in the false and true vacuum using the effective potential Eq.~(\ref{VeffDefn}). These are then given respectively by (cf. \cite{Leitao:2014pda,Espinosa:2010hh,Giese:2020rtr,Tenkanen:2022tly})
\begin{eqnarray}
\label{FOPTFreeEnergy}    \mathcal{F}_{\rm FV}(T) = - \frac{\pi^2}{90}g_\rho T^4 &,&\quad \mathcal{F}_{\rm TV}(T) = - \frac{\pi^2}{90}g_\rho T^4 + \Delta p(T)\,,\\
\label{FOPTPressure}    p_{\rm FV}(T) = \frac{\pi^2}{90}g_\rho  T^4 &,&\quad p_{\rm TV}(T) =  \frac{\pi^2}{90}g_\rho T^4 -\Delta p(T)\,, \\
 \label{FOPTEntropy}   s_{\rm FV}(T) = \frac{2\pi^2}{45}g_\rho T^3 &,&\quad s_{\rm TV}(T) = \frac{2\pi^2}{45}g_\rho T^3 -\Delta s(T)\,,\\
\label{FOPTEnergyDens}    \rho_{\rm FV}(T) = \frac{\pi^2}{30}g_\rho T^4 &,&\quad \rho_{\rm TV}(T) = \frac{\pi^2}{30}g_\rho T^4 -\Delta \rho(T)\,,
\end{eqnarray}
where
\begin{align}
   \Delta p = -\Delta V(T)\,,\quad \Delta s = \frac{d}{dT}\Delta p = -\frac{d\Delta V(T)}{dT}\,,\quad \Delta \rho = T \Delta s - \Delta p = \Delta V(T) - T \frac{d\Delta V(T)}{dT}\,.
\end{align}
By definition, the critical temperature is the moment at which $\Delta p = 0$, so that $\Delta V(T_{\rm c}) = 0$. In terms of the parameters in the quartic potential, 
\begin{eqnarray}
   \label{CritTempExp}  \left(A^2 - \lambda D\right)T_{\rm c}^2 + 2AC T_{\rm c} + \left(C^2 + \lambda D T_0^2\right) = 0\,.
 \end{eqnarray}
We can trade the $D$ parameter in exchange for $T_{\rm c}$ via
 \begin{eqnarray}
     \lambda D = \frac{\left(A+\frac{C}{T_{\rm c}}\right)^2}{1 - \frac{T_0^2}{T_{\rm c}^2}}\,.
 \end{eqnarray}
The derivative of $\Delta p$ is discontinuous at $T = T_{\rm c}$ and is simply attributed to the presence of latent heat, which is a defining characteristic of an FOPT. Denoting the latent heat at the critical temperature by $L_{\rm c}$, where
\begin{eqnarray}
\Delta \rho(T_{\rm c}) = L_{\rm c}\,,
\end{eqnarray}
we have
\begin{eqnarray}
\label{LatentHeatExpression}    L_{\rm c} \equiv -T \frac{d\Delta V(T)}{dT}\Bigg\vert_{T=T_{\rm c}} = T_{\rm c}^4~\frac{8(A+x_C)^3\left(AT_0^2/T_{\rm c}^2 + x_C\right)}{\lambda^3 \left(1-T_0^2/T_{\rm c}^2\right)}\,,\quad x_C \equiv \frac{C}{T_{\rm c}}\,.
\end{eqnarray}
This quantity roughly tells us the amount of energy that will be liberated during the transition from the false vacuum to true vacuum. Note that Eq.~(\ref{LatentHeatExpression}) can be inverted to give
\begin{eqnarray}
 \label{x02Expression}   \frac{T_0^2}{T_{\rm c}^2} = \frac{\lambda^3 (L_{\rm c}/T_{\rm c}^4) - 8x_C (A+x_C)^3}{\lambda^3 (L_{\rm c}/T_{\rm c}^4) + 8 A (A+x_C)^3}\,.
\end{eqnarray}
Since $0 \leq T_0^2/T_{\rm c}^2 \leq 1$, there must be a lower bound for the latent heat, \textit{i.e.},
\begin{eqnarray}
\label{PhysicalLc}    \frac{L_{\rm c}}{T_{\rm c}^4} \geq \frac{8x_C (A+x_C)^3}{\lambda^3}\,.
\end{eqnarray}
\subsection{Physical conditions}
\label{sec:Physicalconditions}
We further impose physical conditions on the thermodynamic quantities, to restrict the class of effective potentials that will be the subject of this study. These conditions should apply for all temperatures within $T_0 \leq T \leq T_{\rm c}$. Generally, $\Delta p = -\Delta V_{\rm eff} \leq 0$ is automatically satisfied, which ensures that the TV regions exert a net outward pressure on the surrounding FV region, to allow nucleated TV bubbles to expand. Furthermore, we require $\Delta \rho \geq 0$ since the FV has a higher energy than the TV. The energy difference between the two phases is a monotonically increasing function of $T$. On the other hand, we also need to ensure that $\rho_{\rm TV}(T) = g_\rho \pi^2/30~T^4 - \Delta \rho(T) \geq 0$ for all $T_0 \leq T \leq T_{\rm c}$. We also require that the square of the sound speed in the TV, taken in isolation, 
\begin{eqnarray}
\label{csTV2}  \boxed{ c_{s\rm ,TV}^2 = \frac{dp_{\rm TV}/dT}{d\rho_{\rm TV}/dT}}\,,
\end{eqnarray}
be positive and subluminal. If $s_{\rm TV} = dp_{\rm TV}/dT \geq 0$, then a necessary condition for $c_{s\rm ,TV}^2 > 0$ is $d\rho_{\rm TV}/dT > 0$; this latter condition and $\rho_{\rm TV}(T_0) \geq 0$ guarantee $\rho_{\rm TV}(T) \geq 0$.
We find it convenient to consider
 \begin{eqnarray}
 \nonumber A, \quad T_{\rm c},\quad \lambda,\quad \frac{L_{\rm c}}{T_{\rm c}^4},\quad x_C
 \end{eqnarray}
 as fundamental parameters. Once these parameters are specified, the effective potential is known. In choosing our benchmark points, we take $A = 10^{-1}, 10^{-2}, 10^{-3}, 10^{-4}$. As for the other parameters, we turn to the physical conditions for guidance. If $\Delta\rho'(T)>0$, it follows that if $\rho_{\rm TV}(T_0) = 0$,
\begin{align}
   \frac{\pi^2}{30}g_\rho \left(\frac{T_0}{T_{\rm c}}\right)^4 = \frac{\Delta\rho(T_0)}{T_{\rm c}^4} \leq \frac{L_{\rm c}}{T_{\rm c}^4}.\,
\end{align}
At $x_C = 0$ we find that
\begin{align}
    \frac{\pi^2}{30}g_\rho \left(\frac{T_0}{T_{\rm c}}\right)^2\left[1-\left(\frac{T_0}{T_{\rm c}}\right)^2\right] \leq \frac{8A^4}{\lambda^3}.\,
\end{align} 
 In the range $0 \leq T_0/T_{\rm c} \leq 1$, the maximum value of the left hand side of the inequality is $g_\rho\pi^2/120$. Then we define $\lambda_0$ as a reference scale for $\lambda$ via
 \begin{align}
     \frac{\pi^2}{120}g_\rho \equiv \frac{8A^4}{\lambda_0^3} \rightarrow \lambda_0 \simeq 0.129 \left(\frac{A}{0.1}\right)^{4/3} \left(\frac{4.5}{g_\rho}\right)^{1/3}\,.
 \end{align}
Note that statements regarding the physical condition are independent of $T_{\rm c}$, which only sets the scale of the phase transition. In the following, we take 
\begin{eqnarray}
\nonumber T_{\rm c} = \unit[10]{keV}\,, \unit[1]{MeV}\,, \unit[10]{MeV}\,, \unit[1]{GeV}\,, \unit[100]{GeV}\,.
\end{eqnarray}
\begin{figure}[t]
    \centering
    \begin{tabular}{cc}
          \includegraphics[scale=0.26]{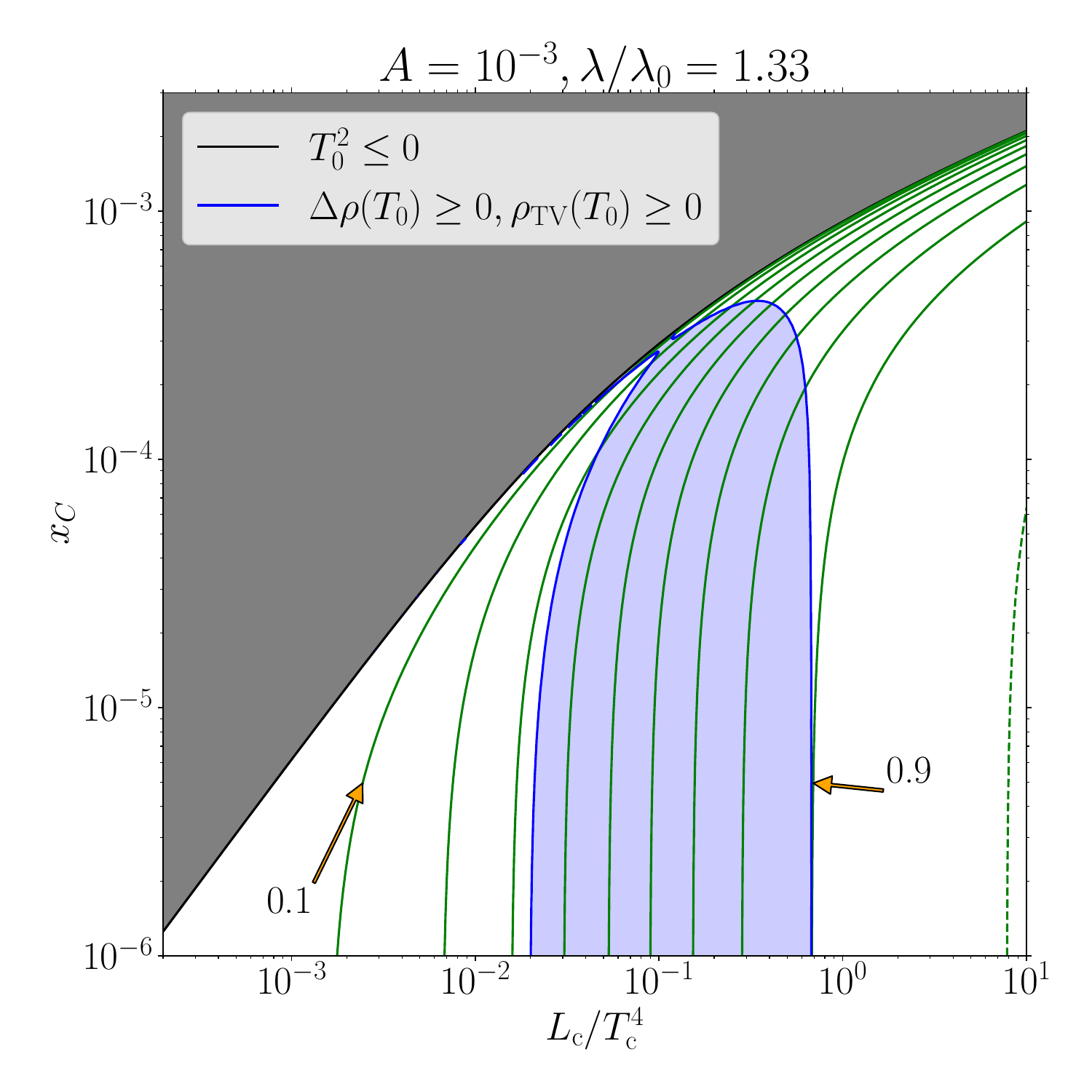} &  \includegraphics[scale=0.26]{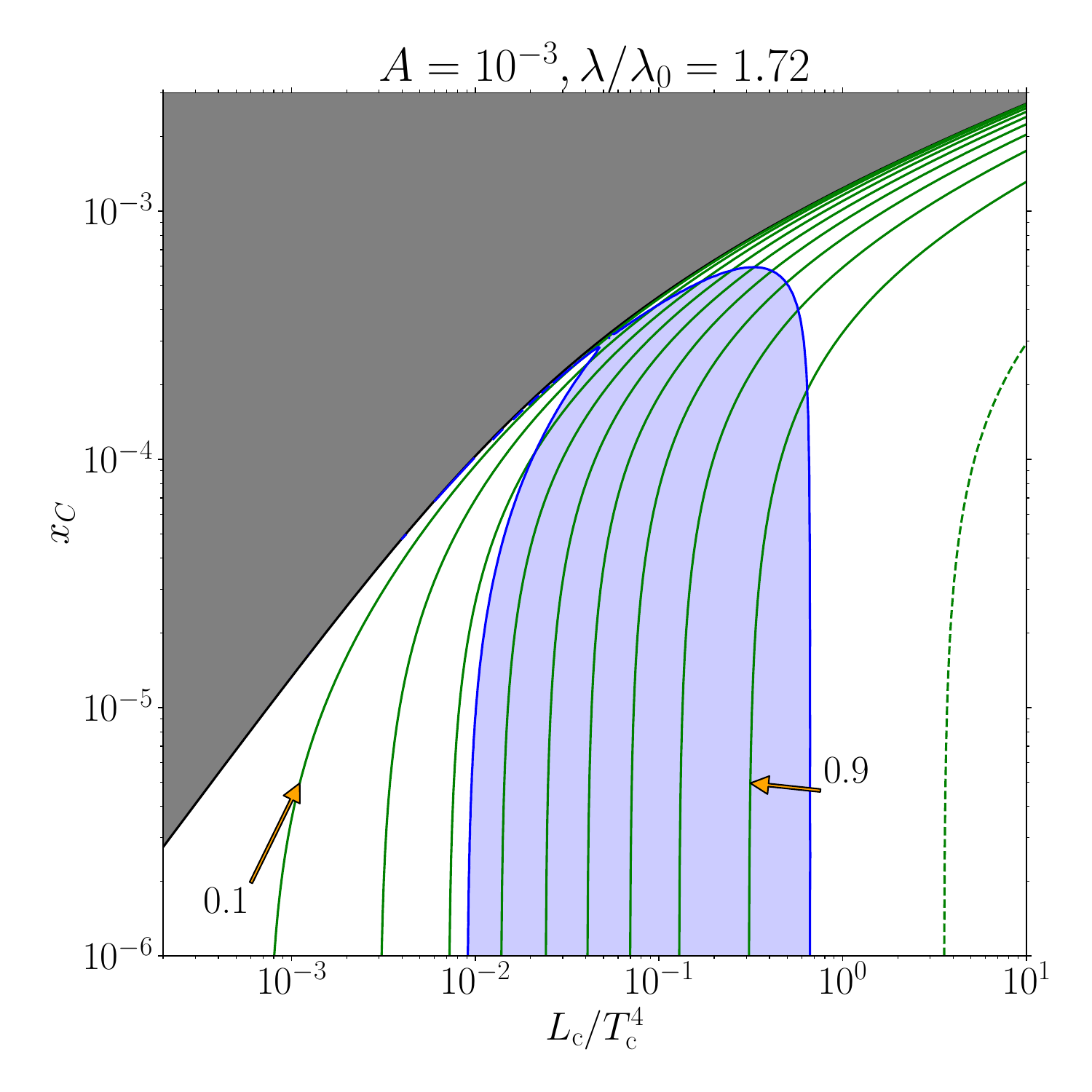} \\
        (a) & (b)\\
         \includegraphics[scale=0.26]{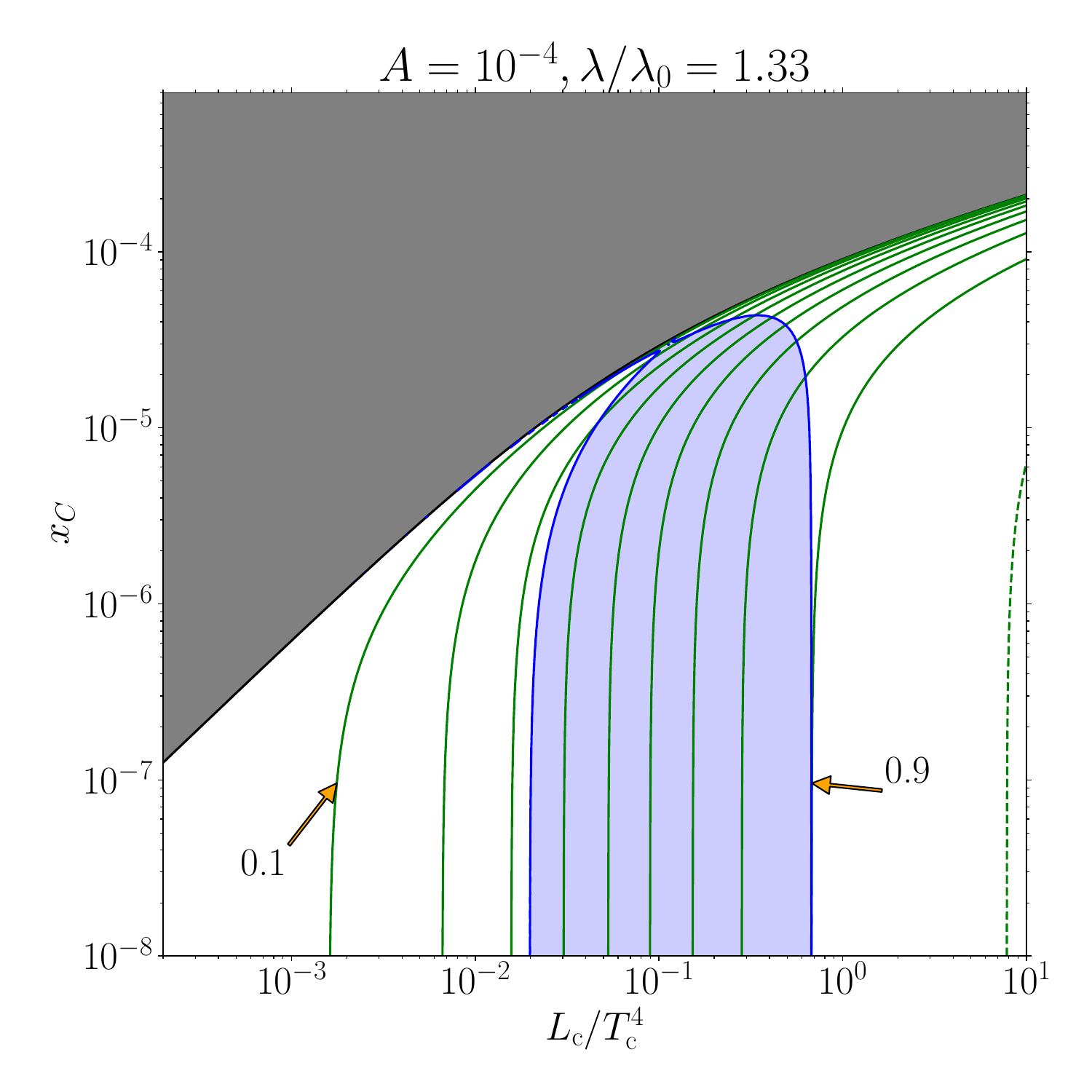} &  \includegraphics[scale=0.26]{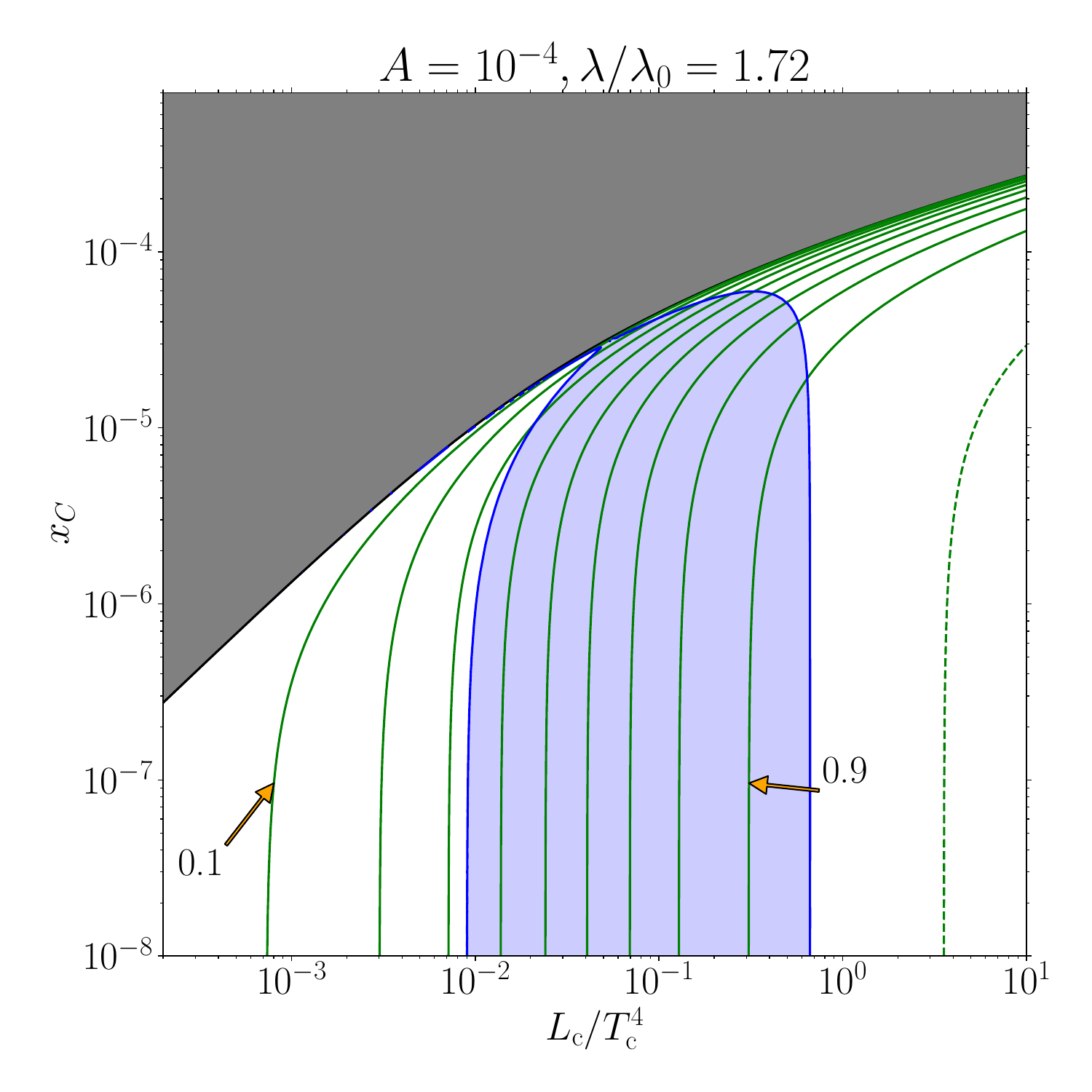}\\
         (c) & (d)
    \end{tabular}
    \caption{The physical region is shaded in blue for $A = 10^{-3}, 10^{-4}$, and two values of $\lambda/\lambda_0 \simeq 1.33, 1.72$. The green curves correspond to contours of constant $T_0/T_{\rm c}$, from 0.1 to 0.9, in steps of 0.1. The dashed green contour corresponds to $T_0/T_{\rm c} = 0.99$. For all cases, $g_\rho = 4.5$.}
    \label{fig:PhysRegion}
\end{figure}
As an illustration, in Fig.~\ref{fig:PhysRegion} we show the physical region shaded in blue, for $A = 10^{-3}, 10^{-4}$, choosing two values of $\lambda/\lambda_0$ for each: 1.32529 and 1.72359. The black line corresponds to $T_0= 0$, so that the region above it leads to $T_0^2<0$, and is thus forbidden. The green contours correspond to contours of constant $T_0/T_{\rm c} = 0.1, 0.2, ...,$ to $0.9$, indicated by solid curves. The lone dashed curve in each panel corresponds to $T_0/T_{\rm c} = 0.99$. We observe that a smaller $x_C$ leads to a wider range of allowed $L_{\rm c}/T_{\rm c}^4$ values. The maximum range of physical $L_{\rm c}/T_{\rm c}^4$ values is achieved when $x_C = 0$, for fixed $A$ and $\lambda$. Note that a larger $\lambda$ leads to a larger area occupied by the physical region, as well as an increase in the maximum $x_C$. Smaller values of $L_{\rm c}/T_{\rm c}^4$ are accommodated, opening up a wider range of physical rescaled latent heat values.

\section{Angular momentum from cosmological perturbations}
\label{sec:AngularMomCosmo}
In our approach, we would like to estimate the angular momentum of a volume enclosing a dark fluid, induced by cosmological perturbations. We assume that these volumes are spherical to simplify our calculations. To set up the starting expression for the angular momentum, we first provide a discussion of some well-known subtleties regarding its definition in GR (cf.~\cite{landau1975classical,thorne2000gravitation,Wald:1984rg,Poisson:2009pwt,Jaramillo:2010ay}). 

\subsection{Angular momentum in general relativity}
\label{subsec:AngularMomGR}
In flat spacetime, the mass $M$ and angular momentum $J^c$ of an isolated system can be constructed from the energy momentum tensor, and are just the components of tensors, expressed in terms of volume integrals, given by
\begin{eqnarray}
    \label{PmuJmunu}P^\mu = \int_\mathcal{V} d^3\vec{x}~T^{0\mu}\,,\quad J^{\mu\nu} = \int_\mathcal{V} d^3\vec{x}~\left(x^\mu T^{0\nu} - x^\nu T^{0\mu}\right)\,,
\end{eqnarray}
Specifically, $M = P^0$ and $J_c = 1/2 ~\epsilon_{abc}J^{ab}$. The above quantities are conserved, owing to the fact that
\begin{eqnarray}
    \partial_\mu T^{\mu\nu} = 0\,.
\end{eqnarray}
Furthermore, one can show that the total mass and angular momentum reduce into surface integrals. This follows from the linearized field equations, expanded about Minkowski spacetime, which can be written as \cite{thorne2000gravitation}
\begin{eqnarray}
\partial_\alpha\partial_\beta H^{\mu\alpha\nu\beta} = 16\pi G_{\rm N} T^{\mu\nu}\,,
\end{eqnarray}
where $H^{\mu\alpha\nu\beta}$ is quantity with the same symmetries as the Riemann tensor, and which also depends on the linearized metric
\begin{align}
    g_{\alpha\beta} \approx \eta_{\alpha\beta} + h_{\alpha\beta}\,,\quad \vert h_{\alpha\beta} \vert \ll 1\,.
\end{align}
The conservation of $T^{\mu\nu}$ trivially follows from the symmetries of $H$ and the linearized field equations. Explicitly, we then have
\begin{eqnarray}
    P^\mu &=& \frac{1}{16\pi G_{\rm N}}\oint_{\partial\mathcal{V}}\partial_i H^{\mu i0j}~n_j dS\,,\\
    J^{\mu\nu} &=& \frac{1}{16\pi G_{\rm N}}\oint_{\partial\mathcal{V}}\left(x^\mu  \partial_\beta H^{\nu \beta 0 j} - x^\nu  \partial_\beta H^{\mu \beta 0 j} + H^{\mu j0\nu} - H^{\nu j0\mu}\right)n_j dS\,.
\end{eqnarray}
Note that even in the linearized case, the integrands of the flux integrals are gauge-dependent, because we have the freedom to choose a coordinate system where the derivatives of the metric vanish. However, the surface integrals are defined at the boundary where the spacetime is Minkowski, and hence the quantities $P^\mu$ and $J^{\mu\nu}$ behave as tensors, and this presents no ambiguity regarding the values of $M$ and $J^c$. In a generic spacetime, however, the energy momentum tensor $T^{\mu\nu}$ is covariantly conserved, \textit{i.e.},
\begin{eqnarray}
    \nabla_\mu T^{\mu\nu} = \frac{1}{\sqrt{-g}}\partial_\mu\left(\sqrt{-g} T^{\mu\nu}\right) + \Gamma^\nu_{\alpha\beta}T^{\alpha\beta} = 0\,,
\end{eqnarray}
where $\nabla_\mu$ is the covariant derivative associated with the Christoffel connection $\Gamma^\nu_{\alpha\beta}$. This does not automatically imply the existence of a global conservation law. Intuitively, this can be traced back to the fact that the gravitational field itself, not just the matter fields, generates energy. Heuristically, the contribution of the gravitational field to the total energy momentum is expected to involve a product of metric derivatives, by analogy with the Newtonian case where the energy density is the square of the gradient of the gravitational potential. Then one can construct a quantity $\tau_{\mu\nu}$, which represents the stress energy of the gravitational field, such that, when combined with the matter stress energy tensor $T^{\mu\nu}$, yields a global conservation law. As an example, Landau and Lifshitz~\cite{landau1975classical} constructed (see also \cite{trautman1962conservation} for a unified treatment in terms of superpotentials \cite{ph1939uber,Goldberg:1953zz,Komar:1958wp,Katz:1996nr,Chang:1998wj}; not to be confused with the superpotential in supersymmetry)
\begin{eqnarray}
    \label{LL}\tau^\alpha_{\rm (LL)}~_\lambda + \sqrt{-g}T^\alpha_\lambda = \partial_\beta\left(\sqrt{-g}U^{\alpha\beta}_\lambda \right)\,,
\end{eqnarray}
where
\begin{eqnarray}
    U^{\alpha\beta}_\lambda \equiv \frac{1}{16\pi G_{\rm N} \sqrt{-g}}g_{\lambda\kappa}\partial_\sigma\left[g(g^{\alpha\sigma}g^{\beta\kappa}-g^{\beta\sigma}g^{\alpha\kappa})\right]\,.
\end{eqnarray}
We make a few comments about $\tau$. Notice that $U$ is antisymmetric in $\alpha$ and $\beta$, so that the left hand side of Eq.~(\ref{LL}) follows a conservation law. We also note that $\tau_{\rm (LL)}$ is not a tensor, but rather a pseudotensor, because it only transforms as a tensor under a subset of general coordinate transformations. Furthermore, as in the case of linearized gravity, there exists a coordinate system such that $U^{\alpha\beta}_\lambda = 0$, so that $\tau_{\rm (LL)}$ is inherently ambiguous and coordinate dependent; nevertheless, when $\tau$ is contracted with the normal of a timelike hypersurface, and then integrated over a volume that extends to spatial infinity, the value is unambiguous, assuming that the spacetime is asymptotically flat. Finally, the form of $\tau$ is not uniquely determined and there are other constructions that lead to a globally conserved quantity. Then from the energy momentum pseudotensor, one can define angular momentum as
\begin{eqnarray}
\label{JDefPseudo}    J_i = \epsilon_{ijk}\int_{\mathcal{V}} d^3\vec{x}~x^j \left[\tau_{\rm (LL)}^{0k} + \sqrt{-g}T^{0k}\right],
\end{eqnarray}
and this can be rewritten as a surface integral, like in the case of linearized gravity. If the spacetime is asymptotically flat, then taking the boundary to spatial infinity ensures that $J_i$ will be meaningful and unambiguous.

An alternative procedure to obtain conserved quantities is to exploit the presence of symmetries in a given spacetime. If there exists a Killing vector $\xi^\mu$ satisfying 
\begin{eqnarray}
    \nabla_\mu \xi_\nu + \nabla_\nu \xi_\mu = 0\,,
\end{eqnarray}
one can construct globally conserved quantities. Note that the quantity $T^\alpha_\beta \xi^\beta$ is covariantly conserved, which implies that
\begin{eqnarray}
    \partial_\alpha\left(\sqrt{-g}T^\alpha_\beta \xi^\beta\right) = 0\,.
\end{eqnarray}
In particular, stationary and axisymmetric spacetimes admit a timelike Killing vector $\xi_{(t)}$, associated with time translation symmetry, and a spacelike Killing vector $\xi_{(\phi)}$, associated with rotation isometry. Then the Komar mass and angular momentum are given by integrals over a codimension-2 surface, corresponding to the boundary of a timelike hypersurface $\Sigma$ with a normal vector $n^\mu$. Explicitly these are given by 
\begin{eqnarray}
    M = -\frac{1}{8\pi G_{\rm N}}\oint_{\partial \Sigma} \nabla^\mu \xi_{(t)}^\nu~dS_{\mu\nu}\,,\quad J = \frac{1}{16\pi G_{\rm N}}\oint_{\partial \Sigma} \nabla^\mu \xi_{(\phi)}^\nu~dS_{\mu\nu}\,,
\end{eqnarray}
where
\begin{eqnarray}
    dS_{\mu\nu} = \left(s_\mu n_\nu - s_\nu n_\mu\right)\sqrt{\gamma}~d^2 y
\end{eqnarray}
is the surface element on $\partial\Sigma$, with spacelike normal $s_\mu$, induced metric $\gamma$, and coordinates $y^a$ installed on it. It is assumed that the spacetime is asymptotically flat and the exterior region is vacuum. In terms of hypersurface integrals, we have
\begin{eqnarray}
   \label{KomarM} M &=& 2\int_\Sigma\left(T_{\mu\nu} - \frac{1}{2}g_{\mu\nu}T\right)n^\mu \xi_{(t)}^\nu~\sqrt{g_3} d^3\vec{x}\,,\\
   \label{KomarJ} J &=& -\int_\Sigma\left(T_{\mu\nu} - \frac{1}{2}g_{\mu\nu}T\right)n^\mu \xi_{(\phi)}^\nu~\sqrt{g_3} d^3\vec{x}\,,
\end{eqnarray}
where $g_3$ is the determinant of the induced metric on 
$\Sigma$. We note that the angular momentum in Eq.~(\ref{KomarJ}) was adopted in \cite{DeLuca:2019buf} to calculate the spin of PBHs in a flat FRW background. In the case of a perturbed FRW background however, there is no Killing vector associated with rotational symmetry to begin with. Nevertheless, globally conserved quantities can still be defined in the case of perturbed spacetimes.

\subsection{Angular momentum in a perturbed FRW spacetime}
\label{subsec:AngularMomFRW}
The discussion in Ref.~\cite{Petrov:1999tbb} will be useful for our task in determining the angular momentum of fluids subjected to cosmological perturbations, where they constructed conserved charges based on the seminal work in Ref.~\cite{Katz:1996nr}. 
Reference~\cite{Petrov:1999tbb} formulated a procedure to obtain superpotentials and conserved quantities, followed by an application of the formalism in the case of perturbed FRW backgrounds. We refer the interested reader to Ref.~\cite{Petrov:1999tbb} for the explicit expressions for the superpotentials and conserved quantities in a general curved spacetime, and instead we only quote the relevant results. Consider a flat FRW background endowed with a metric $\bar{g}_{\mu\nu} = a^2(\eta) \eta_{\mu\nu}$, and $\xi$ are chosen to be the 15 conformal Killing vectors of Minkowski spacetime---4 spacetime translations, 3 from spatial rotations, 3 from Lorentz boosts, 1 from dilation and 4 from special conformal transformations. Taking the time component of the conserved quantity $\hat{I}^\mu$, with an associated superpotential $\hat{I}^{\mu\nu}$, we have
\begin{align}
\label{I0i} 8\pi G_{\rm N} \hat{I}^{0i} &= \frac{1}{2}a^2 \left[\left(2\mathcal{H}\tilde{l}^{0i}-\partial^k q_k^i\right)\xi^0 + q_k^i \partial^k \xi^0 + Q_k^i \xi^k + \tilde{l}^0_k \partial^{[k}\xi^{i]}\right]\,,\\
  \label{I0}  \hat{I}^0 &= \partial_i \hat{I}^{0i}\,, \quad\tilde{l}^{\mu\nu} \equiv \frac{1}{a^2}\left(\sqrt{-g}g^{\mu\nu}-\sqrt{-\bar{g}}\bar{g}^{\mu\nu}\right)\,,
\end{align}
where
\begin{align}
   \mathcal{H} &\equiv \frac{1}{a}\frac{da}{d\eta}\,,\quad q_l^m \equiv \delta_l^m \tilde{l}^{00} - \tilde{l}^m_l\,,\\
   \label{qQDefs} Q_l^m &\equiv \partial^m \tilde{l}_l^0 - \partial_0 q_l^m - \delta_l^m\left[\partial^n \tilde{l}^0_n+\mathcal{H}\left(\tilde{l}^n_n+\tilde{l}^{00}\right)\right]\,.
\end{align}
Note that Eqs.~(\ref{I0i})-(\ref{qQDefs}) are exact expressions, even for ``large" perturbations of the background FRW metric $\bar{g}$. Note that the Latin indices are raised and lowered using the Kronecker delta. Typically, we are interested in the volume integral of $\hat{I}^0$ over a spherical volume at a given conformal time $\eta$. Ref.~\cite{Petrov:1999tbb} demonstrated that the conserved charges are gauge dependent, and these quantities take on simpler forms in the uniform Hubble gauge \cite{Bardeen:1980kt,kodama1984cosmological,Schmid:1998mx,hwang1999relativistic}, where the change in the trace of the extrinsic curvature of surfaces of constant conformal time, under metric perturbations, is zero. In this gauge, the perturbed FRW metric in comoving coordinates can be written as \cite{Bardeen:1980kt,Schmid:1998mx}
\begin{eqnarray}
    ds^2 = -(1+2\Psi) dt^2 +a^2(t)\left[\delta_{ij}(1+2\Phi)+2\partial_i \partial_j \gamma\right]dx^i dx^j\,.
\end{eqnarray}
One can also quickly arrive at the realization that the angular momentum due to cosmological perturbations is a gauge dependent quantity, by noting that the velocity perturbations vanish in the comoving gauge, so that the angular momentum goes to zero. Since we are particularly interested in the angular momentum, the relevant Killing vectors are $\boldsymbol{\xi}_i$ associated with rotation symmetry, with coordinate basis components,
\begin{eqnarray}
    \left(\boldsymbol{\xi}_i\right)_b = \epsilon_{iab}x^a\,.
\end{eqnarray}
The subscript $i$ refers to the direction along the $i$ axis. Then the corresponding superpotential takes a simpler form,
\begin{eqnarray}
    8\pi G_{\rm N} (\hat{I}_i)^{0j} = \frac{1}{2}a^2 \left(Q^{jk} \epsilon_{iak}x^a + \tilde{l}^{0k} \epsilon_{ikj}\right).
\end{eqnarray}
To first order in perturbations, one finds
\begin{eqnarray}
    q_l^m \simeq \tilde{h}^m_l - \delta^m_l \tilde{h}^n_n,\quad Q^m_l \simeq \left(2\mathcal{H}\tilde{h}_{00}-\partial^n\tilde{h}_{0n}\right)\delta^m_l + \partial^m\tilde{h}_{0l}-\partial_0 q^m_l,
\end{eqnarray}
where $a^2 \tilde{h}_{\mu\nu} = g_{\mu\nu} - \bar{g}_{\mu\nu}$. The corresponding density is then
\begin{eqnarray}
 (\hat{I}_i)^0 \simeq a~\delta T^{0k}~\epsilon_{ijk}x^j~\sqrt{\bar{g}_3}\,.
\end{eqnarray}
For a spherical volume $\mathcal{V}$ at constant conformal time $\eta$, the conserved charge, which is just the angular momentum, is
\begin{eqnarray}
   \label{JaFirstOrder} J_c = \int_{\mathcal{V}}a~\delta T^{0k}\epsilon_{kcl}x^l~\sqrt{\bar{g}_3} d^3\vec{x}\,.
\end{eqnarray}
We emphasize here that the angular momentum obtained in Eq.~(\ref{JaFirstOrder}), through the formalism of \cite{Petrov:1999tbb}, coincides with the starting expression in \cite{DeLuca:2019buf,Harada:2020pzb}. The latter obtained Eq.~(\ref{JaFirstOrder}) by assuming that the hypersurface integral version of the Komar formula for angular momentum, Eq.~(\ref{KomarJ}), still holds even in the perturbed FRW case by simply replacing $T_{\mu\nu}$ with the perturbed energy momentum tensor.

We proceed with the calculation of the angular momentum, for a spherical volume with comoving radius $x_0$ enclosing some fluid. Adopting Cartesian coordinates for the background FRW spacetime, we have
\begin{eqnarray}
    \sqrt{\bar{g}_3} = a^3.
\end{eqnarray}
If we assume that the perturbation to the momentum flow is a gradient of some potential $\psi$, then the integral over the spherical volume is
\begin{eqnarray}
\label{SphericalFirstOrderZero}    \int_{\mathcal{V}} \epsilon^{kal} x_l \partial_k \psi~d^3\vec{x} = \oint_{\partial\mathcal{V}} \epsilon^{kal}x_0 n_l n_k \psi~dS = 0\,.
\end{eqnarray}
It was pointed out by \cite{Peebles:1969jm,Mirbabayi:2019uph} that one must consider the second-order contribution to angular momentum in the case of spherical volumes. The determinant of the spatial part of the metric, to first order in perturbations, is given by
\begin{eqnarray}
    g_3 \approx (a^3)^2 (1+6\Phi+2\nabla^2\gamma)\,.
\end{eqnarray}
We take
\begin{eqnarray}
    \delta T_{0j} = \partial_j \psi \equiv \bar{\rho} \left(1+\delta\right) a v_j\,,
\end{eqnarray}
where $\delta$ is the fluid density contrast, and $v_j$ is a dimensionless velocity term, which arises from a potential, \textit{i.e.},
\begin{eqnarray}
    v_j = -\partial_j \theta\,.
\end{eqnarray}
Then the nonvanishing part of the integral is
\begin{eqnarray}
   \label{startingJc} \boxed{J_c \approx a~\bar{\rho} a^4  \epsilon_{ijc}\int_{x<x_0} d^3\vec{x}~x^i v^j \left(\delta +3 \Phi + \nabla^2 \gamma\right)}\,.
\end{eqnarray}
Note that the integrand consists of products of two first-order quantities. Aside from the product of the density contrast and fluid velocity, which is present in \cite{Peebles:1969jm} and \cite{Mirbabayi:2019uph}, we also have the following contributions: the products of the fluid velocity and gravitational potential, and the fluid velocity and $\nabla^2 \gamma$.

To determine the evolution of the angular momentum, it is convenient to first expand the perturbations in terms of their Fourier modes. We can express the angular momentum as an integral over the momenta $\vec{k}$ and $\vec{k}'$, and the integrand can be expressed as a product of a geometric form factor, associated with the spherical shape of the integration volume in comoving coordinates, and a product of the density and velocity potential. Defining
\begin{eqnarray}
    \hat{\psi}_k \equiv i\hat{k}\cdot \vec{v}_k = k\theta_{k}\,,
\end{eqnarray}
we have
\begin{eqnarray}
    \label{JSphereCenter}\vec{J} &=& -a~4\pi \bar{\rho} R^4 x_0\int d\Pi_k d\Pi_{k'} (\vec{k}\times \vec{k}') ~\delta_{\rm eff}(\vec{k})~\frac{\hat{\psi}(\vec{k}')}{k'}~\mathcal{F}(\vert\vec{k}+\vec{k}'\vert x_0)\,,
\end{eqnarray}
where
\begin{align}
\label{DefDeltaeff}\delta_{\rm eff}(\vec{k},\eta) &\equiv \delta(\vec{k},\eta) + 3\Phi(\vec{k},\eta) - k^2\gamma(\vec{k},\eta)\,,\\
    R(\eta) &= x_0 a(\eta)\,,\quad d\Pi_k \equiv \frac{d^3\vec{k}}{(2\pi)^3}\,,\quad \mathcal{F}(z) \equiv \frac{3z \cos z - (3-z^2)\sin z}{z^5}\,.
\end{align}
The quantity $R$ is the physical radius of the false vacuum bubble with comoving size $x_0$, at conformal time $\eta$. Following \cite{Peebles:1969jm}, we subtract the contribution from the translational motion of the sphere's center of mass (CM). We find that the angular momentum about the CM is
\begin{eqnarray}
    \vec{J}_{\rm CM}(\eta) &=& 4\pi \bar{\rho} R^5 \int d\Pi_k d\Pi_{k'}~(\vec{k}\times \vec{k}')~\delta_{\rm eff}(\vec{k},\eta)\\
    &\times& \left[\mathcal{F}(\vert\vec{k}+\vec{k}'\vert x_0) - 3\mathcal{F}(kx_0)\mathcal{G}(k'x_0)\right]\frac{\hat{\psi}(\vec{k}',\eta)}{k'}\,,
    \label{JCMmaster}
\end{eqnarray}
where the function $\mathcal{G}$ is defined as \begin{eqnarray}
    \mathcal{G}(y) &\equiv& \frac{\sin y - y \cos y}{y^3}\,.
\end{eqnarray}
Since the density and velocity perturbations are seeded by initial Gaussian curvature perturbations, the conditional probability, $P(\vec{J}\vert x_0)$, 
 of obtaining an angular momentum $\vec{J}$ for any sphere with comoving radius $x_0$ is Gaussian with zero mean, and can be constructed entirely from the two-point function $\langle J^a J^b\rangle$. This transforms properly as a tensor under spatial rotations and can be shown to be diagonal, so we can write
 \begin{eqnarray}
     P\left(\vec{J}\Big\vert x_0, \eta\right) = \frac{3}{J_{\rm CM,rms}^2(x_0,\eta)\sqrt{2\pi}}\exp\left[-\frac{3\vec{J}^2}{2J_{\rm CM,rms}^2(x_0,\eta)}\right]\,.
 \end{eqnarray}
 A similar statement can be made about the spin parameter,
 \begin{eqnarray}
     \vec{s} \equiv \frac{\vec{J}}{G_{\rm N} M^2}\,.
 \end{eqnarray}
For a spherical volume with physical radius $R$, the mass of the enclosed fluid $f$ is $M = 4\pi/3~\bar{\rho}_f R^3$. An explicit expression for the RMS value of the CM angular momentum is given by
 \begin{empheq}[box=\fbox]{align}  
\label{JCMrmsmaster} J_{\rm CM,rms}^2 &= \left(\frac{3}{4}A_{\rm s} MR\right)^2 \tilde{x}_0^2 \mathcal{C}(\tilde{x}_0, x_0)\,,\\
 \nonumber   \mathcal{C}(\tilde{x}_0, x_0) &\equiv A_{\rm s}^{-2}\int_0^\infty dz~z \int_0^\infty dz'~z'\\
 \label{CFunctionMaster}&\times\int_{-1}^1 dx~(1-x^2)~W^2(z,z';\tilde{x}_0)\Delta_\mathcal{R}^2(k)\Delta_\mathcal{R}^2(k')\,,\\
\label{DefWFunc}    W(z,z';\tilde{x}_0) &\equiv \frac{U_{\rm eff}(k)U_\psi(k')\left(\mathcal{F}_{k+k'}-3\mathcal{F}_k\mathcal{G}_{k'}\right)}{z'} - (z\leftrightarrow z')\,,
\end{empheq}
where 
\begin{eqnarray}
     z \equiv \frac{1}{\sqrt{3}}\frac{k}{\mathcal{H}}\,,\quad z' \equiv \frac{1}{\sqrt{3}}\frac{k'}{\mathcal{H}}\,,\quad x \equiv \hat{k}\cdot\hat{k}'\,.
\end{eqnarray}
We have also introduced the rescaled form of the curvature power spectrum,
\begin{eqnarray}
    \Delta_\mathcal{R}^2(k) \equiv \frac{k^3 P_\mathcal{R}(k)}{2\pi^2}\,.
\end{eqnarray}
The quantity $P_\mathcal{R}(k)$ is essentially the mean squared value of the curvature fluctuation $\mathcal{R}_k$, which should be thought of as a Gaussian random variable with zero mean. More precisely,
\begin{eqnarray}
    \label{CurvPertPk}\langle \mathcal{R}_k \mathcal{R}_{k'}\rangle = (2\pi)^3 P_\mathcal{R}(k) \delta^{(3)}(\vec{k}-\vec{k}')\,,
\end{eqnarray}
with
\begin{eqnarray}
    \label{PlanckPR} P_\mathcal{R}(k) = \frac{2\pi^2}{k^3}~A_{\rm s}\left(\frac{k}{k_{\rm s}}\right)^{n_{\rm s}-1},\,
\end{eqnarray}
where $A_{\rm s}$ is the amplitude of superhorizon curvature perturbations, $k_{\rm s}$ is the tilt scale, and $n_{\rm s}$ is the spectral index of scalar perturbations. The measured values are given by~\cite{Planck:2018vyg}:
\begin{eqnarray}
    A_{\rm s} &=& \left(2.196 \pm 0.060\right) \times 10^{-9}\,,\\
k_{\rm s} &=& \unit[0.05]{Mpc^{-1}} = \unit[3.205 \times 10^{-31}]{eV}\,,\quad n_{\rm s} = 0.9603 \pm 0.0073\,. 
\end{eqnarray}
In Eq.~(\ref{CFunctionMaster}) it is understood that, \textit{e.g.},~$\mathcal{F}_{k+k'} \equiv \mathcal{F}(\vert \vec{k}+\vec{k}'\vert x_0)$, $\mathcal{F}_k \equiv \mathcal{F}(kx_0)$ and $\mathcal{G}_k \equiv \mathcal{G}(kx_0)$. We have introduced the parameter,
\begin{eqnarray}
    \label{DefxTilde}\tilde{x}_0(\eta) \equiv \sqrt{3} \mathcal{H}x_0\,,
\end{eqnarray}
to refer to the size of the spherical volume, with comoving radius $x_0$, relative to the size of the Hubble horizon, at conformal time $\eta$ when the comoving Hubble parameter is $\mathcal{H}$. To factor out the amplitude of the curvature perturbation, which carries the statistical information about the perturbations, we have introduced the transfer functions
\begin{eqnarray}
    U_{Q_{\rm I}}(k) \equiv \frac{Q_{\rm I}(k)}{\mathcal{R}_k}\,,
\end{eqnarray}
where $Q_{\rm I}$ is any perturbation; for example, $U_{\hat{\psi}}(k) \equiv \hat{\psi}_k/\mathcal{R}_k$, $U_{\rm{eff}}(k) \equiv \delta_{\rm eff}(k)/\mathcal{R}_k$.

\subsection{Spin of primordial black holes}
Our calculation of the angular momentum and spin of false vacuum remnants is motivated by the desire to determine the spin of PBHs formed by the direct collapse of FV bubbles or by the collapse of Fermi balls. On the other hand, the spin of PBHs from cosmological perturbations has been discussed extensively in the literature, mainly in Refs.~\cite{DeLuca:2019buf,Mirbabayi:2019uph,Harada:2020pzb}, in the usual context of PBH formation through the collapse of overdensities. These sizable density perturbations in the cosmological fluid are seeded by rare peaks in the curvature power spectrum. Such enhancements in the power spectrum may be realized in certain inflationary models~\cite{Bartolo:2018rku,Clesse:2018ogk,Vaskonen:2020lbd}. However, there seems to be an apparent disagreement between the two main approaches to this mechanism. Reference~\cite{DeLuca:2019buf} argued that it is appropriate to consider ellipsoidal volumes in describing overdensities. This follows from an earlier work by \cite{heavens1988tidal} 
in which the overdensities are expanded to quadratic order about the peak and the overdensity within the volume is required to be at least a certain fraction of the peak value, which is one of the criteria to ensure the collapse into PBHs; then, in general, the boundary of this volume is triaxial. The relevant phase space of triaxial configurations and velocity shear has 16 dimensions, where the phase space variables are: overdensity; first and second derivatives of the overdensity; and the first derivative of the velocity field. The joint probability distribution over this configuration space is essentially Gaussian \cite{Bardeen:1985tr}, where the covariance matrix, which contains the two point correlators of the phase space variables, is ultimately determined by the fluid transfer functions and curvature perturbations. In choosing the gauge, they considered constant mean curvature slicing \cite{Shibata:1999zs,Harada:2013epa,Harada:2015yda}, which is equivalent to the uniform Hubble gauge, and worked in the subhorizon limit for the fluid perturbations. By considering ellipsoidal volumes, \cite{DeLuca:2019buf} showed that, starting from the Komar expression, angular momentum is a quantity that is \textit{first order} in perturbations.

In contrast, the authors of Ref.~\cite{Mirbabayi:2019uph} considered spherical volumes in their PBH spin calculation, which they have justified by referring to \cite{Bardeen:1985tr}. Firstly, they assumed that curvature perturbations $\zeta$ have rare high peaks in position space. Once these perturbations enter the horizon, they collapse into PBHs. In position space, one can construct contours of constant curvature, as shown in Figure 1 of Ref. \cite{Mirbabayi:2019uph}. The ellipticity of those contours is proportional to $\zeta_{\rm rms}/\zeta_0$, where $\zeta_0$ is the amplitude of the peak. Sending $\zeta_{\rm rms} \rightarrow 0$ brings the ellipticity to zero. The angular momentum is calculated using Eq.~(\ref{JDefPseudo}), at a time when the PBH has already formed. It was argued, in a manner similar to Eq.~(\ref{SphericalFirstOrderZero}), that the 
first-order contribution to the angular momentum is zero for a spherical volume, and thus the leading nonvanishing contribution is at \textit{second order} in perturbations. The size of the integration volume is chosen carefully: the 3-volume around the PBH must be sufficiently large such that the spacetime geometry generated by the PBH is close to being Minkowski, but small enough that it does not see the curvature due to Hubble. These criteria can be easily realized at late times, but still during the radiation dominated era of the Universe, where the Hubble horizon has grown to a larger size. They then calculated the flux of angular momentum from acoustic waves through this 3-volume, and identified it with the torque on the PBH. Because they were interested in the initial spin of PBHs, they extrapolated their result to early times. It is worth noting that they worked in the Newtonian gauge, used the subhorizon limit for the evolution of perturbations, and considered an enhanced power spectrum with a narrow Gaussian peak.
\subsection{Angular momentum due to uniform pressure}
One may wonder about the angular momentum on a closed surface due to the torque generated by a uniform fluid pressure. A simple argument can be formulated to demonstrate that the result is zero. We start by considering a generic, smooth surface, defined parametrically as
\begin{eqnarray}
    \vec{r} = \vec{\xi}(u,v)\,.
\end{eqnarray}
The area element on the surface is
\begin{eqnarray}
    d\vec{A} = \partial_u \vec{\xi}\times \partial_v \vec{\xi}~dudv\,,
\end{eqnarray}
where $\partial_u \equiv \hat{u}\cdot \vec{\nabla}$, $\partial_v \equiv \hat{v}\cdot \vec{\nabla}$. If $\Delta\vec{\xi} \equiv \vec{\xi} - \langle\vec{r}_c\rangle$ is the position vector on the surface, relative to the center of mass position $\langle\vec{r}_c\rangle$ of the bulk fluid enclosed by the surface, then the infinitesimal torque is then given by
\begin{eqnarray}
    d\vec{\tau}^i &=& p \left[\Delta \vec{\xi} \times \left(\partial_u \vec{\xi}\times \partial_v \vec{\xi}\right)\right]^i~dudv\\
    &=& \frac{p}{2}\left[\partial_v \left(\Delta \vec{\xi}\right)^2 \partial_u \Delta \xi^i - \partial_u \left(\Delta \vec{\xi}\right)^2 \partial_v \Delta \xi^i\right]~dudv\,.
\end{eqnarray}
Here we have taken $p$ to be the uniform pressure on the surface. The above expression can be shown to be a total derivative, \textit{i.e.},
\begin{eqnarray}
    d\vec{\tau}^i = \frac{p}{2}\vec{\nabla}\times \left[\Delta \xi^i \vec{\nabla}\left(\Delta \vec{\xi}\right)^2\right]\cdot d\vec{A}\,.
\end{eqnarray}
One can use Stokes' theorem to integrate the torque over the closed surface, and show that the result is zero.

\section{Evolution of cosmological perturbations}
\label{sec:CosmoPertUHG}
The angular momentum of a spherical volume containing several fluid components can be computed once the perturbations on each fluid component and the gravitational potentials $\Phi$ and $\gamma$ are known. We identify the periods in cosmological history that are crucial for us: the conformal time $\eta=\eta_0 \simeq 0$ at which all perturbations are superhorizon, the critical point $\eta=\eta_{\rm c}$, and the onset of percolation $\eta=\eta_*$. In what follows, quantities evaluated at the percolation time are indicated with a subscript *. Here we aim to calculate only the perturbations, and eventually the angular momentum, during $[0, \eta_{\rm c}]$ and $[\eta_{\rm c}, \eta_*]$. The subsequent evolution, which will result in either the direct collapse to a PBH or the formation of a FB before collapsing into a PBH, is described by different dynamics beyond the scope of our study.

Following Ref.~\cite{Schmid:1998mx}, and working in the UHG, we can track the evolutions of $\delta$, $\hat{\psi}$, and $\Phi$. For a fluid component $f$, the equation of state and sound speeds, defined as
\begin{eqnarray}
    w_f \equiv \frac{p_f}{\rho_f}\,,\quad c_{s,f}^2 \equiv \left(\frac{\partial p_f}{\partial \rho_f}\right)_s = \frac{dp_f/dT}{d\rho_f/dT}\,,
\end{eqnarray}
must be known a priori. The zeroth order continuity equation gives
\begin{eqnarray}
\label{ZerothOrder}    \sum_f \left[\rho_f' + 3\mathcal{H}(1+w_f) \rho_f\right] = 0\,,
\end{eqnarray}
while the first order perturbations evolve via
\begin{eqnarray}
\label{ContinuityEq}    \frac{\delta_f'}{\mathcal{H}} - \frac{k}{\mathcal{H}}\hat{\psi}_f + 3(1+w_f) \Psi_k + 3(c_{s,f}^2 - w_f)\delta_f &=& 0\,,\\
   \label{EulerEq} \frac{\hat{\psi}_f'}{\mathcal{H}}+(1-3w_f)\hat{\psi}_f+c_{s,f}^2 \frac{k}{\mathcal{H}}\delta_f+(1+w_f)\frac{k}{\mathcal{H}}\Psi_k &=& 0\,.
\end{eqnarray}
Note that in Eqs.~(\ref{ZerothOrder})-(\ref{EulerEq}), the prime denotes differentiation with respect to the conformal time $\eta$. In Fourier space, the analog of the Poisson equation which gives the gravitational potential $\Psi_k$ is
\begin{eqnarray}
  \label{PoissonEq}  \left[\left(\frac{k}{\mathcal{H}}\right)^2+\frac{9}{2}(1+w)\right]\Psi_k + \frac{3}{2}(1+3c_s^2) \delta = 0\,,
\end{eqnarray}
where
\begin{eqnarray}
    w \equiv \frac{\sum_f w_f \rho_f}{\sum_f \rho_f}\,,\quad c_s^2 \equiv \frac{\sum_f c_{s,f}^2 \rho_f'}{\sum_f \rho_f'}\,,\quad \delta \equiv \frac{\sum_f \delta_f \rho_f}{\sum_f \rho_f}\,.
\end{eqnarray}
The gravitational potential $\Phi$ can be obtained from the linearized Einstein field equations in the UHG via \begin{eqnarray}
    \nabla^2 \Phi = -4\pi G_{\rm N} a^2 \sum_f \rho_f \delta_f = -\frac{3}{2}\mathcal{H}^2 \frac{\sum_f \rho_f \delta_f}{\sum_f \rho_f}\,,\quad \mathcal{H} \equiv \frac{a'}{a}\,,
\end{eqnarray}
where we used the Friedmann equation in the last equation. The metric perturbation $\gamma$ in Fourier space can be obtained from
\begin{eqnarray}
   \label{gammaPertEvol} \left(-k^2 \gamma_k\right)' = \frac{9}{2}\frac{\mathcal{H}^2}{k} \frac{\sum_f \rho_f \hat{\psi}_f}{\sum_f \rho_f}\,,
\end{eqnarray}
and we couple Eq.~(\ref{gammaPertEvol}) with Eqs.~(\ref{ContinuityEq}) and (\ref{EulerEq}). We note that if the perturbations $\delta_f$ and $\theta_f$ across fluids are of the same order, then the fluid component that contributes dominantly to the total background energy density is expected to contribute the most to $\Phi$ and $\gamma$. Typically it is assumed that the dominant fluid component is the SM plasma, but this condition may be relaxed.

\subsection{Evolution of perturbations: $\eta_0$ to $\eta_{\rm c}$}
\label{subsec:EvolPert0ToCrit}
\begin{figure}[t]
    \centering
    \includegraphics[scale=0.32]{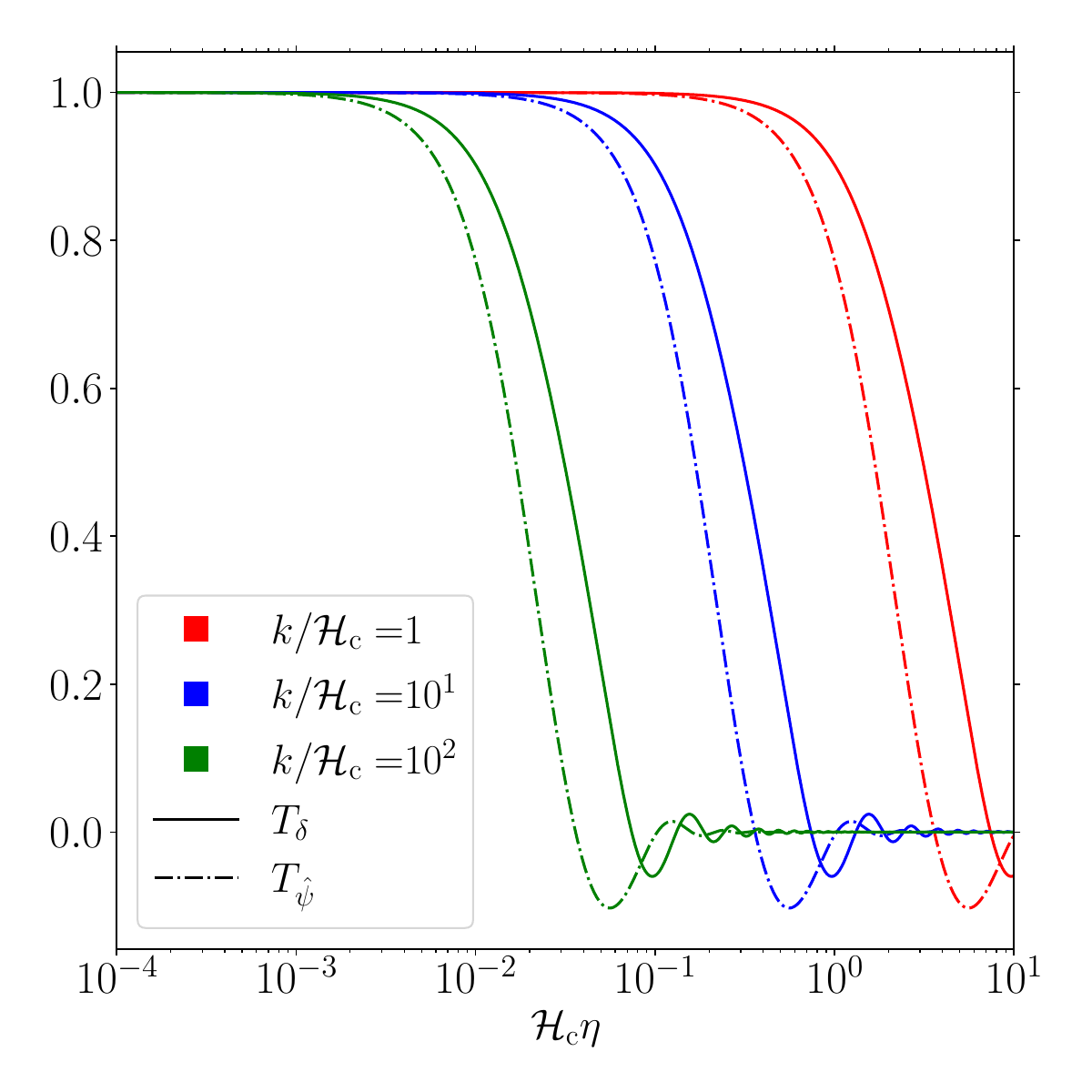} 
 \caption{The transfer functions $T_\delta$ and $T_{\hat{\psi}}$ in a radiation dominated background cosmology, for different perturbation modes.}
    \label{fig:PerturbationsExact}
\end{figure}

In the radiation dominated era of the early Universe, we have
\begin{eqnarray}
    \mathcal{H} = \frac{1}{\eta}\,,\quad 0 < \eta < \eta_{\rm c}\,.
\end{eqnarray}
Note that the SM fluid component has $w = c_s^2 = 1/3$. On the other hand, the $\chi$ and $\phi$ fluid components constitute a tightly coupled system and should be treated as a single fluid component, similar to the baryon-photon fluid. In the following, we refer to this fluid as the dark plasma $\rm{D} = \chi+\phi$. Before the onset of the FOPT, the dark plasma still has $w = c_s^2 = 1/3$. Noting that the SM plasma has a temperature $T_{\rm SM}$ that is assumed to be much larger than the dark plasma temperature $T$, we have $\rho_{\rm SM} \gg \rho_{\rm D}$. Thus, $\delta \approx \delta_{\rm SM}$, so that perturbations in the SM plasma mainly dictate the evolution of $\Psi_k$. Then
\begin{eqnarray}
    \Psi_k \simeq -\frac{\delta_{\rm SM}}{2+(k\eta)^2/3}\,.
\end{eqnarray}
The evolution of the SM radiation fluid perturbations are given by
\begin{eqnarray}
    \delta_{\rm SM}' &=& k \hat{\psi}_{\rm SM} - \frac{4}{\eta}\Psi_k\,,\\
\hat{\psi}_{\rm SM}' &=& -\frac{k}{3} \delta_{\rm SM} - \frac{4}{3}k \Psi_k\,.
\end{eqnarray}
The initial conditions are set by the primordial curvature perturbation $\mathcal{R}_k(0)$ associated with the comoving wavenumber $k$, when modes are superhorizon. In the case where the background Hubble evolution is determined solely by a component with constant $w$, and with a constant sound speed $c_s^2 = w$, the evolution equations for the perturbations admit exact solutions. For the SM plasma with $w = c_s^2 = 1/3$, one can show that the density and velocity perturbations in the UHG are \cite{Hwang:1993ai}
\begin{eqnarray}
    \label{DensPsi}\delta_{\rm SM}(k,\eta) = 2\mathcal{R}_k(0) z^2 T_\delta(z)\,,\quad \hat{\psi}_{\rm SM}(k,\eta) = \frac{2\mathcal{R}_k(0)}{3\sqrt{3}}z^3 T_{\hat{\psi}}(z)\,,
\end{eqnarray}
where
\begin{eqnarray}
 \label{Tdeltapsi}   T_\delta(z) \equiv \frac{2(2\sin z - z\cos z)}{z(2+z^2)}\,,\quad T_{\hat{\psi}}(z) \equiv \frac{6[2z \cos z + (z^2 - 2)\sin z]}{z^3(2+z^2)}\,,\quad z \equiv \frac{k\eta}{\sqrt{3}}\,.
\end{eqnarray}
It also follows that
\begin{eqnarray}
\label{PhiExact}    \Phi_k \simeq \frac{1}{2z^2} \delta_{\rm SM}(k) = \mathcal{R}_k(0) T_\delta(z)\,.
\end{eqnarray}
Observe that the perturbations in the UHG, given by Eqs.~(\ref{DensPsi}) and (\ref{PhiExact}) are nonsingular in the superhorizon regime $z \rightarrow 0$, and this is another advantage of adopting the UHG, as pointed out by, \textit{e.g.},~\cite{Bardeen:1980kt} and \cite{Schmid:1998mx}. In fact, as shown in Fig.~\ref{fig:PerturbationsExact}, the transfer functions are close to unity at early times, \textit{i.e.},~when the modes are superhorizon. As for $\nabla^2\gamma$ in Fourier space, we can obtain it by integrating the ordinary differential equation (ODE),
\begin{eqnarray}
    \frac{d}{dz}\left(-k^2 \gamma_k\right) = \frac{3\sqrt{3}}{2}\frac{\hat{\psi}_{\rm SM}(z)}{z^2}\,,
\end{eqnarray}
where we set the initial condition to be such that $-k^2 \gamma_k = 0$ at $z=0$.  
When the mode enters the horizon, \textit{i.e.},~when $k \sim \mathcal{H} = 1/\eta$, the perturbations start to oscillate as they feel the presence of the radiation fluid with a nonzero sound speed. One can similarly write down the evolution equations for $\delta_{\rm D}$ and $\hat{\psi}_{\rm D}$, and find that
\begin{eqnarray}
    \delta_{\rm D} \simeq \delta_{\rm SM},\quad \hat{\psi}_{\rm D} \simeq \hat{\psi}_{\rm SM}\,.
\end{eqnarray}
\subsection{Evolution of perturbations: $\eta_{\rm c}$ to $\eta_*$}
\label{subsec:EvolPertCritToPerc}
\subsubsection{Background evolution}
\label{sec:BckgCosmoEvolVac}
From the critical temperature up to percolation, the usual cosmological evolution of the Universe is modified by FOPT dynamics. The FOPT is triggered mainly by the effective potential given in Eq.~(\ref{VeffDefn}) developing a nonzero vacuum expectation value.

During the FOPT, the equation of state of the dark fluid is modified, which will affect the evolution of the density and velocity perturbations in the dark sector. This requires us to first track the background cosmological evolution during the FOPT. Furthermore, tracking the background cosmological evolution will allow us to determine relevant physical quantities at the percolation time. The total energy density that drives the cosmological expansion is given by
\begin{eqnarray}
    \rho =\rho_{\rm SM}(t) + \rho_{\rm D}(t)\,,\quad \rho_{\rm D}(t) \equiv \rho_{\rm FV}(T(t)) - \left[1 - F(t)\right] \Delta \rho(t)\,,
\end{eqnarray}
where $F(t)$ is the fraction of the Universe in the false vacuum. We can write the energy density of the SM radiation component as
\begin{eqnarray}
    \rho_{\rm SM} &=& \frac{\pi^2}{30}g_{\rho\rm,SM}(T_{\rm SM}) T_{\rm SM}^4\,,\quad \rho_{\rm SM,c} = \frac{\pi^2}{30}g_{\rho\rm, SM,c}T_{\rm SM,c}^4\,.
\end{eqnarray}
The value of $g_{\rho,\rm SM}(T_{\rm SM})$ is taken from Ref.~\cite{Drees:2015exa}. The reduced Hubble parameter, which is obtained by normalizing the Hubble parameter relative to its value at the critical temperature, is given by
\begin{eqnarray}
    h^2(t) \equiv \frac{H^2(t)}{H^2(t_{\rm c})} = \frac{\rho_{\rm SM}+\rho_{\rm FV} - (1-F)\Delta\rho}{\rho_{\rm SM,c}+\rho_{\rm FV,c}}\,.
\end{eqnarray}
Assuming that the bubbles expand as in a \textit{detonation} process, the shock front trails behind the expanding bubble wall, so that reheating occurs only in the true vacuum regions. Only when the bubbles collide is there a sudden release of latent heat. Since we do not model the dynamics of bubble collisions, we simply take 
\begin{eqnarray}
 \label{TDarkDilute}   \dot{T} = -HT\,,
\end{eqnarray}
 which is consistent with the continuity equation in the period before the latent heat is suddenly released. As for the false vacuum fraction $F(t)$, it can be obtained by knowing the nucleation rate $\Gamma(T)$, the bubble wall velocity $v_{\rm w}$, and the scale factor for all $t'$ such that $t_{\rm c} \leq t' \leq t$. Then
\begin{eqnarray}
\label{DefFVFrac}    -\ln F(t) = \frac{4\pi}{3} \int_{t_{\rm c}}^t dt'~\Gamma(T(t'))~a^3(t') r^3(t,t')\,,\quad r(t,t') \equiv \int_{t'}^t dt''~\frac{v_{\rm w}(t'')}{a(t'')}\,.
\end{eqnarray}
The wall velocity is as an important ingredient in tracking the false vacuum fraction. In particular, earlier 
work~\cite{Heckler:1994uu,Megevand:2000da,Leitao:2014pda} incorporated the evolution of the bubble wall velocity, where reheating effects in the case of deflagrations will slow down the advancing bubble wall. Recent work, \textit{e.g.},~\cite{Ai:2021kak,Ai:2023see}, provided prescriptions to obtain the bubble wall velocity in a model-independent manner. Here we take a simplified approach, following Ref.~\cite{kampfer1988phenomenologicalPaper2}, where 
\begin{eqnarray}
\label{vwPrescription}    v_{\rm w} = 1 - \left(\frac{T}{T_{\rm c}}\right)^{n_{\rm w}}\,,\quad n_{\rm w} = 8\,.
\end{eqnarray}
We adopt this temperature-dependent velocity profile to satsify the physical condition that if a phase boundary existed at $T_{\rm c}$, it would be stationary because $\Delta p=0$ when the two phases are in equilibrium. Equivalently, $v_{\rm w} \propto \Delta V(T) \to 0$ as $T \to T_{\rm c}$ from below. We have checked that quantities that do not explicitly depend on the wall velocity, such as the percolation temperature and inverse FOPT duration, marginally depend on $n_{\rm w}$. We will find that the spin of FV bubbles scales roughly as the inverse square of the wall velocity. As we remarked in Section \ref{sec:DarkSectorModel}, the nucleation rate depends mainly on the particle physics model. The exact connection lies in the following expression
\begin{eqnarray}
    \Gamma(T) = T^4 \left[\frac{S_3(T)}{2\pi T}\right]^{3/2}\exp\left[-\frac{S_3(T)}{T}\right]\,,
\end{eqnarray}
where $S_3(T)$ is the bounce action associated with the scalar field configuration that satisfies the equation of motion. In the case of quartic effective potentials for a single real scalar field, a semianalytic expression has been formulated in Ref.~\cite{Adams:1993zs}, which we adopt in this work. For our effective potential, we take $S_3$ to be
\begin{eqnarray}
\label{S3SemiAnalytic}    S_{3}(T) &\equiv& \frac{\pi a(T)}{\bar{\lambda}^{3/2}}\frac{8\sqrt{2}}{81}\left[2-\delta_3(T)\right]^{-2}\sqrt{\frac{\delta_3(T)}{2}}\left[\beta_1 \delta_3(T)+\beta_2 \delta_3^2(T)+\beta_3 \delta_3^3(T)\right]\,,\\
    \beta_1 &=& 8.2938\,,\quad \beta_2 = -5.5330\,,\quad \beta_3 = 0.8180\,,
\end{eqnarray}
where
\begin{eqnarray}
    \delta_3(T) \equiv \frac{8\bar{\lambda}b(T)}{a^2(T)}\,,\quad \bar{\lambda} \equiv \frac{\lambda}{4}\,,\quad a(T) = AT+C,\quad b(T) \equiv D\left(T^2 - T_0^2\right)\,,\quad D>0\,.
\end{eqnarray}
We can now formulate the system of coupled first order ODEs which describe the background evolution in the false vacuum during the FOPT. The ODEs track $x \equiv H_{\rm c}(t-t_{\rm c}), y \equiv T/T_{\rm c}, y_{\rm SM} \equiv T_{\rm SM}/T_{\rm SM,c}, F$, the derivatives of $F$, and $\tilde{\eta} \equiv \mathcal{H}_{\rm c} (\eta-\eta_{\rm c})$. We set $a = 1$ at the critical point. This then implies that $H_{\rm c} = \mathcal{H}_{\rm c}$. Taking the scale factor $a$ to be the independent variable, one can show that
\begin{empheq}[box=\fbox]{align}
   \label{dxda} \frac{dx}{da} &= \frac{1}{ah}\,,\\
   \frac{dy}{da} &=-\frac{y}{a}\,,\\
   \frac{dy_{\rm SM}}{da} &= -\frac{y_{\rm SM}}{a}\frac{1}{1+\frac{1}{3}\frac{d\ln g_{s\rm, SM}}{d\ln T_{\rm SM}}}\,,\\
        \label{g3}\frac{d(\ln F)}{da} &= \frac{g_1}{a}v_{\rm w}~\frac{1}{ah}\,,\\
\label{gRecursive}\frac{dg_{i+1}}{da} &= \frac{g_i}{a}v_{\rm w}~\frac{1}{ah}\,,\quad 1 \leq i \leq 2\\    
      \label{g1}\frac{dg_3}{da} &= -8\pi a^3 \frac{\Gamma(T_{\rm c} y)}{H_{\rm c}^4}~\frac{1}{ah}\,,\\
        \label{detada} \frac{d\tilde{\eta}}{da} &= \frac{1}{a^2 h}\,,
\end{empheq}
where
\begin{eqnarray}
    g_n(t) \equiv -\frac{4\pi}{3}~\frac{3!}{(3-n)!} H_{\rm c}^{4-n} \int_{t_{\rm c}}^t dt'~\frac{\Gamma(T(t'))}{H_{\rm c}^4}a^3(t')~r^{3-n}(t,t')\,.
\end{eqnarray}
The $g_n$s are dimensionless quantities that are related to the derivatives of $\ln F$, and can also be defined recursively via Eq.~(\ref{gRecursive}). A similar system of ODEs, in particular that converts the integro-differential equation into a system of ODEs, was obtained in Ref.~\cite{kampfer1988phenomenologicalPaper1} and recently in Ref.~\cite{Flores:2024lng}. Note that our system of ODEs has one less equation than Ref.~\cite{Flores:2024lng}, if we do not include Eq.~(\ref{detada}) in the counting, which only tracks the elapsed conformal time from the critical point in units of $H_{\rm c}$. The initial conditions are
\begin{eqnarray}
   x = 0\,, \quad y = 1\,,\quad F = 1\,,\quad g_i = 0\,,
\end{eqnarray}
for $a = 1$. We check that the physical volume in the false vacuum regions is decreasing, at least at the percolation time. This can be expressed as \cite{Turner:1992tz,Athron:2022mmm}
\begin{eqnarray}
\frac{d}{dt}(a^3 F) &=& H F a^3 \left[3 + \frac{1}{H}\frac{d(\ln F)}{dt} \right]\\
&=& H F a^3 \left(3 + \frac{g_3 v_{\rm w}}{ah}\right) < 0\,.
\end{eqnarray}
\begin{figure}[t]
    \centering
    \begin{tabular}{cc}
        \includegraphics[scale=0.25]{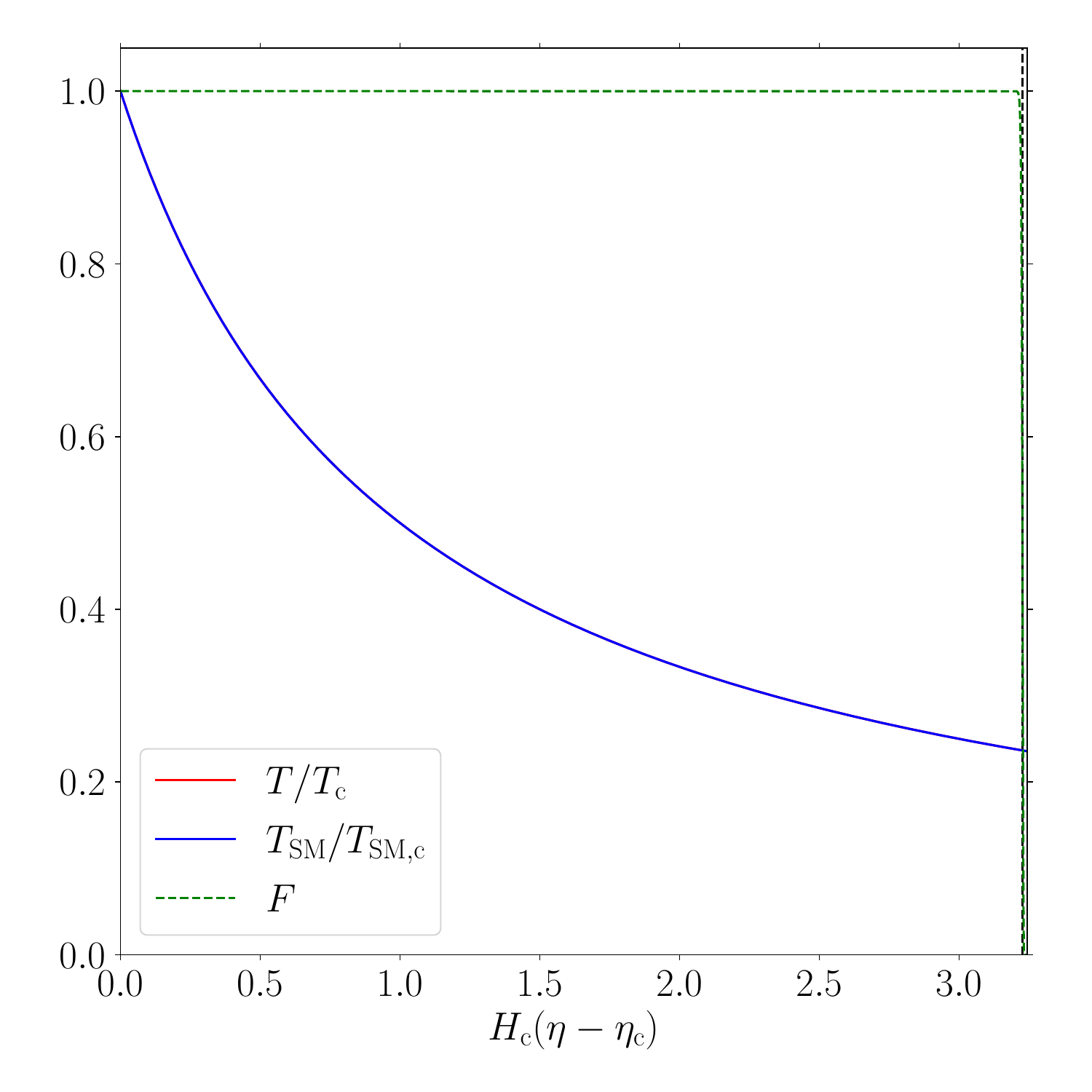} &  \includegraphics[scale=0.25]{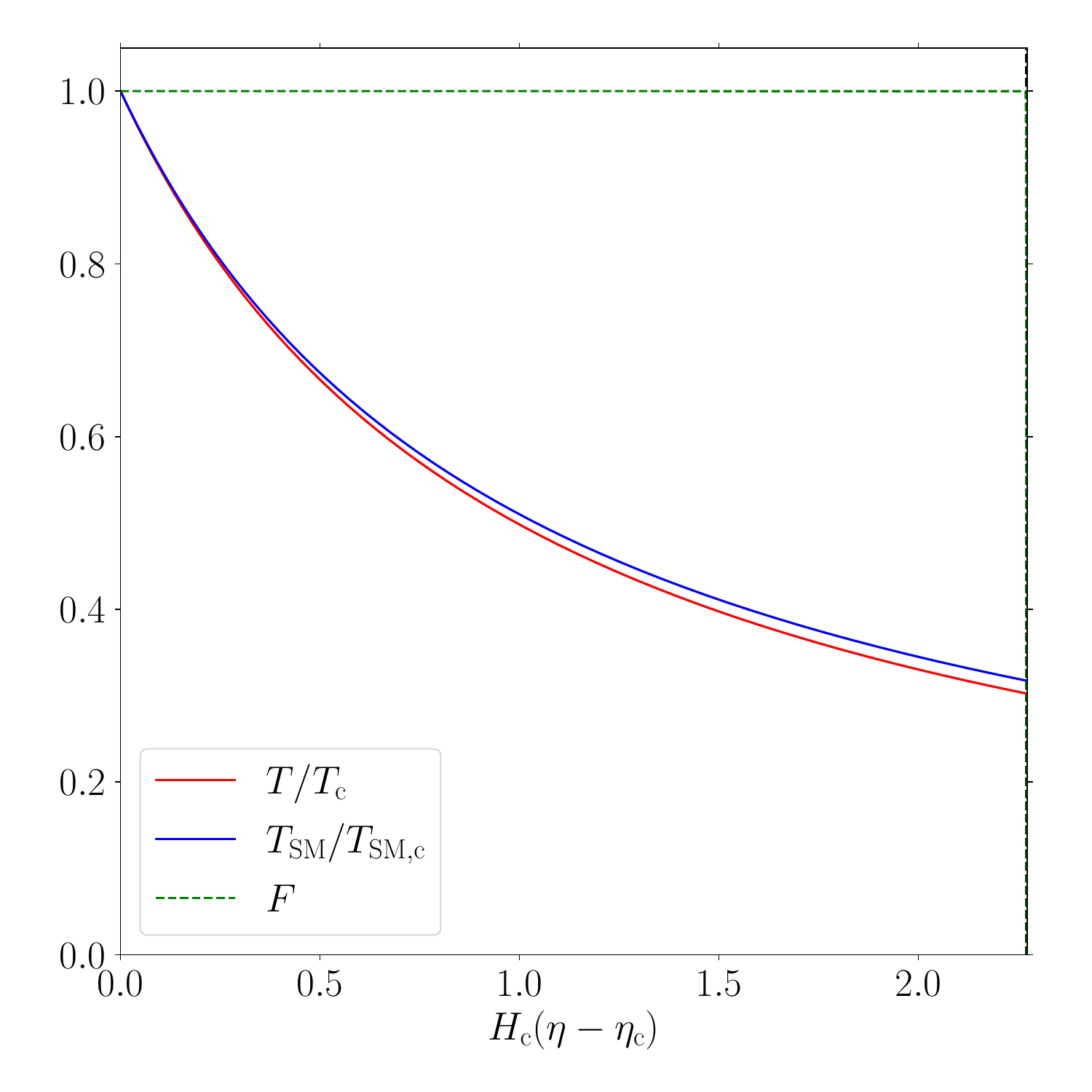}\\
          \includegraphics[scale=0.25]{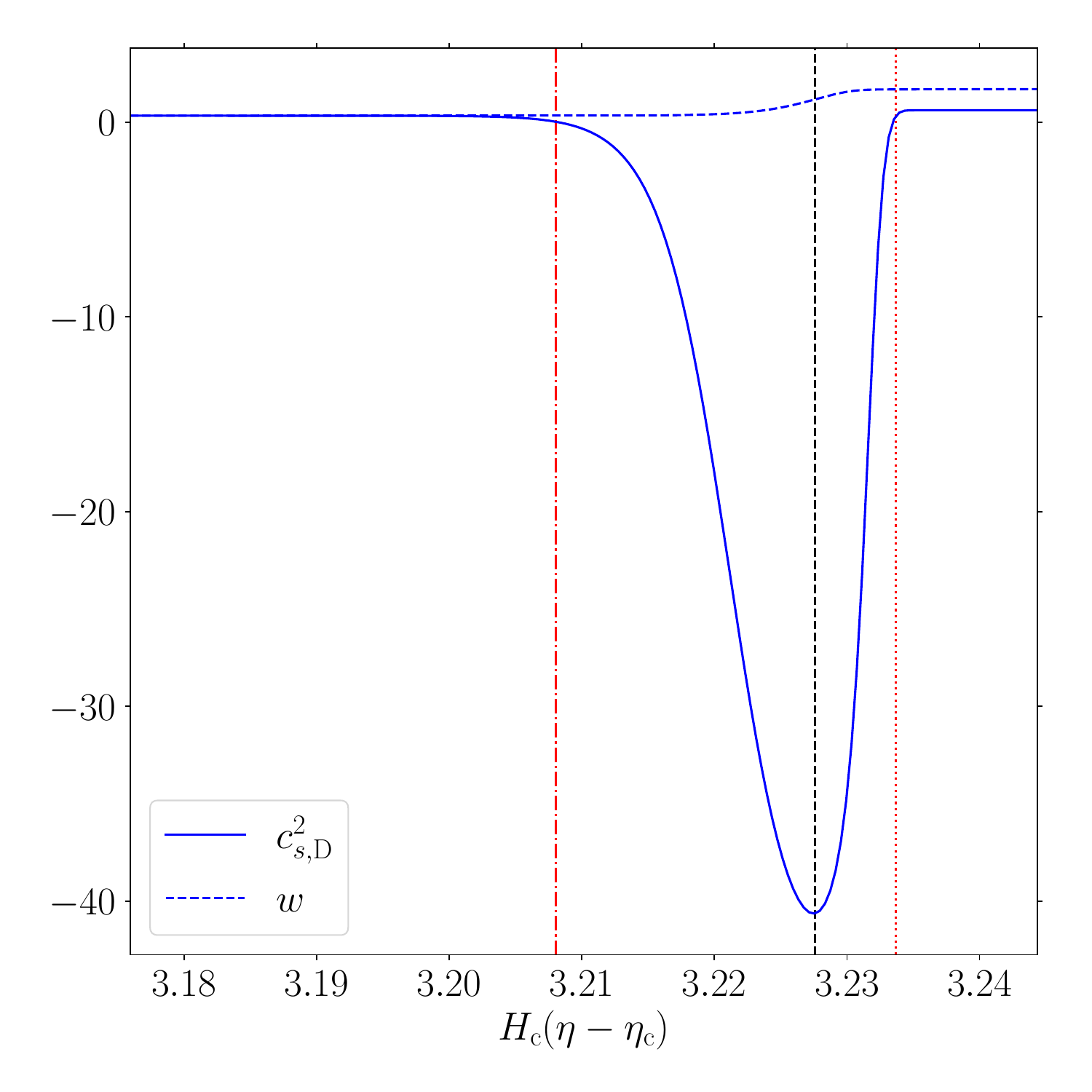} &  \includegraphics[scale=0.25]{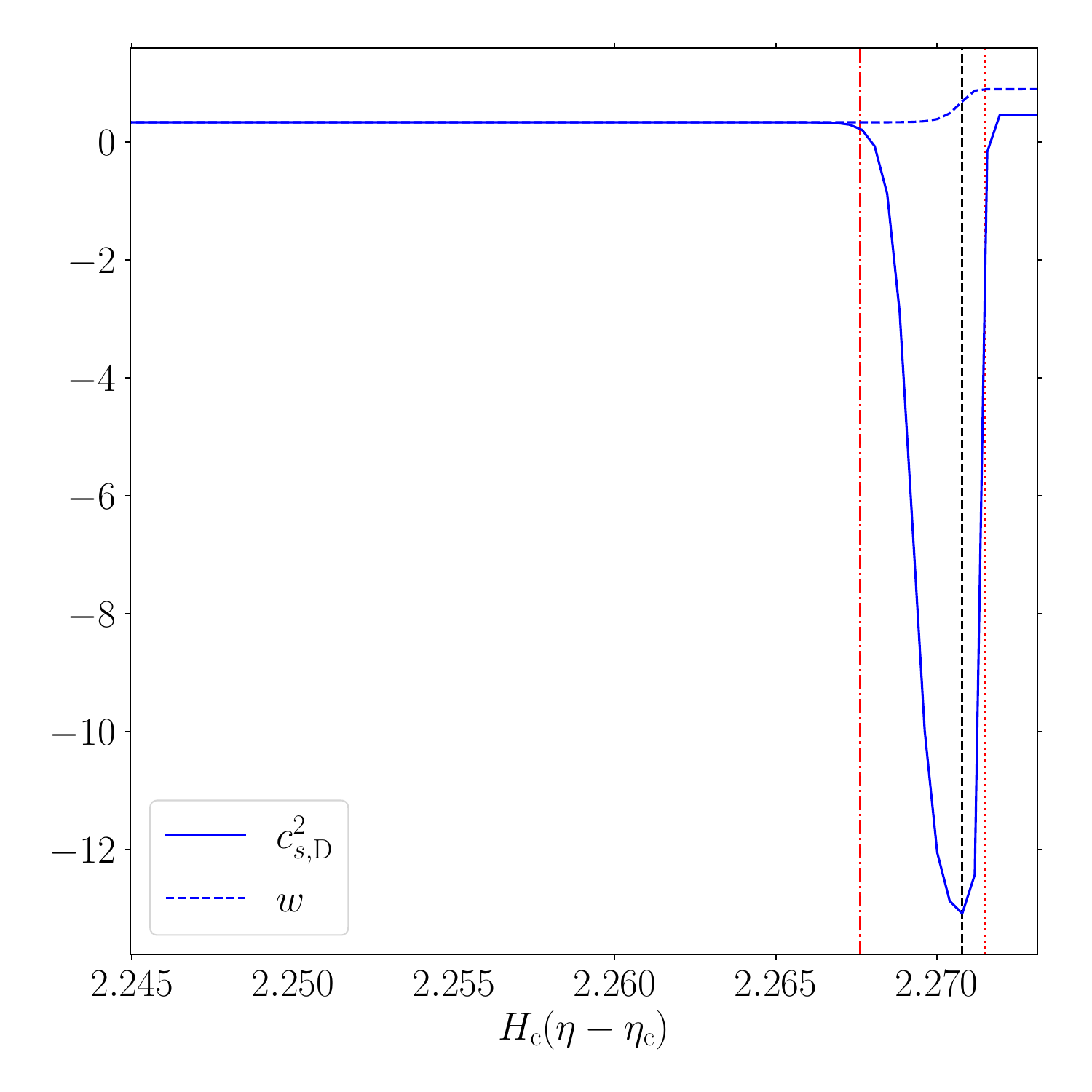}\\
          (a) & (b) 
    \end{tabular}
    \caption{Evolution of the dark/visible sector temperatures and FV fraction (top row), and of $c_{s\rm,D}^2$ and $w_{\rm D}$ (bottom row), for BP-2 (column a) and BP-4 (column b). The vertical black dashed lines indicate the percolation temperature in all panels, while the vertical
dot-dashed and dotted lines respectively indicate $\tilde{\eta}_i$ and $\tilde{\eta}_f$ in the bottom panels.}
    \label{fig:SampleEvol}
\end{figure}
For illustration purposes, in Fig.~\ref{fig:SampleEvol} and in the following figures, we feature two specific FOPT scenarios: BP-2 and BP-4 from Table~\ref{table:FinalBPsOld}. For each case, the evolution of the temperature (solid) and false vacuum fraction (dashed) versus the rescaled elapsed conformal time $H_{\rm c}(\eta - \eta_{\rm c})$, are shown in the top row plots. The labeling is based on Table \ref{table:FinalBPsOld}, where we list the benchmark FOPT scenarios of interest.
\begin{table}[t]
\resizebox{\textwidth}{!}{\begin{tabular}{ccccccc}\toprule
~&{\bf BP-1}&{\bf BP-2}&{\bf BP-3}&{\bf BP-4}&{\bf BP-5}&{\bf BP-6}\\\midrule
$A$& $10^{-3}$& $10^{-3}$& $10^{-3}$& $10^{-4}$& $10^{-4}$& $10^{-4}$\\
$x_C$& $1.1063\times10^{-4}$& $2.4432\times10^{-4}$& $2.4131\times10^{-4}$& $1.712\times10^{-5}$& $3.97\times10^{-7}$& $8.19\times10^{-6}$\\
$\lambda$& $3.6917\times10^{-4}$& $3.6917\times10^{-4}$& $4.8063\times10^{-4}$& $1.7136\times10^{-5}$& $1.7136\times10^{-5}$& $2.2309\times10^{-5}$\\
$L_{\rm c}/T_{\rm c}^4$& $5.0456\times10^{-2}$& $9.0072\times10^{-2}$& $3.998\times10^{-2}$& $7.2479\times10^{-2}$& $2.8264\times10^{-2}$& $1.9733\times10^{-2}$\\
$T_{\rm c}$ (MeV)& $10^{3}$& $10^{-2}$& $10$& $10^{3}$& $10$& $10^{5}$\\\midrule
$r_{T\rm ,c}$& 0.4& 0.4& 0.4& 0.4& 0.1& 0.4\\
$T_*$ (MeV)& $3.4306\times10^{2}$& $2.3654\times10^{-3}$& 2.4216& $3.0246\times10^{2}$& 3.8815& $3.3936\times10^{4}$\\
$T_{\rm SM,*}$ (MeV)& $8.9392\times10^{2}$& $5.9135\times10^{-3}$& 6.2448& $7.9351\times10^{2}$& $4.1986\times10^{1}$& $8.5999\times10^{4}$\\
$\Delta N_{\rm eff}$& $3.4901\times10^{-3}$& 0.2496& $4.7392\times10^{-2}$& $3.4165\times10^{-3}$& $1.0787\times10^{-4}$& $2.5327\times10^{-3}$\\
$\sqrt{3}H_* R_*$& $1.1716\times10^{-3}$& $1.7169\times10^{-3}$& $2.3446\times10^{-3}$& $2.9002\times10^{-4}$& $2.2558\times10^{-4}$& $3.4178\times10^{-4}$\\
$M_*/M_\odot$& $2.88\times10^{-14}$& 0.2359& $8.73\times10^{-8}$& $5.44\times10^{-16}$& $3.8\times10^{-15}$& $5.3\times10^{-20}$\\
$\alpha_*$& $1.2\times10^{-5}$& $8.1792\times10^{-4}$& $3.19\times10^{-5}$& $5.1\times10^{-5}$& $2.71\times10^{-7}$& $4.99\times10^{-6}$\\
$\beta_*/H_*$& $2.2962\times10^{3}$& $1.5653\times10^{3}$& $1.1432\times10^{3}$& $9.309\times10^{3}$& $1.3535\times10^{4}$& $8.1307\times10^{3}$\\
$v_{\rm w,*}$& 0.9998& 1& 1& 0.9999& 0.9995& 0.9998\\
$s_{\rm rms,*}/(v_{\rm eff,*}/A_{\rm s})$& 7.3349& 0.1379& 0.2581& $1.2304\times10^{2}$& $1.1132\times10^{4}$& $1.0651\times10^{2}$\\
$\tilde{\eta}_* - \tilde{\eta}_i$& $1.2187\times10^{-2}$& $2.5631\times10^{-2}$& $3.2774\times10^{-2}$& $3.8714\times10^{-3}$& $2.08\times10^{-3}$& $3.9741\times10^{-3}$\\
$\tilde{\eta}_*$& 1.8899& 3.2276& 3.1085& 2.2708& 1.5419& 1.9378\\\midrule
$c_{s\rm ,D*}^2$& -$1.664\times10^{1}$& -$4.0616\times10^{1}$& -$4.0832\times10^{1}$& -$1.3081\times10^{1}$& -6.9385& -$1.4892\times10^{1}$\\
$\mathcal{C}_*$& 0.8866& $3.555\times10^{-2}$& $1.866\times10^{-3}$& $3.0621\times10^{1}$& $5.5423\times10^{1}$& 2.2367\\
$v_{\rm eff,*}/A_{\rm s}$& $8.2735\times10^{-4}$& $2.4278\times10^{-4}$& $7.6\times10^{-5}$& $1.2037\times10^{-3}$& $1.2595\times10^{-3}$& $3.8337\times10^{-4}$\\
$k_{\rm BH}/\mathcal{H}_{\rm c}$& $7.3322\times10^{2}$& $1.7558\times10^{2}$& $1.8489\times10^{2}$& $1.7493\times10^{3}$& $5.9323\times10^{3}$& $2.5752\times10^{3}$\\
$k_{\rm cut}/\mathcal{H}_{\rm c}$& $5.6035\times10^{2}$& $1.7729\times10^{2}$& $1.3711\times10^{2}$& $1.7669\times10^{3}$& $3.7758\times10^{3}$& $1.8512\times10^{3}$\\
$k_{\rm BH}/k_{\rm cut}$& 1.3085& 0.9904& 1.3485& 0.9900& 1.5711& 1.3911\\
$s_{\rm rms}$& $6.0685\times10^{-3}$& $3.3471\times10^{-5}$& $1.96\times10^{-5}$& 0.1481& $1.4021\times10^{1}$& $4.0831\times10^{-2}$\\\bottomrule
\end{tabular}}
\caption{\label{table:FinalBPsOld}Selection of benchmark points from our scan.}
\end{table}
We mark the moment of percolation with a vertical dashed line. In both cases, we have taken $r_{T\rm,c} = 0.4$. For these two cases, the percolation temperatures are such that $T_*/T_{\rm c} < 0.9$, and the phase transition strength $\alpha_*$ is at least $10^{-7}$. Here we define $\alpha_*$ as
\begin{eqnarray}
\label{alphaBag}\alpha_* &\equiv& \frac{1}{\rho_{\rm R, tot}}\left(1 - T\frac{d}{dT}\right)\Delta V(T)\Bigg\vert_{T = T_*},\quad \rho_{\rm R,tot}(T_*) = \frac{\pi^2}{30}T_*^4\left[g_\rho + \frac{g_{\rho,{\rm SM}}(T_*)}{r_{T\rm, *}^4}\right]\,.
\end{eqnarray}
 The conformal time elapsed from the critical point to the percolation time can be estimated by first noting that $T \propto 1/a$ and $a \propto \eta$ in the radiation-dominated era. Then
\begin{eqnarray}
H_{\rm c}(\eta_* - \eta_{\rm c}) \approx \frac{1 - T_*/T_{\rm c}}{T_*/T_{\rm c}}\,.
\end{eqnarray}
From the above estimate, a lower percolation temperature implies a longer conformal time duration between the critical point and percolation. The percolation temperature, in turn, can be estimated from Eq.~(\ref{DefFVFrac}) by approximating
\begin{align}
    \label{GammaBeta}\Gamma(T(t')) \approx \Gamma(t) \exp\left[\beta(t)(t-t')\right],\quad \frac{\beta(t)}{H} \equiv \frac{1}{H}\frac{\dot{\Gamma}}{\Gamma} \simeq T \frac{d}{dT}\left(\frac{S_3}{T}\right)\,,
\end{align}
where $\beta(t)$ is the instantaneous inverse time scale of the FOPT at time $t$. Assuming that the scale factor is approximately constant and $\beta(t)$ varies slowly within the period of interest, we have~\cite{Enqvist:1991xw}
\begin{align}
 \label{EstimateFVFraction} -\ln F(t) &\approx 8\pi \frac{v_{\rm w}^3}{\beta^4(t)} \Gamma(t)\,.
\end{align}
At percolation, we then have
\begin{eqnarray}
\label{EstimatePerc} 1 \approx 8\pi v_{\rm w,*}^3 \frac{\Gamma_*}{\beta_*^4} \Rightarrow \exp\left[\frac{S_3(T_*)}{T_*}\right] \approx 8\pi\left[\frac{S_3(T_*)}{T_*}\right]^{3/2}\frac{v_{\rm w,*}^3}{(\beta_*/H_*)^4}\left(\frac{T_*}{H_*}\right)^4\,.
\end{eqnarray}
Typically, the bounce action at percolation for an FOPT with $T_{\rm c} = \unit[10]{MeV}$ and $r_{T\rm,c} = 0.4$ is $S_3/T \sim 150$. In Fig.~\ref{fig:SampleS3TboH} we plot $S_3/T$ as solid curves, where we feature BP-2 (red) and BP-4 (blue). These benchmark points correspond to critical temperatures of \unit[10]{keV} and \unit[1]{GeV}, respectively. The actual values of $S_3(T_*)/T_*$ for BP-2 and BP-4 are, respectively, 194.7 and 133.5. In general, the bounce action can change depending on the scale of $T_{\rm c}$. Based on the estimate in Eq.~(\ref{EstimatePerc}), the bounce action can increase, relative to the value at $T_{\rm c} = \unit[10]{MeV}$, by $4\ln\left[\left(T_{\rm c}/\unit[10]{MeV}\right)\right]$, where we approximated $T_* \simeq T_{\rm c}$.
\begin{figure}[t]
    \centering
        \includegraphics[scale=0.3]{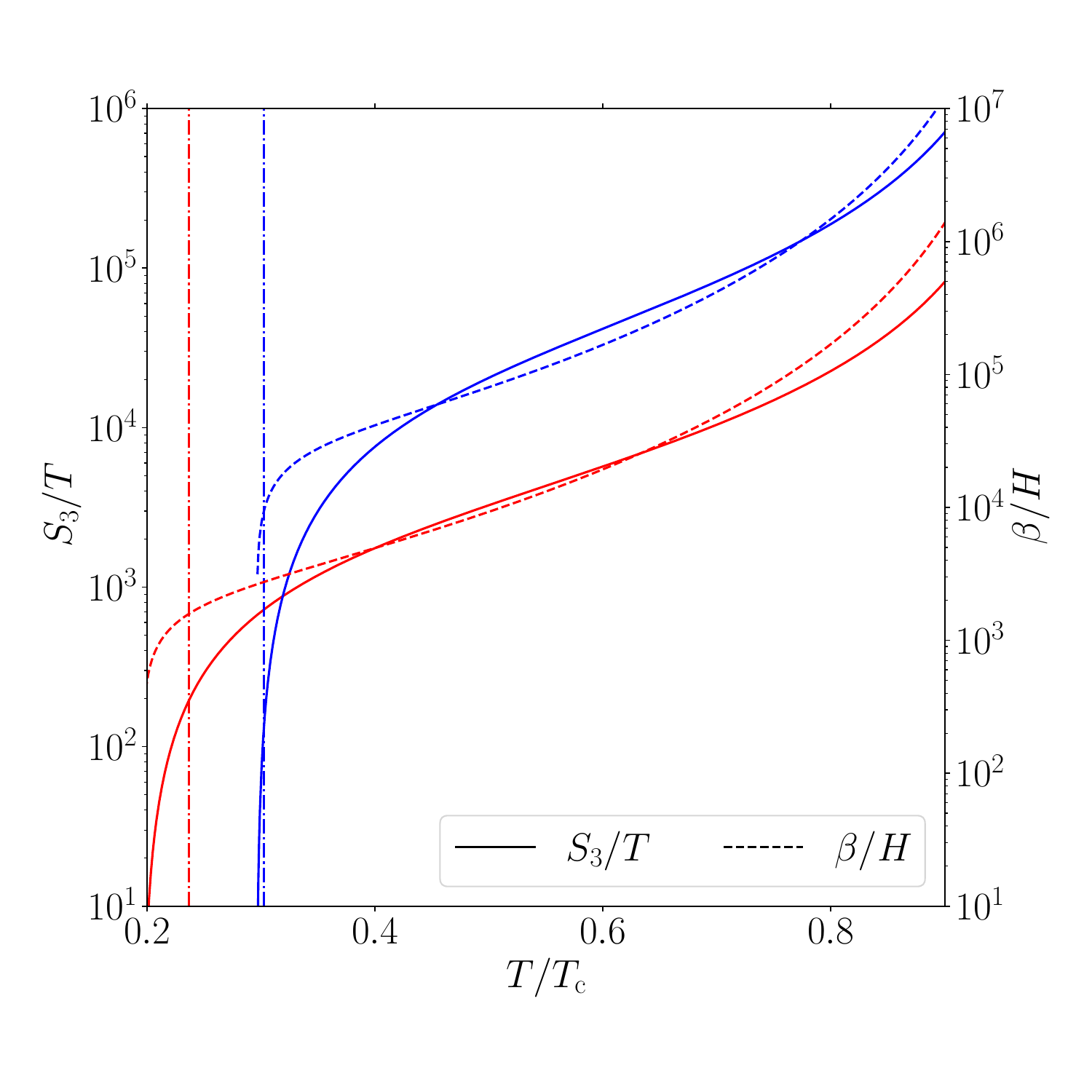}
    \caption{$S_3/T$ and $\beta/H$, for BP-2 (red) and BP-4 (blue). The vertical dot-dashed lines indicate the percolation temperature for each benchmark point.}
    \label{fig:SampleS3TboH}
\end{figure}

In order to account for the presence of the FV and TV, the thermodynamic properties of the dark fluid are determined by the average of the FV and TV contributions, weighted by the respective volume fractions, \textit{i.e.},
\begin{eqnarray}
    \nonumber p_{\rm D} = Fp_{\rm FV}+(1-F)p_{\rm TV}\,,\quad s_{\rm D} = F s_{\rm FV}+(1-F) s_{\rm TV}\,,\quad \rho_{\rm D} = F\rho_{\rm FV}+(1-F)\rho_{\rm TV}\,.\\
\label{DefAveThermoD}    ~
\end{eqnarray}
The above prescription, which uses volume fractions as weights to account for the contribution of each phase, is valid over sufficiently large scales where FV and TV features are indistinguishable. The dark sector equation of state and the sound speed are then respectively given by
\begin{eqnarray}
  \label{DefwDcs2D} \boxed{w_{\rm D} = \frac{p_{\rm D}}{\rho_{\rm D}},\quad c_{s\rm,D}^2 = \frac{dp_{\rm D}/dt}{d\rho_{\rm D}/dt} = \frac{s_{\rm D} - \left(\frac{\dot{F}}{H}\right)\frac{\Delta p}{T}}{T~\frac{\partial s_{\rm D}}{\partial T}-\left(\frac{\dot{F}}{H}\right)\frac{\Delta\rho}{T}}}\,.
\end{eqnarray}
Note that in the limit where the system is predominantly in one of the vacuum configurations, $\dot{F}/H \rightarrow 0$, so that
\begin{eqnarray}
    w_{\rm D} \rightarrow \frac{p_i}{\rho_i},\quad c_{s\rm,D}^2 \rightarrow c_{s,i}^2 \equiv \left(\frac{d\ln s_i}{d\ln T}\right)^{-1},\quad i = \rm FV, TV.
\end{eqnarray}
During the short period when the false vacuum fraction drops rapidly, \textit{i.e.}, when the system quickly transforms from the false to the true vacuum configuration, we have $\vert \dot{F}\vert/H \gg 1$, and
\begin{eqnarray}
    c_{s\rm,D}^2 \approx \frac{\Delta p}{\Delta \rho}.
\end{eqnarray}
Note that the sound speed squared can be negative since $\Delta p$ and $\Delta \rho$ have opposite signs when $T < T_{\rm c}$. Sufficiently short wavelength perturbations whose scales are comparable with the duration of the sudden drop in $F$, are expected to grow.
We emphasize that $c_{s,i}^2$ in this limit refers to the sound speed in the FV or the TV, when each is taken in isolation. This is the sound speed that has been computed in, \textit{e.g.}, Refs.~\cite{Espinosa:2010hh,Tenkanen:2022tly}. However, during the phase transition, the sound speed $c_{s\rm,D}^2$ in the dark plasma may deviate from the value suggested by $c_{s\rm ,FV}^2 = 1/3$, or $c_{s\rm ,TV}^2$, not necessarily equal to 1/3. For example, if the false vacuum fraction decreases sharply, the denominator in Eq.~(\ref{DefwDcs2D}) can be large so that $c_{s\rm,D}^2$ can become negative. A similar situation occurs in the physical system considered in Ref.~\cite{Schmid:1998mx}, where the system undergoes prolonged phase coexistence at $T \simeq T_{\rm c}$, keeping the pressure constant, which leads to a zero sound speed in the long-wavelength limit. On the other hand, Ref.~\cite{Tenkanen:2022tly} mentioned in a footnote that the sound speed does not vanish at the critical point, which is contrary to the result in Ref.~\cite{Schmid:1998mx}. This is not very surprising, since the former considered the sound speeds in each phase taken in isolation, {\it i.e.}, $c_{s,\rm FV}^2$ and $c_{s\rm, TV}^2$. For the sake of clarity, in Table \ref{tab:Definitions} we explicitly define quantities, such as the equation of state and sound speed squared, that refer to the dark sector; SM sector; and the average over the dark and SM sectors.

\begin{table}[t]
    \centering
    \begin{tabular}{ccc}\toprule
        Quantity & Description\\\midrule
        $p_{\rm D}, \rho_{\rm D}$ & Dark sector pressure, density averaged over the false and true vacua\\
        $w_{\rm D}$ & Dark sector equation of state\\
        $w_{\rm SM}$ & Visible sector equation of state\\
        $w$ & Equation of state, averaged over the dark and visible sectors\\
        $c_{s\rm,D}^2$ & Dark sector sound speed squared\\      
        $c_{s\rm,SM}^2$ & Visible sector sound speed squared\\
        $c_s^2$ & Sound speed squared, averaged over the dark and visible sectors\\\midrule
        $\delta_{\rm D}, \hat{\psi}_{\rm D}$ & Density and velocity contrasts in the dark sector, respectively\\
        $\delta_{\rm SM}, \hat{\psi}_{\rm SM}$ & Density and velocity contrasts in the visible sector, respectively\\
        $\delta$ & Density contrast, averaged over the dark and visible sectors\\
        $\delta_{\rm eff}$ & Effective density contrast in the angular momentum, Eq.~(\ref{startingJc}) \\\bottomrule
    \end{tabular}
    \caption{Summary of the main quantities relevant in tracking the background cosmology and perturbations.}
    \label{tab:Definitions}
\end{table}

Like the background evolution history, the evolution of the dark fluid equation of state and the sound speed qualitatively differ between the two benchmark cases in Fig.~\ref{fig:SampleEvol}. For example, we observe that the period when $c_{s\rm,D}^2$ deviates significantly from $1/3$ is shorter for BP-4 than for BP-2. We can estimate the duration $\eta_f - \eta_i$ when the dark plasma sound speed deviates significantly from either $c_{s\rm ,FV}^2$ and $c_{s\rm ,TV}^2$ as follows. We take $\eta_i$ and $\eta_f$ to be the instants in time when the fractional rate of change in the FV fraction $-\dot{F}/(HF)$, are, respectively, 1 and 10. These can be calculated numerically.
However, an estimate can be obtained from Eq.~(\ref{EstimateFVFraction}) assuming that $v_{\rm w}$ and $\beta$ are slowly varying functions of $t$:
\begin{align}
\label{FracChangeFV}\frac{-\dot{F}}{HF} &\approx 8\pi \frac{v_{\rm w}^3}{\beta^3(t)}\frac{\Gamma(t)}{H(t)}\,.
\end{align}
The duration for which $c_{s\rm,D}^2$ exhibits a dip is 
\begin{eqnarray}
H_{\rm c}(\eta_f - \eta_i) \approx \frac{1}{T_f/T_{\rm c}} - \frac{1}{T_i/T_{\rm c}}\,.
\end{eqnarray}
To estimate $T_i$ and $T_f$, we note that $v_{\rm w}$, $\beta$, $\Gamma$, and $H$ are functions of the dark sector temperature, and the Hubble parameter can be approximated as
\begin{eqnarray}
H \approx H_{\rm c}\left(\frac{\rho_{\rm SM}}{\rho_{\rm SM,c}}\right)^{1/2} \approx H_{\rm c}\left(\frac{T_{\rm SM}}{T_{\rm SM,c}}\right)^2 \approx H_{\rm c}\left(\frac{T}{T_{\rm c}}\right)^2\,.
\end{eqnarray} 
Then
\begin{eqnarray}
    8\pi \frac{v_{\rm w}^3(T_i)}{(\beta/H)^3(T_i)}\frac{\Gamma(T_i)}{H_{\rm c}^4}\left(\frac{T_{\rm c}}{T_i}\right)^8 \simeq 1,\quad 8\pi \frac{v_{\rm w}^3(T_f)}{(\beta/H)^3(T_f)}\frac{\Gamma(T_f)}{H_{\rm c}^4}\left(\frac{T_{\rm c}}{T_f}\right)^8 \simeq 10\,.
\end{eqnarray}

In the bottom panels of Fig.~\ref{fig:SampleEvol}, we indicate the estimates for $\eta_i$ and $\eta_f$ as vertical dashed and dotted lines, respectively, and overlay them on the $c_{s\rm,D}^2$ curves. For these two BPs, we can clearly see that our definitions of $\eta_i$ and $\eta_f$ provide us with a robust method to determine the width of the dip in $c_{s\rm,D}^2$. As for the equation of state, note that $w_{\rm D}$ changes with conformal time via
\begin{eqnarray}
   \label{wDPrime} w_{\rm D}' = -3\mathcal{H}(1+w_{\rm D})(c_{s\rm,D}^2-w_{\rm D})\,,
\end{eqnarray}
which directly follows from the definition of sound speed and from the conservation of the dark fluid stress energy tensor. Eq.~(\ref{wDPrime}) can be rewritten as
\begin{eqnarray}
    \frac{d}{d(\ln a)}\ln(1+w_{\rm D}) = -3(c_{s\rm,D}^2-w_{\rm D})\,.
\end{eqnarray}
Thus, the increase in $w_{\rm D}$ is expected to occur whenever $c_{s\rm,D}^2 < w_{\rm D}$. Note that the equation of state at the end of the FOPT $w_{\rm end} \simeq \frac{p_{\rm TV}}{\rho_{\rm TV}}$ may, in general, not relax to the radiation equation of state $w_{\rm rad} = 1/3$, as illustrated by the dashed curves in the bottom row of Fig.~\ref{fig:SampleEvol}.
 
\subsubsection{Parameter selection and scans over physical FOPT scenarios}
We have shown in Section~\ref{sec:Physicalconditions} that the physical conditions restrict the allowed regions in the $(L_{\rm c}/T_{\rm c}^4,x_C)$ plane, for given $A$ and $\lambda$. We observed that for a given $A$, a larger $\lambda$ allows part of the physical region to reach values of $T_0/T_{\rm c}$ close to 1. Since the destabilization temperature $T_0$ determines the lower limit of the percolation temperature $T_*$, there are situations where $T_*$ is guaranteed to be extremely close to $T_{\rm c}$. We avoid such scenarios, which correspond to small potential barriers. The maximum possible $T_0/T_{\rm c}$ allowed by the physical conditions can be obtained by setting $\rho_{\rm TV}(T_0) = 0$ at $x_C = 0$:
\begin{align}
    \frac{\lambda}{\lambda_0(A)} = \left(\frac{9}{128}\right)^{1/3}\left(\frac{9T_0^2/T_{\rm c}^2-1}{1-T_0^2/T_{\rm c}^2}\right)^{1/3}.
\end{align}
For any given value of $A$, if we set the maximum $T_0/T_{\rm c} = 0.9, 0.95, 0.99, 0.995, 0.999$, they correspond, respectively, to $\lambda/\lambda_0 \simeq 1.33, 1.72, 3.02, 3.82, 6.55$.
\begin{figure}[t]
    \centering
    \begin{tabular}{cc}
                \includegraphics[scale=0.26]{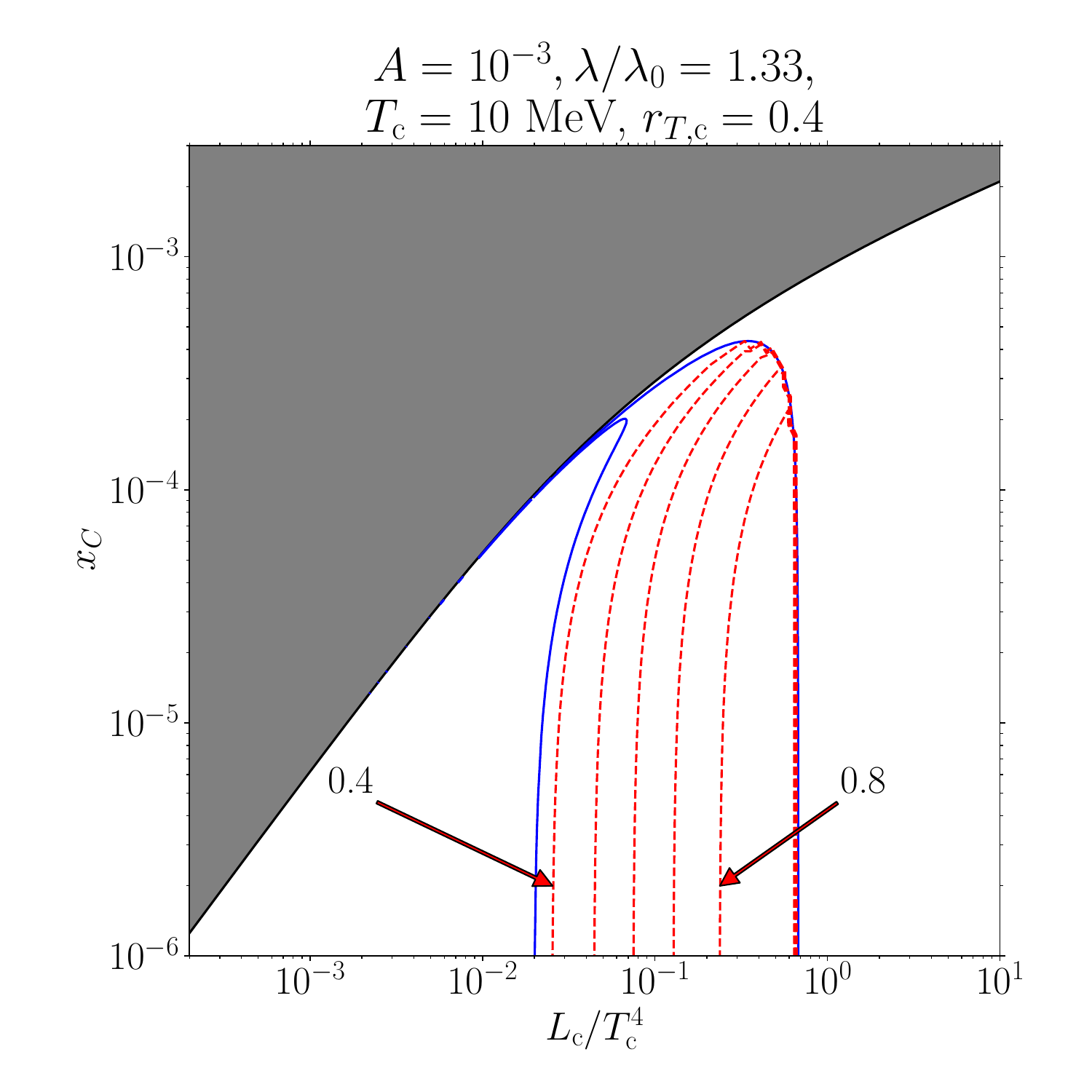} &  \includegraphics[scale=0.26]{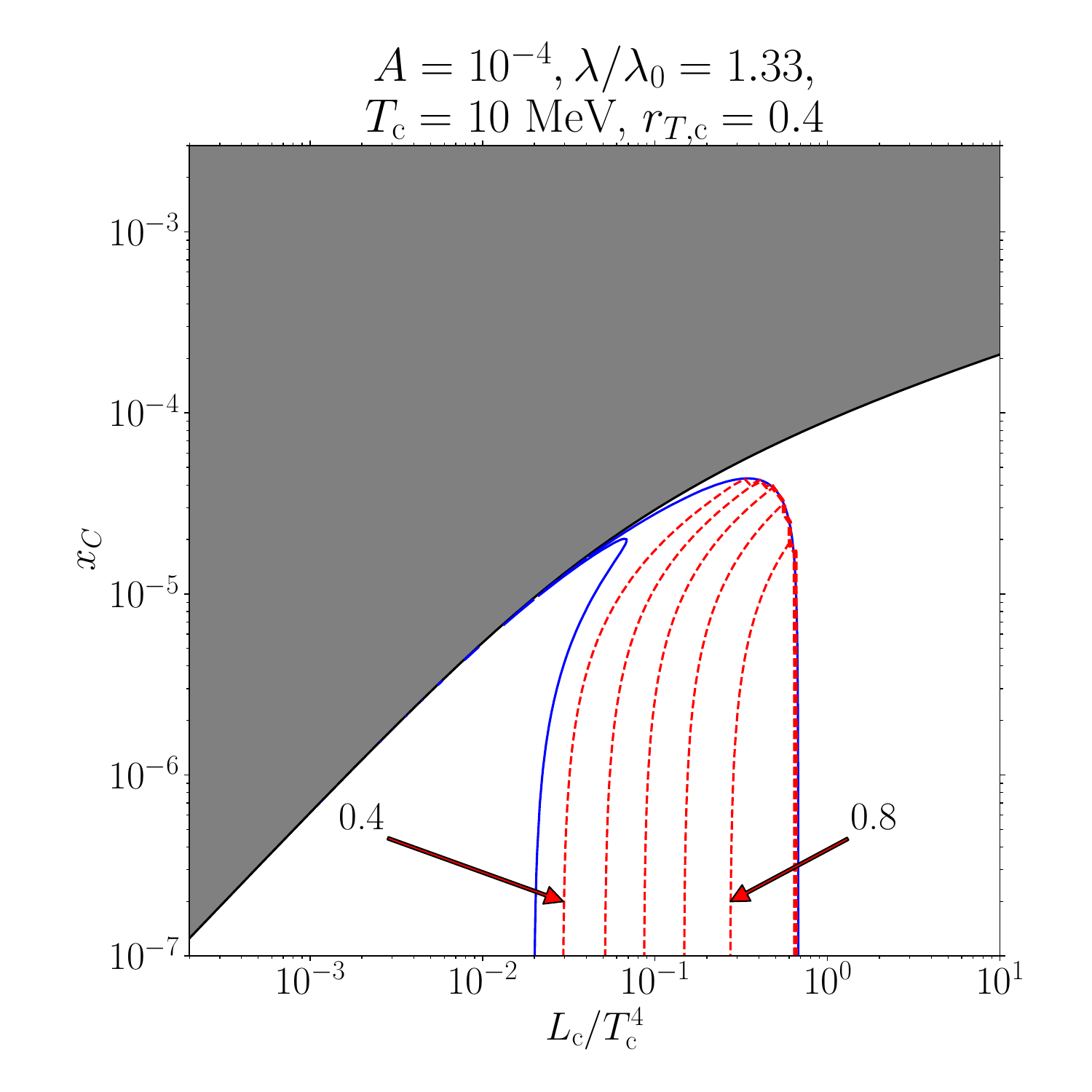}\\
          (a) & (b) \\
          \includegraphics[scale=0.26]{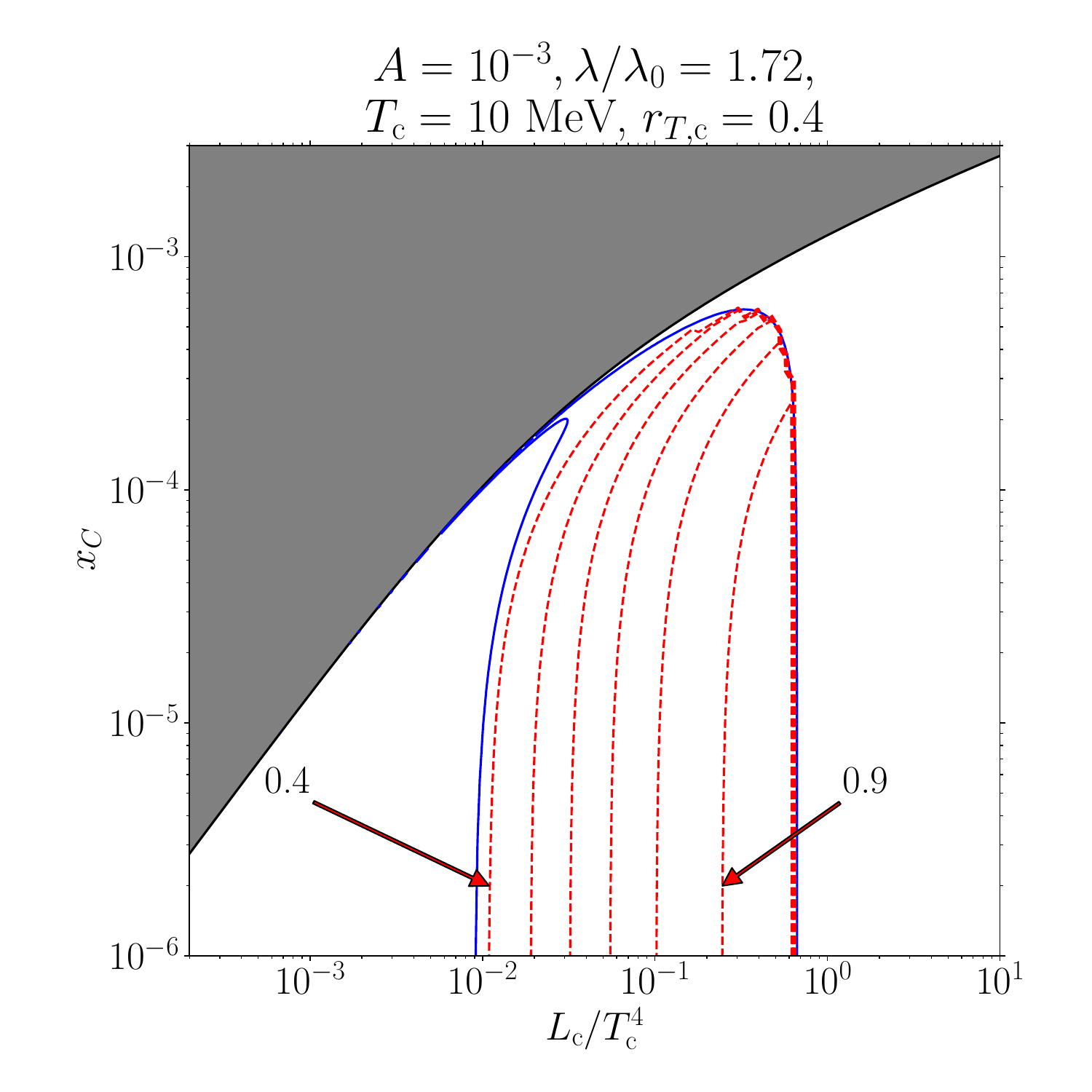} &  \includegraphics[scale=0.26]{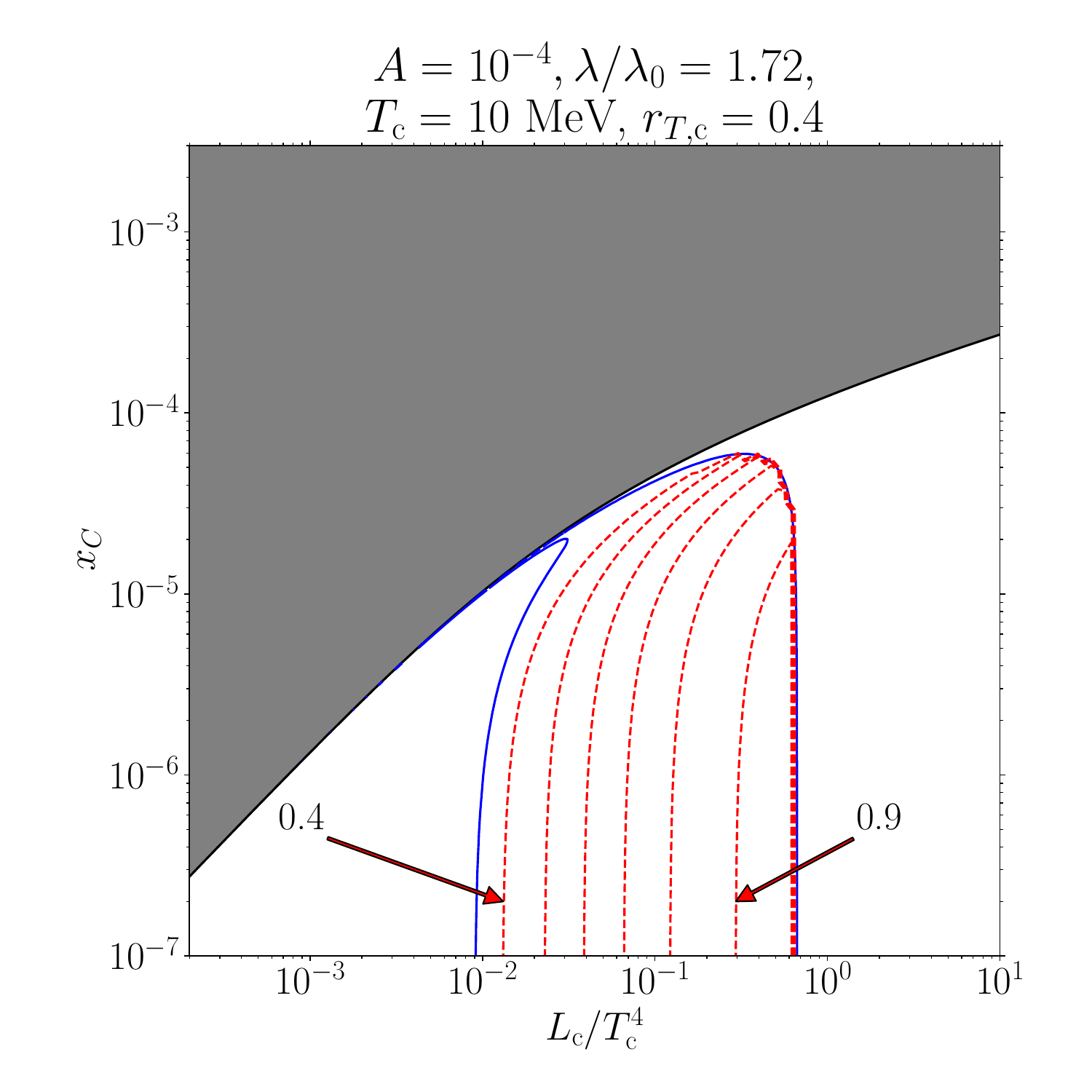}\\
          (c)&(d)
    \end{tabular}
    \caption{Contours constant of $T_*/T_{\rm c}$ (red) for $A = 10^{-3}, 10^{-4}$, for $\lambda/\lambda_0 \simeq 1.33$ and $\lambda/\lambda_0 \simeq 1.72$, and $T_{\rm c} = \unit[10]{MeV}$, $r_{T\rm,c} = 0.4$. The blue contour marks the boundary of the physical region. }
    \label{fig:ScanTPerc}
\end{figure}

Restricting our attention to physical FOPT scenarios, we would like to determine how the fundamental parameters that characterize the FOPT determine physically relevant quantities, such as the percolation temperature and the strength of the phase transition. This requires us to specify $T_{\rm c}$ and $r_{T\rm,c}$, and then solve the system of ODEs that describe the background evolution. Firstly, in Fig.~\ref{fig:ScanTPerc} we show scans of $T_*/T_{\rm c}$. Here we take $A = 10^{-3}, 10^{-4}$. For each $A$, we choose $\lambda = 3.6917 \times 10^{-4}, 1.7136 \times 10^{-5}$, respectively, for fixed $\lambda/\lambda_0 \simeq 1.33$ (and $\lambda  = 4.806 \times 10^{-4}, 2.231 \times 10^{-5}$, respectively, for fixed $\lambda/\lambda_0 \simeq 1.72$). We commit to $T_{\rm c} = \unit[10]{MeV}$ and $r_{T\rm,c} = 0.4$, unless we specify otherwise. For a fixed $x_C$, we find that the percolation temperature $T_*$ tends to become closer to the critical temperature, along the direction of rescaled latent heat. This is expected behavior, knowing that $T_0$ also approaches $T_{\rm c}$ along this direction (see Fig.~\ref{fig:PhysRegion}). We have checked that the general trend on $T_*$ still holds for different $T_{\rm c}$ and $r_{T\rm,c}$; also, the contours for fixed $A$ and $\lambda/\lambda_0$ change only marginally. Thus, we find that imposing the condition that $T_*$ is at most $0.95 T_{\rm c}$ restricts us from going close to the edge of the physical region towards the maximum rescaled latent heat.

\begin{figure}[t]
    \centering
    \begin{tabular}{cc}
            \includegraphics[scale=0.26]{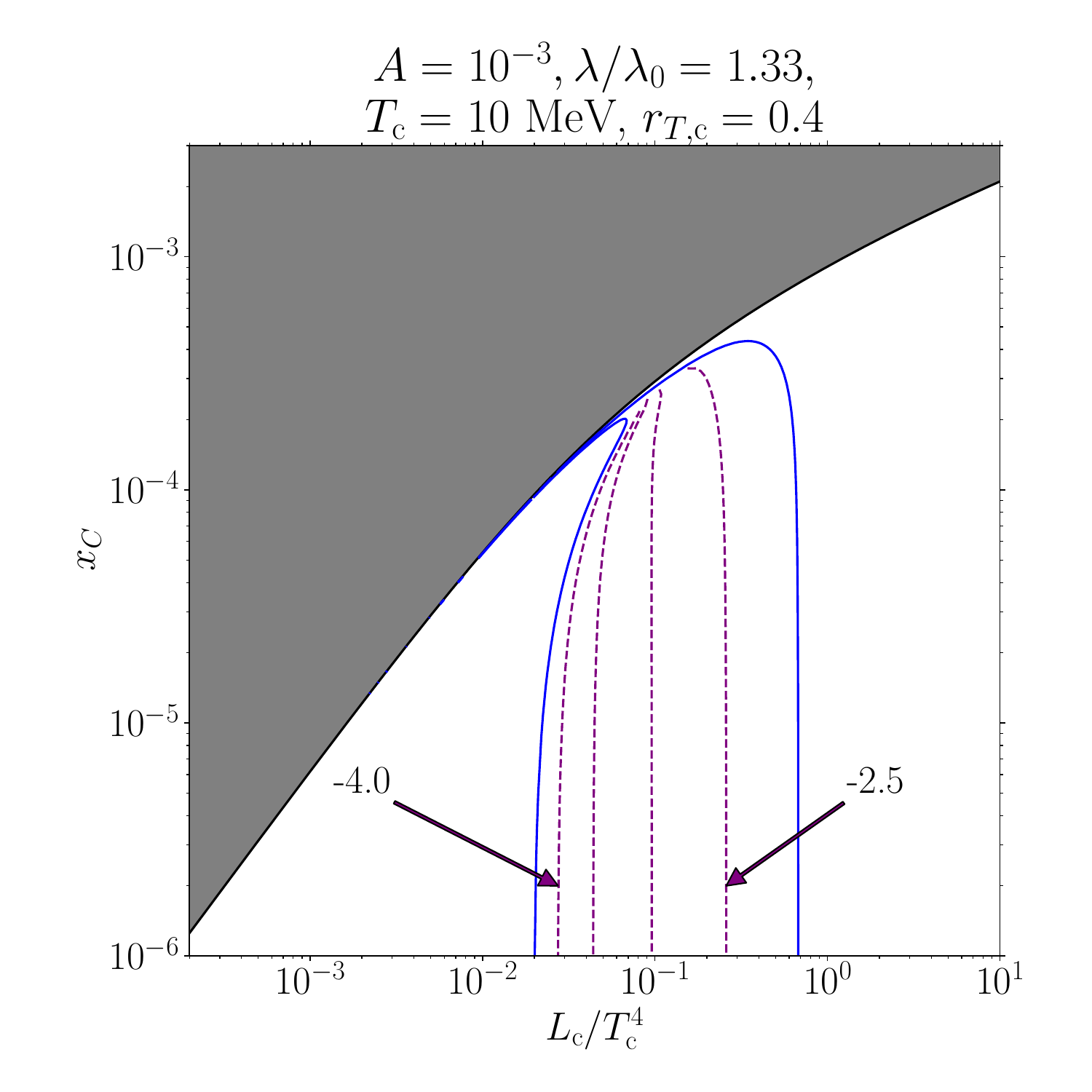} &  \includegraphics[scale=0.26]{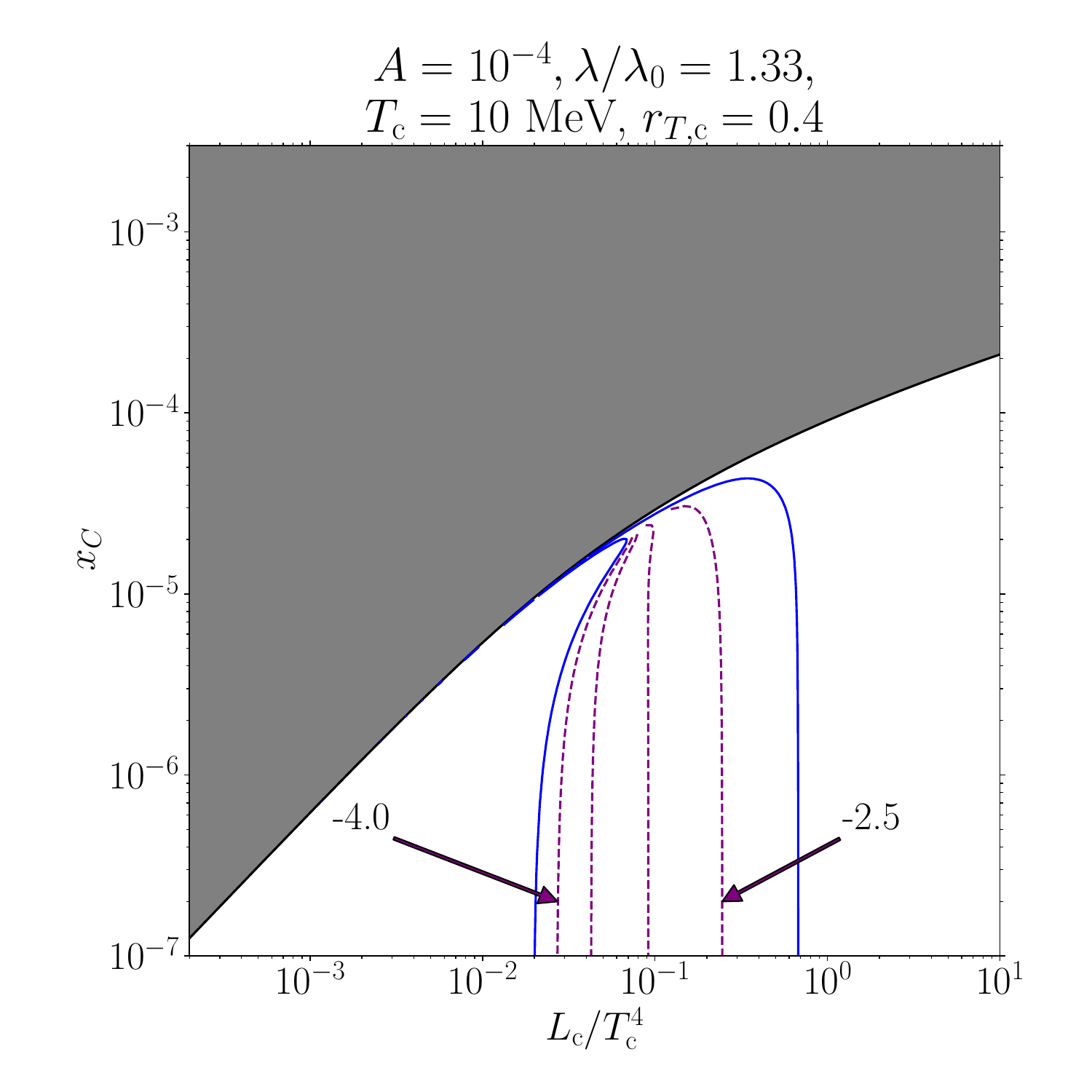}\\
          (a) & (b)\\
                      \includegraphics[scale=0.26]{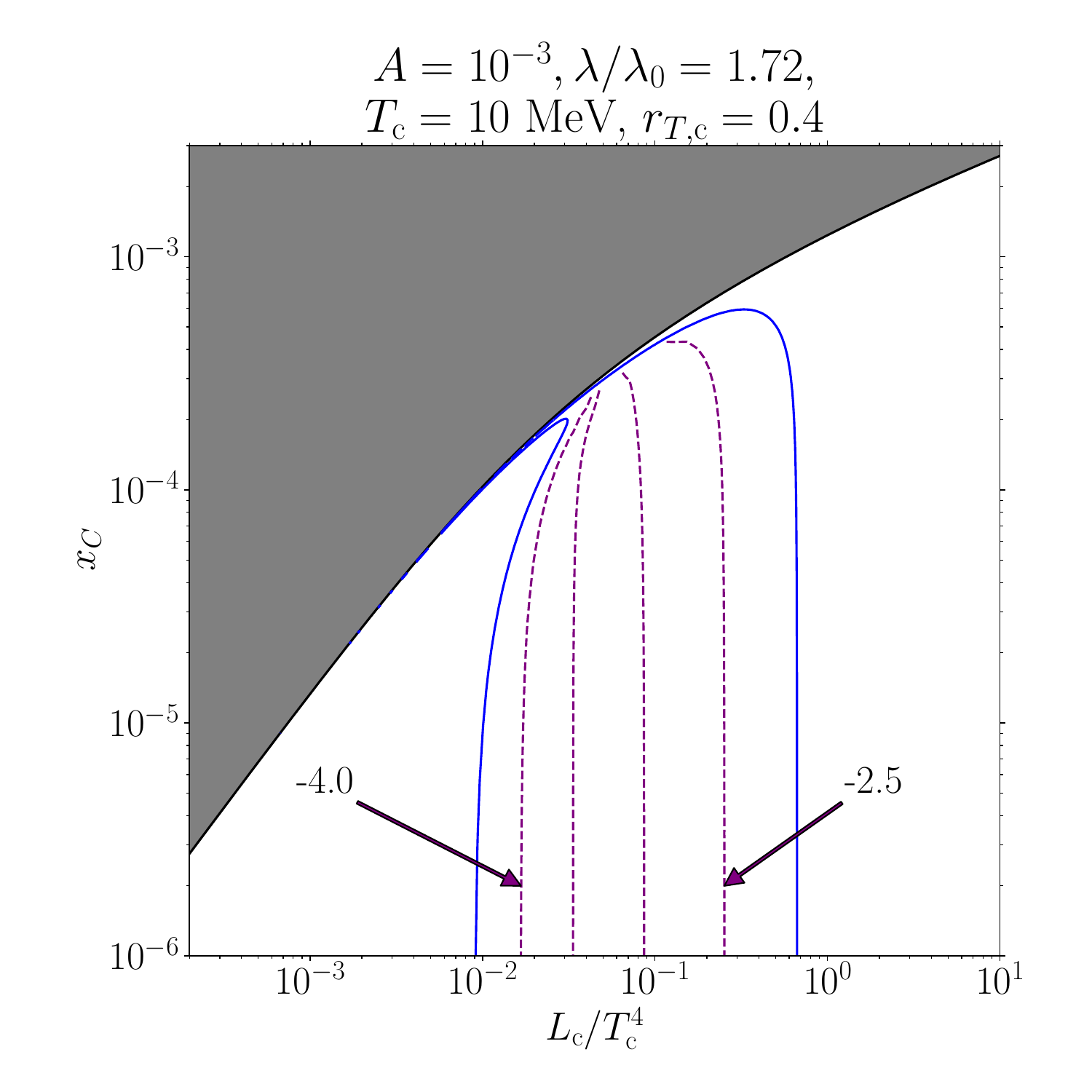} &  \includegraphics[scale=0.26]{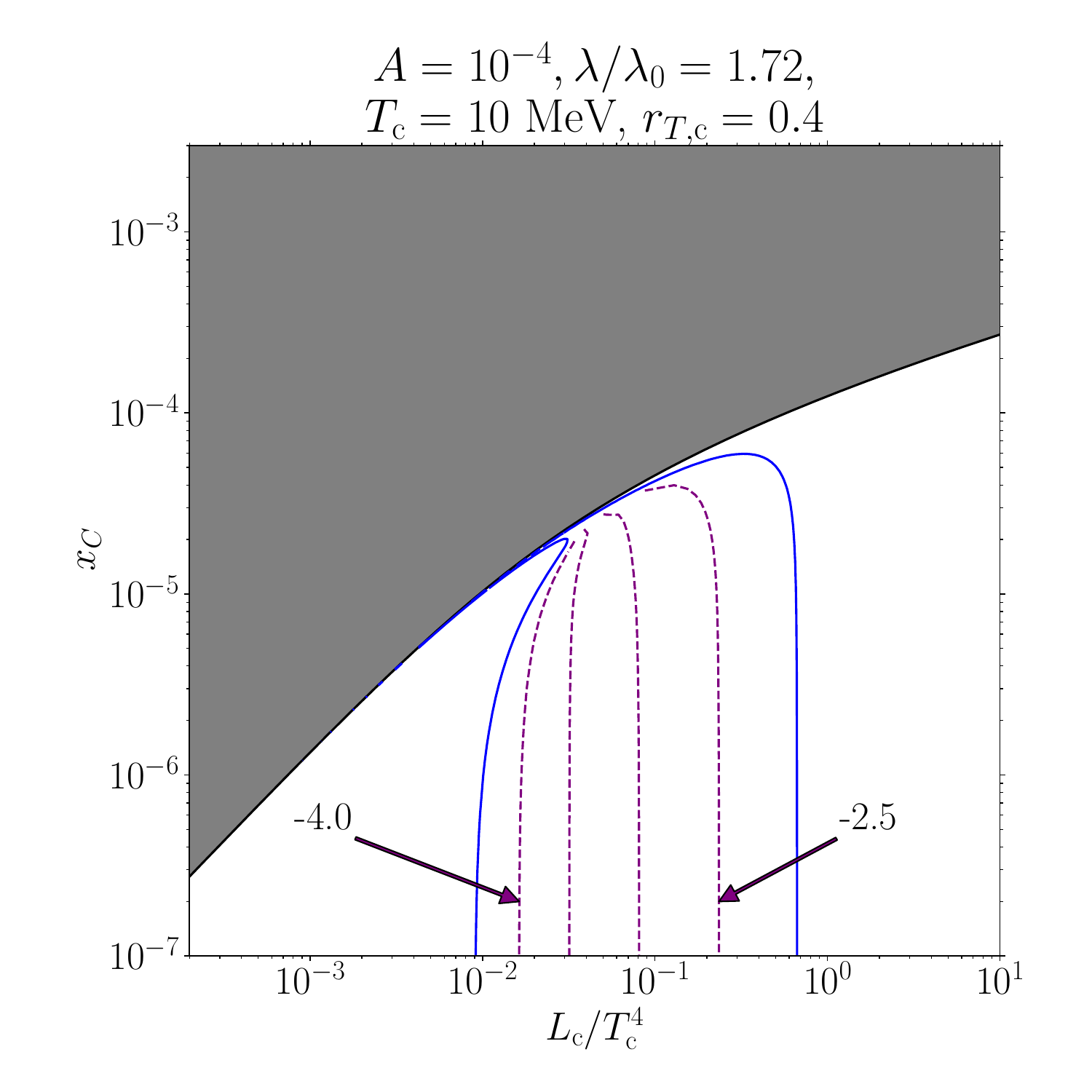}\\
          (c) & (d)
    \end{tabular}
    \caption{Contours of constant of $\log_{10} \alpha_*$ (purple) in steps of 0.5 for $A = 10^{-3}, 10^{-4}$, for fixed $\lambda/\lambda_0 \simeq 1.33$ and $\lambda/\lambda_0 \simeq 1.72$, and $T_{\rm c} = \unit[10]{MeV}$, $r_{T\rm,c} = 0.4$. The blue contour marks the boundary of the physical region.}
    \label{fig:ScanalphaBag}
\end{figure}

In Fig.~\ref{fig:ScanalphaBag} we show scans of the FOPT strength, defined in Eq.~(\ref{alphaBag}). To understand the scans, we estimate $\alpha_*$ by noting that we are working in the regime where $\rho_{\rm SM} \gg \rho_{\rm D}$. Then 
\begin{eqnarray}
\rho_{\rm tot, *} \approx \rho_{\rm SM}(T_{\rm SM,*}) \approx \frac{\pi^2}{30}\frac{g_{\rho,{\rm SM}}(T_{\rm SM,*})}{r_{T\rm,*}^4}T_*^4\,,
\end{eqnarray}
gives
\begin{eqnarray}
\label{alphaStarApprox}\alpha_* \approx \frac{30}{\pi^2 g_{\rho,\rm SM}(T_{\rm SM,*})}\frac{r_{T\rm,*}^4}{(T_*/T_{\rm c})^4}\frac{\Delta\rho(T_*)}{T_{\rm c}^4}\,,
\end{eqnarray}
where we made the approximation that the SM sector contributes dominantly to the total radiation energy density.
The trend in percolation temperature in Fig.~\ref{fig:ScanTPerc}, combined with the estimate for $\alpha_*$ in Eq.~(\ref{alphaStarApprox}), are consistent with the scans of $\alpha_*$ in Fig.~\ref{fig:ScanalphaBag}.
   If we assume instantaneous reheating immediately after the onset of percolation, then the reheating temperature is 
\begin{align}
    \frac{\pi^2}{30}g_\rho T_*^4 - (1 - F_*) \Delta \rho(T_*) = \frac{\pi^2}{30}g_\rho T_{\rm rh}^4 - \Delta \rho(T_*) \Rightarrow T_{\rm rh} = T_* \left[1 + \frac{30}{\pi^2 g_\rho}F_*\frac{\Delta \rho(T_*)}{T_*^4}\right]^{1/4}\,.
\end{align}
Then applying separate entropy conservation in the SM sector and the dark sector, we find the
effective number of extra massless neutrinos at the end of Big Bang Nucleosynthesis (BBN) to be
\begin{align}
    \Delta N_{\rm eff} \equiv N_{\rm eff}-3.044 = \frac{4}{7}\left(\frac{11}{4}\right)^{4/3}g_\rho r_{T\rm,c}^4\left[\frac{g_{s,\rm SM}(T_{\rm SM, BBN})}{g_{s,\rm SM}(T_{\rm SM,*})}\right]^{4/3}\left[1 + \frac{30}{\pi^2 g_\rho}F_*\frac{\Delta \rho(T_*)}{T_*^4}\right] \leq 0.5\,,
\end{align}
where $g_{s\rm, SM}$ is the number of relativistic degrees of freedom associated with the entropy density in the SM sector, and the upper bound on $\Delta N_{\rm eff}$
arises from a combination of Planck and other data~\cite{Planck:2018vyg}.
Taking BBN to end at $T_{\rm SM, BBN} = \unit[0.01]{MeV}$, we set $g_{s\rm, SM}\left(T_{\rm SM, BBN}\right) \simeq 3.93872$. The upper limit of the term in square brackets is $1 + F_*$, and follows from the physical condition that $\rho_{\rm TV}(T) = g_\rho T^4~\pi^2/30 - \Delta \rho(T) \geq 0$, which implies that $ \Delta \rho(T_*)/\left(\pi^2/30~g_\rho T_*^4\right) \leq 1$. This leads to a conservative upper limit of
\begin{align}
r_{T\rm,c} < 0.598\left[\frac{g_{s, \rm SM}\left(T_{\rm SM,*}\right)}{10}\right]^{1/3}\left(\frac{4.5}{g_\rho}\right)^{1/4}\,.
\end{align}

\subsubsection{Perturbations during an FOPT}
Once we know the background evolution during the FOPT, we can determine the evolution of each Fourier mode $k/\mathcal{H}_{\rm c}$ by solving Eqs.~(\ref{ContinuityEq}) and (\ref{EulerEq}). In the following discussion, we assume that the single-fluid picture of the dark plasma holds for the perturbation modes of interest. The expressions for the average pressure, entropy density, and energy density, in the dark sector, are provided in Eq.~(\ref{DefAveThermoD}); the equation of state and sound speed are given in Eq.~(\ref{DefwDcs2D}).

Using the continuity and Euler equations in Eqs.~(\ref{ContinuityEq}) and (\ref{EulerEq}) we explicitly write the system of equations we need to solve. We have
\begin{eqnarray}
\label{FOPTdeltaD}\frac{\delta_{\rm D}'}{\mathcal{H}} - \frac{k}{\mathcal{H}}\hat{\psi}_{\rm D} + 3(1+w_{\rm D}) \Psi_k + 3(c_{s,\rm D}^2 - w_{\rm D})\delta_{\rm D} &=& 0\,,\\
\label{FOPTpsiD}    \frac{\hat{\psi}_{\rm D}'}{\mathcal{H}}+(1-3w_{\rm D})\hat{\psi}_{\rm D}+c_{s,\rm D}^2 \frac{k}{\mathcal{H}}\delta_{\rm D}+(1+w_{\rm D})\frac{k}{\mathcal{H}}\Psi_k &=& 0\,,\\
    \label{FOPTdeltaSM}\frac{\delta_{\rm SM}'}{\mathcal{H}} - \frac{k}{\mathcal{H}}\hat{\psi}_{\rm SM} + 4 \Psi_k &=& 0\,,\\
    \label{FOPTpsiSM}\frac{\hat{\psi}_{\rm SM}'}{\mathcal{H}}+\frac{1}{3} \frac{k}{\mathcal{H}}\delta_{\rm SM}+\frac{4}{3}\frac{k}{\mathcal{H}}\Psi_k &=& 0\,.
\end{eqnarray}
The gravitational potential $\Psi_k$ is
\begin{eqnarray}
 \left[\left(\frac{k}{\mathcal{H}}\right)^2+\frac{9}{2}(1+w)\right]\Psi_k + \frac{3}{2}(1+3c_s^2) \delta = 0\,,
\end{eqnarray}
where
\begin{eqnarray}
    \delta \equiv \frac{\rho_{\rm D}\delta_{\rm D} + \rho_{\rm SM}\delta_{\rm SM}}{\rho_{\rm D} + \rho_{\rm SM}},\quad w \equiv \frac{\rho_{\rm D} w_{\rm D} + \rho_{\rm SM} w_{\rm SM}}{\rho_{\rm D} +\rho_{\rm SM}},\quad c_s^2 \equiv \frac{\rho_{\rm D}' c_{s\rm,D}^2 + \rho_{\rm SM}' c_{s\rm,SM}^2}{\rho_{\rm D}' + \rho_{\rm SM}'}\,.
\end{eqnarray}
The gravitational potential $\Phi_k$ is \begin{eqnarray}
    -k^2 \Phi_k = -\frac{3}{2}\mathcal{H}^2 \delta_k\,.
\end{eqnarray}
In the regime where $\rho_{\rm D} \ll \rho_{\rm SM}$, we have $\vert  \rho_{\rm D}'\vert \simeq \vert 3\mathcal{H}\rho_{\rm D} (1+w_{\rm D})\vert \ll \vert \rho_{\rm SM}'\vert$. These imply that
\begin{eqnarray}
    \delta \approx \delta_{\rm SM},\quad w \approx w_{\rm SM} \approx c_s^2 \approx c_{s,\rm SM}^2 = 1/3\,,
\end{eqnarray}
so that
\begin{eqnarray}
    \Psi_k \approx -\frac{\delta_{{\rm SM,}k}}{2 + \frac{1}{3}\left(\frac{k}{\mathcal{H}}\right)^2}\,.
\end{eqnarray}
The gravitational potential $\Phi$ appearing in the spatial part of the metric, which is relevant for the angular momentum calculation, is just
\begin{eqnarray}
    \Phi_k \simeq \frac{3}{2}\left(\frac{k}{\mathcal{H}}\right)^{-2} \delta_{{\rm SM,}k}\,.
\end{eqnarray}
\begin{figure}[t]
    \centering
   \includegraphics[scale=0.3]{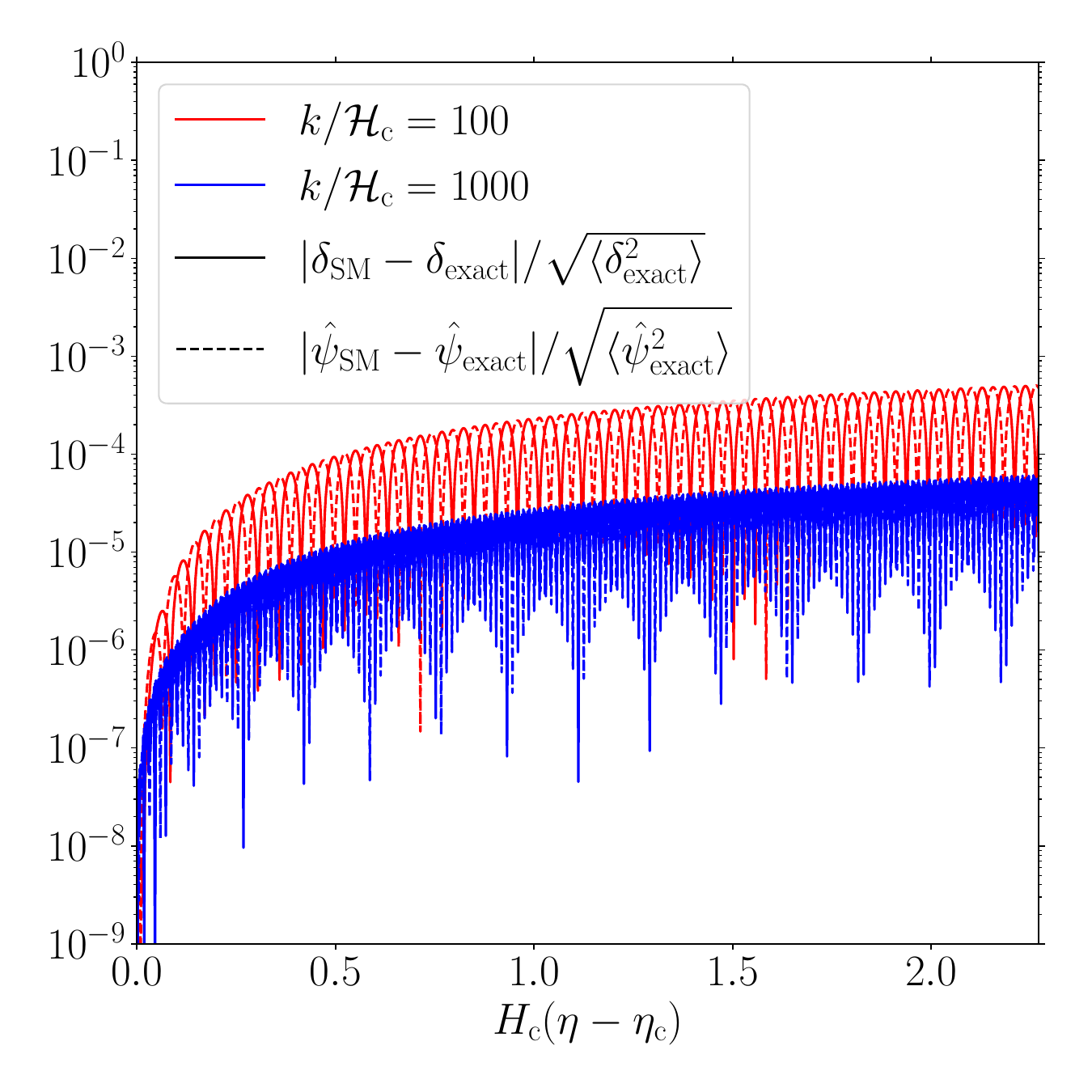}
 \caption{Comparison between exact $\delta$ and $\hat{\psi}$ evolution versus numerically calculated SM perturbations for BP-4 and $k/\mathcal{H}_{\rm c} = 100, 1000$.}
    \label{fig:ComparisonSMPerturbations}
\end{figure}
We first examine the behavior of the SM perturbations. In Fig.~\ref{fig:ComparisonSMPerturbations}, we compare the numerical solutions for $\delta_{\rm SM}$ and $\hat{\psi}_{\rm SM}$ from solving the UHG perturbation equations during the FOPT, and the result that one would obtain from the analytic expressions in Eq.~(\ref{DensPsi}), that are valid during the period when the dominant component of the Universe is a pure radiation fluid. Choosing BP-4, we plot the difference between the numerical result and the analytic result, normalized with respect to the RMS value of the analytic result within the period between the critical point and the percolation time, for the modes $k/\mathcal{H}_{\rm c} = 100$ and $k/\mathcal{H}_{\rm c} = 1000$. The difference is negligible, mainly because the equation of state and sound speed in the SM plasma are unaffected by the FOPT in the dark sector. To a good approximation, the analytic expressions for the perturbations in Eq.~(\ref{DensPsi}) can be extrapolated even during the period of FOPT for this benchmark case.

\subsubsection{Superhorizon and subhorizon limits}
Let us obtain the limiting behaviors of $\delta_{\rm D}$ and $\hat{\psi}_{\rm D}$ in the \textit{superhorizon} and \textit{subhorizon} limits. In the case $k/\mathcal{H} \ll 1$ and $\rho_{\rm SM} \gg \rho_{\rm D}$, the system of equations satisfied by the perturbations in the dark fluid reduce to
\begin{eqnarray}
  \label{ContunitySuperH}\delta_{\rm D}' &\approx& k \hat{\psi}_{\rm D} + \mathcal{H}\left(1+3c_{s\rm,D}^2\right)\delta_{\rm SM} -3\mathcal{H}\left(c_{s\rm,D}^2-w_{\rm D}\right)\delta_{\rm D}\,,\\
  \label{EulerSuperH}\hat{\psi}_{\rm D}' &\approx& -k c_{s\rm,D}^2 \hat{\psi}_{\rm D} + \frac{k}{3}\left(1+3c_{s\rm,D}^2\right)\delta_{\rm SM}-\mathcal{H}\left(1-3w_{\rm D}\right)\hat{\psi}_{\rm D}\,.
\end{eqnarray}
Since $z \ll 1$, we may take $T_\delta(z) \approx 1$ and $d T_\delta/dz \approx 0$. We find that
\begin{eqnarray}
    \frac{d}{dz}\delta_{\rm D} &\approx& \sqrt{3}\hat{\psi}_{\rm D} - \frac{3(c_{s\rm,D}^2-w_{\rm D})}{z}\delta_{\rm D}\,,\\
    \frac{d}{dz}\hat{\psi}_{\rm D} &\approx& -\frac{(1-3w_{\rm D})}{z}\hat{\psi}_{\rm D} - \sqrt{3}c_{s\rm,D}^2\delta_{\rm D}\,,
\end{eqnarray}
where we have dropped terms containing positive powers of $z$ and higher. We can compare these with \cite{Schmid:1998mx}, which found that the evolution of the perturbations in the superhorizon limit is independent of the sound speed. This applies in our case if the dominant fluid component, which mainly dictates the Hubble expansion, is the dark fluid. One can show that the continuity equation can be rewritten as
\begin{eqnarray}
    \frac{d}{dz}\ln\left(\frac{\delta_{\rm D}}{1+w_{\rm D}}\right) \approx \sqrt{3}\frac{\hat{\psi}_{\rm D}}{\delta_{\rm D}}\,.
\end{eqnarray}
Similarly, it follows from Eqs.~(\ref{ContunitySuperH}) and (\ref{EulerSuperH}) that
\begin{eqnarray}
    \frac{d}{dz}\left(\frac{\hat{\psi}_{\rm D}}{\delta_{\rm D}}\right) \approx \frac{\left(3c_{s\rm,D}^2-1\right)}{z}\left(\frac{\hat{\psi}_{\rm D}}{\delta_{\rm D}}\right) - \sqrt{3}c_{s\rm,D}^2 - \sqrt{3}\left(\frac{\hat{\psi}_{\rm D}}{\delta_{\rm D}}\right)^2\,.
\end{eqnarray}
From the initial conditions for $\hat{\psi}_{\rm D}$ and $\delta_{\rm D}$, we have $(\hat{\psi}_{\rm D}/\delta_{\rm D})_{\rm c} \propto z_{\rm c}$. To leading order, we have
\begin{eqnarray}
    \frac{\hat{\psi}_{\rm D}}{\delta_{\rm D}} \approx C_0~z,\quad C_0 \equiv \frac{\sqrt{3}c_{s\rm,D}^2}{3c_{s\rm,D}^2-2}\,.
\end{eqnarray}
Then we have
\begin{eqnarray}
\label{SuperHorizonDelta} \delta_{\rm D} \approx \delta_{\rm D,c}\frac{1+w_{\rm D}}{1+w_{\rm D,c}}\exp\left[\frac{\sqrt{3}C_0}{2}(z^2-z_{\rm c}^2)\right] \approx \delta_{\rm D,c}\frac{1+w_{\rm D}}{1+w_{\rm D,c}}\left[1+\frac{\sqrt{3}C_0}{4}(z^2-z_{\rm c}^2)\right]\,.
\end{eqnarray}
On the other hand, the approximate solutions in the \textit{subhorizon} regime can be worked out by starting from the subhorizon limit of the evolution equations for the perturbations
\begin{eqnarray}
    \label{deltapsiDSubH}  \delta_{\rm D}' \approx k\hat{\psi}_{\rm D} - 3\mathcal{H}(c_{s\rm,D}^2-w_{\rm D})\delta_{\rm D},\quad \hat{\psi}_{\rm D}' \approx -k c_{s\rm,D}^2 \delta_{\rm D} - \mathcal{H}(1-3w_{\rm D})\hat{\psi}_{\rm D}\,.
\end{eqnarray}
Note that the $\Psi_k$ term which involves the weighted sum of all density perturbations, can be dropped for $k/\mathcal{H} \gg 1$. We can use Eq.~(\ref{deltapsiDSubH}), together with (\ref{wDPrime}), to write down a single second order ODE for $\delta_{\rm D}$, given by
\begin{eqnarray}
  \nonumber \delta_{\rm D}'' + \left(3c_{s\rm,D}^2-6w_{\rm D}+1\right)\mathcal{H}\delta_{\rm D}'+\left[k^2 c_{s\rm,D}^2 + 9\mathcal{H}^2\left(c_{s\rm,D}^2-w_{\rm D}\right) + 3\mathcal{H}(c_{s\rm,D}^2)'\right]\delta_{\rm D} \approx 0\,.\\
  \label{deltaD2ndOrder}~
\end{eqnarray} 
A similar expression was found in Ref.~\cite{Schmid:1998mx}, apart from the extra terms appearing in the $\delta_{\rm D}$ term. We then consider the regime where the perturbation is deep in the subhorizon limit, so that the scale of the perturbation is much shorter than the scale of the time variation of the sound speed and of $c_{s\rm,D}^2-w_{\rm D}$, {\it i.e.},~
\begin{eqnarray}
    k^2 c_{s\rm,D}^2 \gg \mathcal{H}~\frac{1}{c_{s\rm,D}^2}\left\vert\left(c_{s\rm,D}^2\right)'\right\vert,\quad k^2 c_{s\rm,D}^2 \gg \mathcal{H}^2\left(c_{s\rm,D}^2-w_{\rm D}\right)\,.
\end{eqnarray}
Following \cite{Schmid:1998mx}, we introduce
\begin{eqnarray}
   \label{DeltaToU} \delta_{\rm D} = \frac{H_{\rm c}^2 (1+w_{\rm D})^{1/2}}{a^2 \rho_{\rm D}^{1/2}}u_{\rm D}\,,
\end{eqnarray}
to eliminate the drag term. The ODE satisfied by $u_{\rm D}$ is then
\begin{eqnarray}
    u_{\rm D}'' + k^2 c_{s\rm,D}^2(\eta)~u_{\rm D} \approx 0\,,
\end{eqnarray}
where we essentially took the zeroth order coefficient to be dominated by the time dependent frequency $\omega(\eta) \equiv k c_{s\rm,D}(\eta)$. Then we approximate $c_{s\rm,D}^2$ by a piecewise linear function with
\begin{eqnarray}
\label{cs2DLinApprox}c_{s\rm,D}^2 \approx \begin{cases}
\frac{1}{3},~0 \leq \tilde{\eta} \leq \tilde{\eta}_i\\
~\\
\frac{1}{3} + \left(\frac{c_{s\rm, D*}^2-1/3}{\tilde{\eta}_* - \tilde{\eta}_i}\right)\left(\tilde{\eta}-\tilde{\eta}_i\right),~\tilde{\eta}_i < \tilde{\eta} \leq \tilde{\eta}_*\,.
\end{cases}
\end{eqnarray}
In the interval $0 \leq \tilde{\eta} \leq \tilde{\eta}_i$ the solution is simply a superposition of sines and cosines given by
\begin{eqnarray}
\label{ApproxSubH1}\delta = \delta_{\rm c} \cos\left(\frac{1}{\sqrt{3}}\frac{k}{\mathcal{H}_{\rm c}}\tilde{\eta}\right)+\sqrt{3}\hat{\psi}_{\rm c} \sin\left(\frac{1}{\sqrt{3}}\frac{k}{\mathcal{H}_{\rm c}}\tilde{\eta}\right),\quad \hat{\psi} = \frac{\delta'}{k}\,,
\end{eqnarray}
where $\delta_{\rm c}$ and $\hat{\psi}_{\rm c}$ are the perturbations evaluated at the critical point, assuming the analytic expression for perturbations evolving in a pure radiation background. On the other hand, in the interval $\tilde{\eta}_i < \tilde{\eta} \leq \tilde{\eta}_*$, the ODE admits a general solution in terms of Airy functions, so that we have
\begin{eqnarray}
\label{ApproxSubH2}
\left(\begin{matrix}
\delta\\
(k/\mathcal{H}_{\rm c})\hat{\psi}
\end{matrix}\right) = M(k/\mathcal{H}_{\rm c},u)\left(\begin{matrix}
    c_1\\
    c_2
\end{matrix}\right),\quad u \equiv \frac{-d_1 - d_2 \tilde{\eta}}{(-d_2)^{2/3}}\,,
\end{eqnarray}
where
\begin{eqnarray}
d_1 = \frac{1}{3}\left(\frac{k}{\mathcal{H}_{\rm c}}\right)^2\left[1 -3\tilde{\eta}_i \left(\frac{c_{s\rm, D*}^2 - 1/3}{\tilde{\eta}_* - \tilde{\eta}_i}\right)\right]\,,\quad d_2 = \left(\frac{k}{\mathcal{H}_{\rm c}}\right)^2\left(\frac{c_{s\rm, D*}^2 - 1/3}{\tilde{\eta}_* - \tilde{\eta}_i}\right)\,,
\end{eqnarray}
and
\begin{eqnarray}
   M(k/\mathcal{H}_{\rm c},u) \equiv \left(\begin{matrix}
    Ai(u)&Bi(u)\\
    \left(-d_2\right)^{1/3} Ai'(u)& \left(-d_2\right)^{1/3}Bi'(u)
\end{matrix}\right)\,.
\end{eqnarray}
The coefficients $c_1$ and $c_2$ are obtained by matching the solutions at $\tilde{\eta} = \tilde{\eta}_i$, which results in
\begin{align}
    \left(\begin{matrix}
        c_1\\
        c_2
    \end{matrix}\right) &= M^{-1}(k/\mathcal{H}_{\rm c},u_i)\left(\begin{matrix}
        \delta(\tilde{\eta}_i)\\
        (k/\mathcal{H}_{\rm c})\hat{\psi}(\tilde{\eta}_i)\\
    \end{matrix}\right),\\
    \left(\begin{matrix}
        \delta(\tilde{\eta}_i)\\
        (k/\mathcal{H}_{\rm c})\hat{\psi}(\tilde{\eta}_i)\\
    \end{matrix}\right) &= T(k)\left(\begin{matrix}
        \delta_{\rm c}\\
        \sqrt{3}\hat{\psi}_{\rm c}
    \end{matrix}\right)
    ,\quad T(k) \equiv \left(\begin{matrix}
        \cos\left(\frac{1}{\sqrt{3}}\frac{k}{\mathcal{H}_{\rm c}}\right)& \sin\left(\frac{1}{\sqrt{3}}\frac{k}{\mathcal{H}_{\rm c}}\right)\\
        -\frac{k}{\sqrt{3}\mathcal{H}_{\rm c}}\sin\left(\frac{1}{\sqrt{3}}\frac{k}{\mathcal{H}_{\rm c}}\right)& \frac{k}{\sqrt{3}\mathcal{H}_{\rm c}}\cos\left(\frac{1}{\sqrt{3}}\frac{k}{\mathcal{H}_{\rm c}}\right)
    \end{matrix}\right)\,.
\end{align}
\subsubsection{Numerical scheme}
\label{sec:numericalscheme}
To determine the transfer functions during the FOPT, we first numerically solve the background evolution equations, Eqs.~(\ref{dxda})-(\ref{detada}), and then work out the perturbation equations, given in Eqs.~(\ref{FOPTdeltaD})-(\ref{FOPTpsiSM}) for a given mode $k/\mathcal{H}_{\rm c}$ and for a fluid component (SM or dark sector), up to the percolation time. Since the perturbations are crucial for the calculation of the angular momentum of FV remnants, it is necessary to perform checks on our numerical scheme. In particular we compare our numerical transfer functions for the dark sector density and velocity perturbations, {\it i.e.},
\begin{eqnarray}
    \nonumber U_{Q_{\rm{I}}} \equiv \frac{Q_{\rm I}}{\mathcal{R}_k}\,,\quad Q_{\rm I} = \delta_{\rm D},\hat{\psi}_{\rm D}\,,
\end{eqnarray}
with (a) the approximate solutions in the subhorizon limit, given by Eqs.~(\ref{ApproxSubH1}) 
and~(\ref{ApproxSubH2}); and (b) the corresponding SM perturbations for modes with $k/\mathcal{H}_{\rm c} \gg 1$. 

\begin{figure}[t]
    \centering
    \resizebox{\linewidth}{!}{\begin{tabular}{cc}
\includegraphics[scale=0.26]{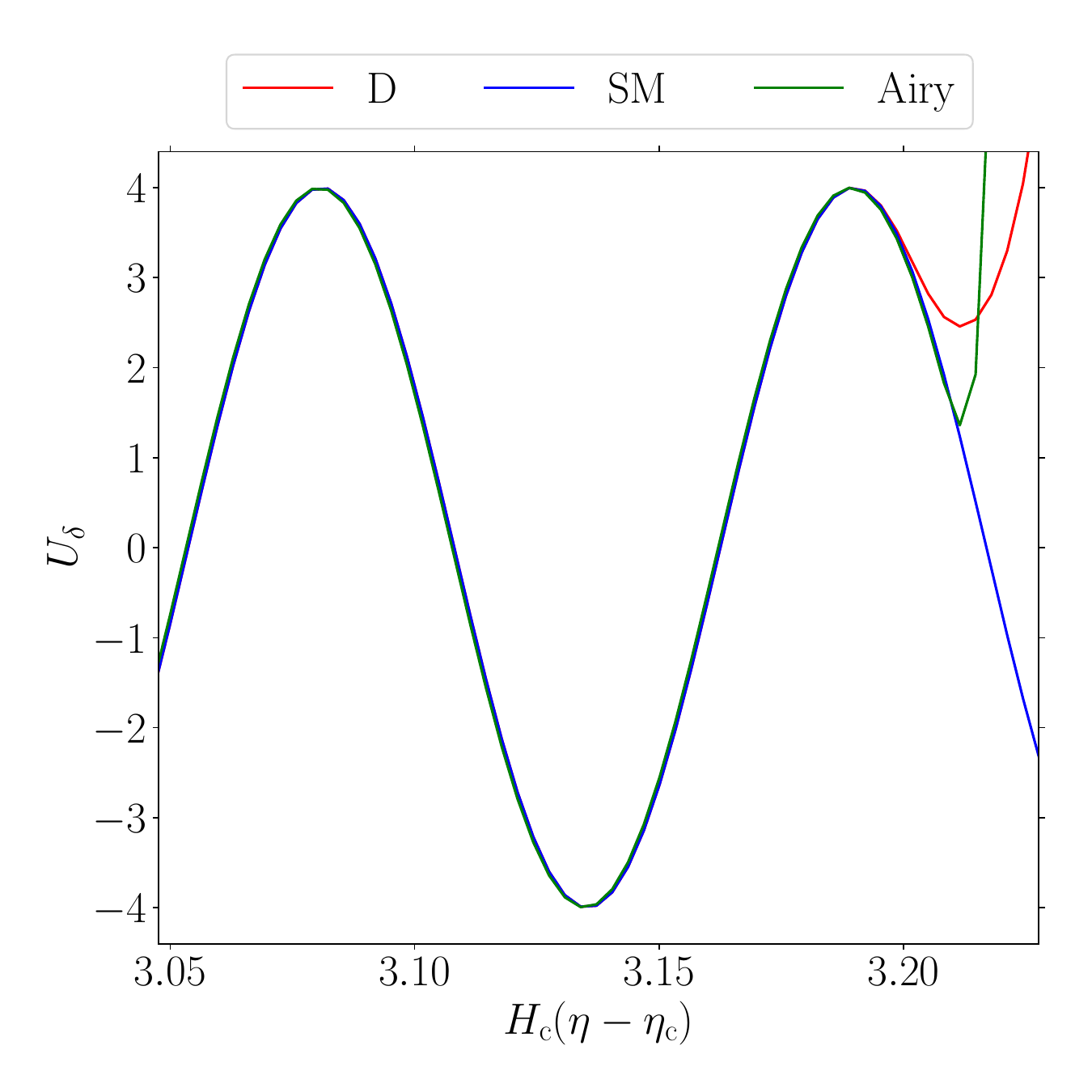} & \includegraphics[scale=0.26]{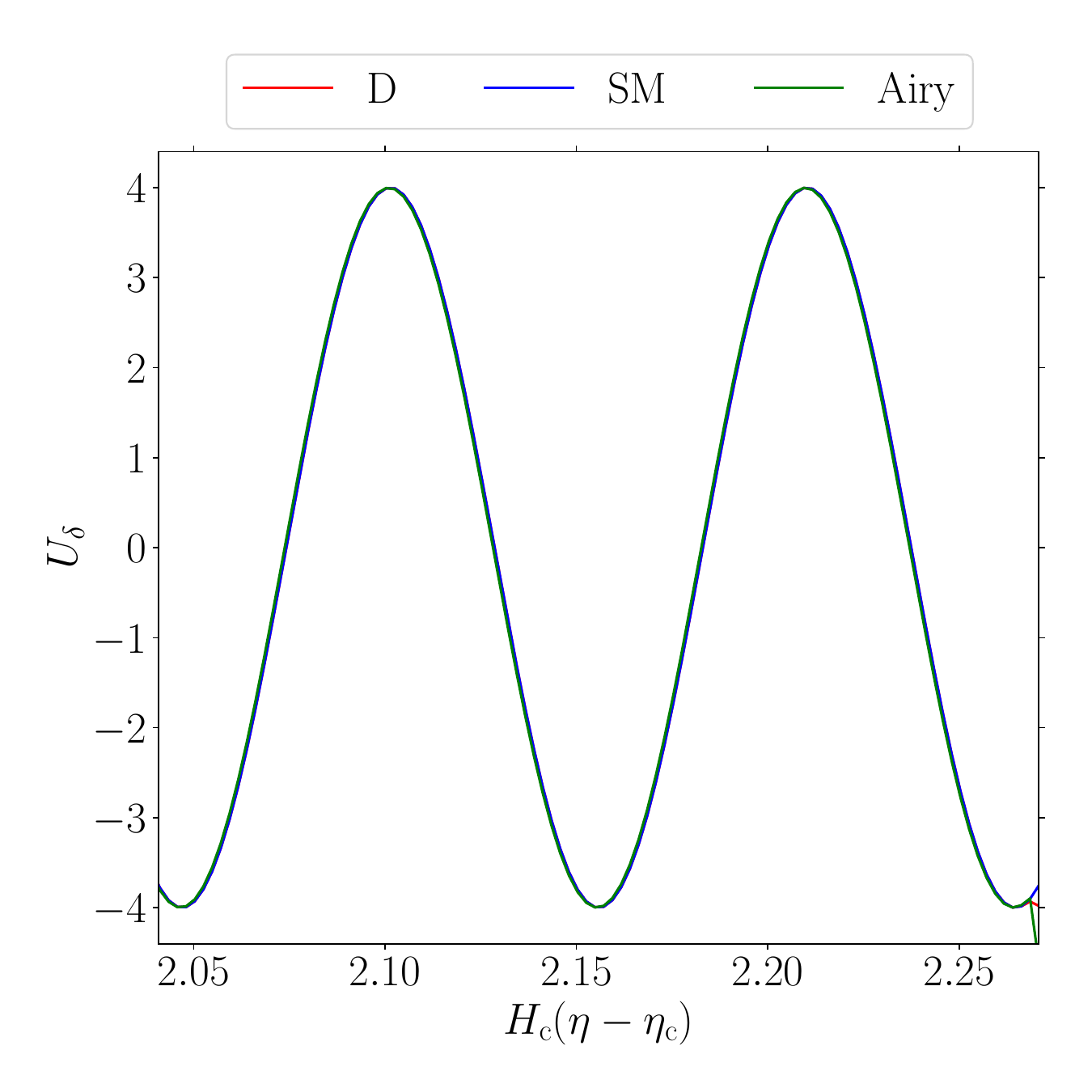} \\
\includegraphics[scale=0.26]{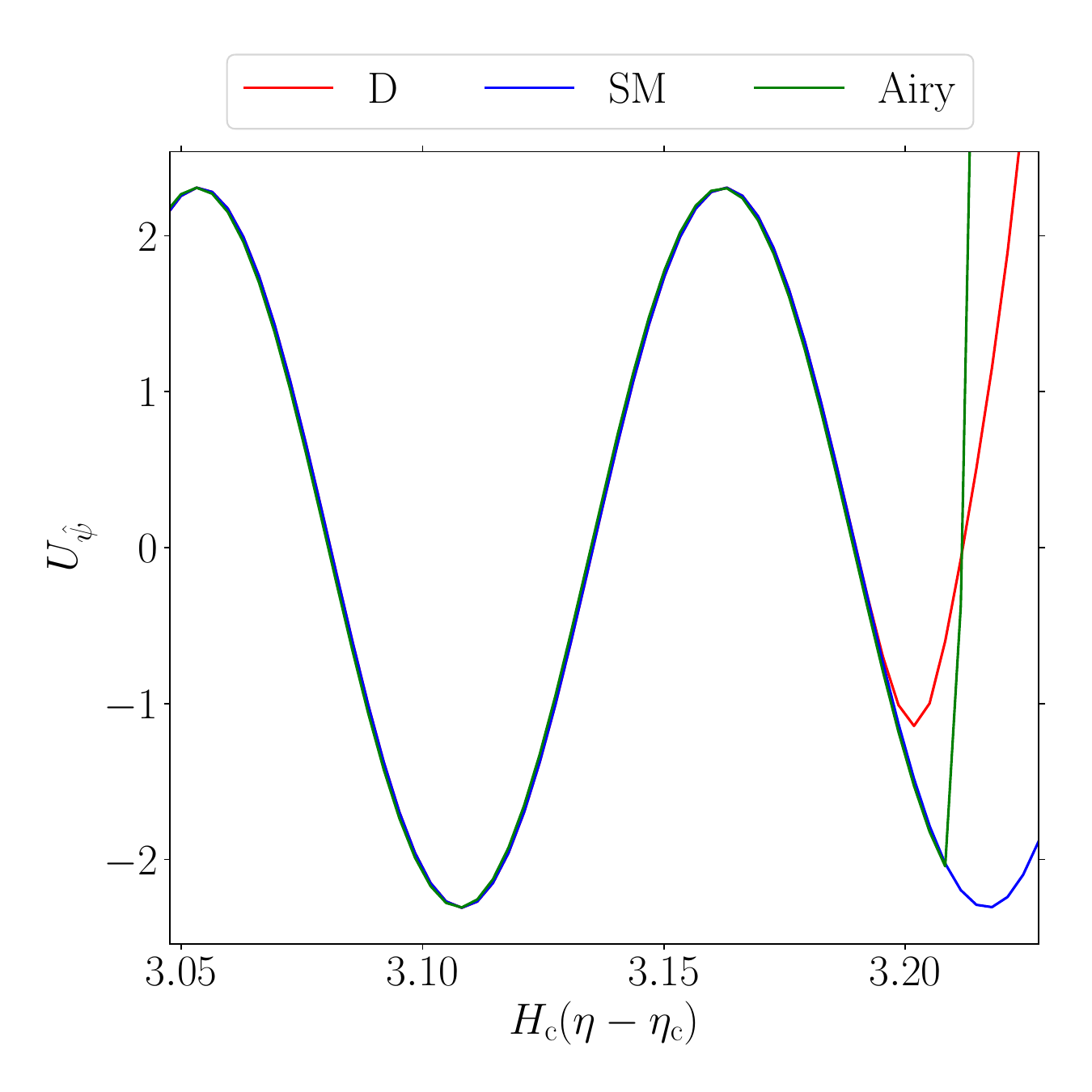} & \includegraphics[scale=0.26]{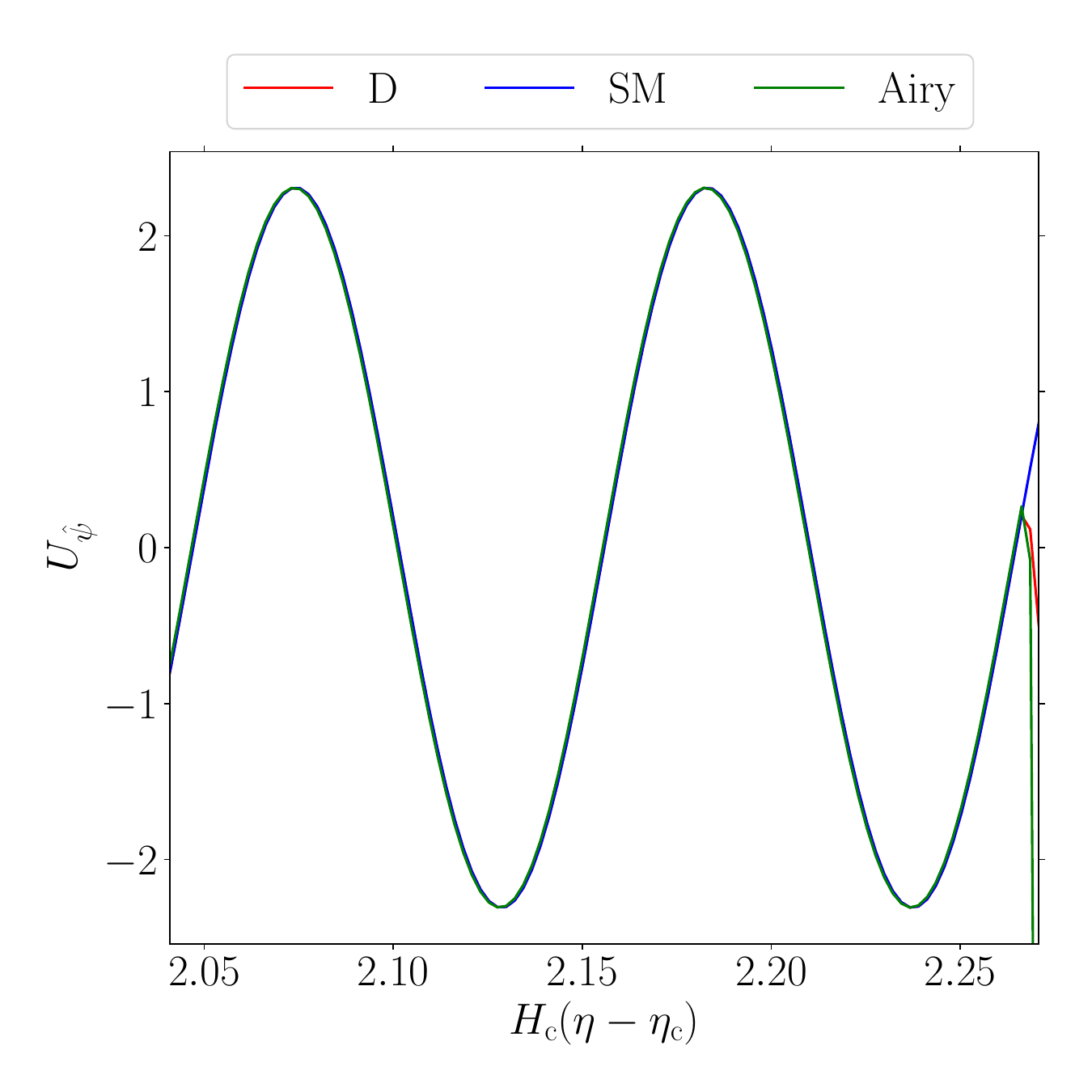} \\
(a) & (b)
    \end{tabular}}
      \caption{\label{fig:NicePert} The density (top) and velocity (bottom) perturbations for BP-2 (column a) and BP-4 (column b) for $k/\mathcal{H}_{\rm c}=10^2$.}
\end{figure}

In Fig.~\ref{fig:NicePert} we show plots of $U_\delta$ (top row) and $U_{\hat{\psi}}$ (bottom row) as functions of the rescaled elapsed conformal time, for BP-2 and BP-4, as in Fig.~\ref{fig:SampleEvol}, where we fix $k/\mathcal{H}_{\rm c} = 100$.
 In all panels, the solid red curves correspond to the dark sector perturbation, the blue correspond to the corresponding SM perturbation, and the green curves correspond to the approximate solution in the subhorizon limit, which we denote as the Airy function solution. In all cases, we observe excellent agreement between the Airy function approximation, the SM perturbation, and the numerical solution, in the time interval when the sound speed is close to $1/\sqrt{3}$. On the other hand, during the period when the sound speed squared quickly drops, the Airy function approximation captures the behavior of the numerical solution very well.
\begin{figure}[t]
    \centering
\resizebox{\linewidth}{!}{\begin{tabular}{cc}
 \includegraphics[scale=0.42]{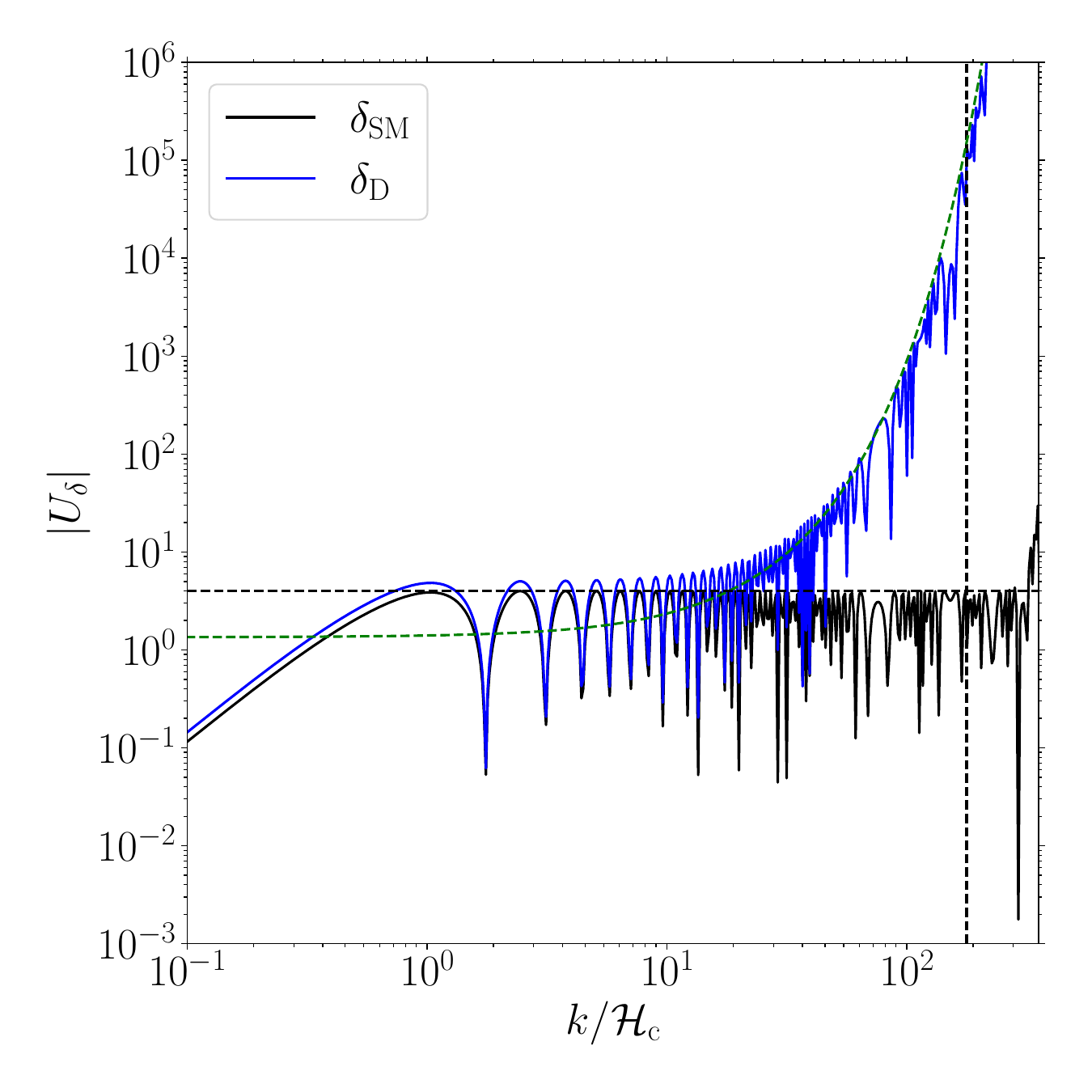}&\includegraphics[scale=0.42]{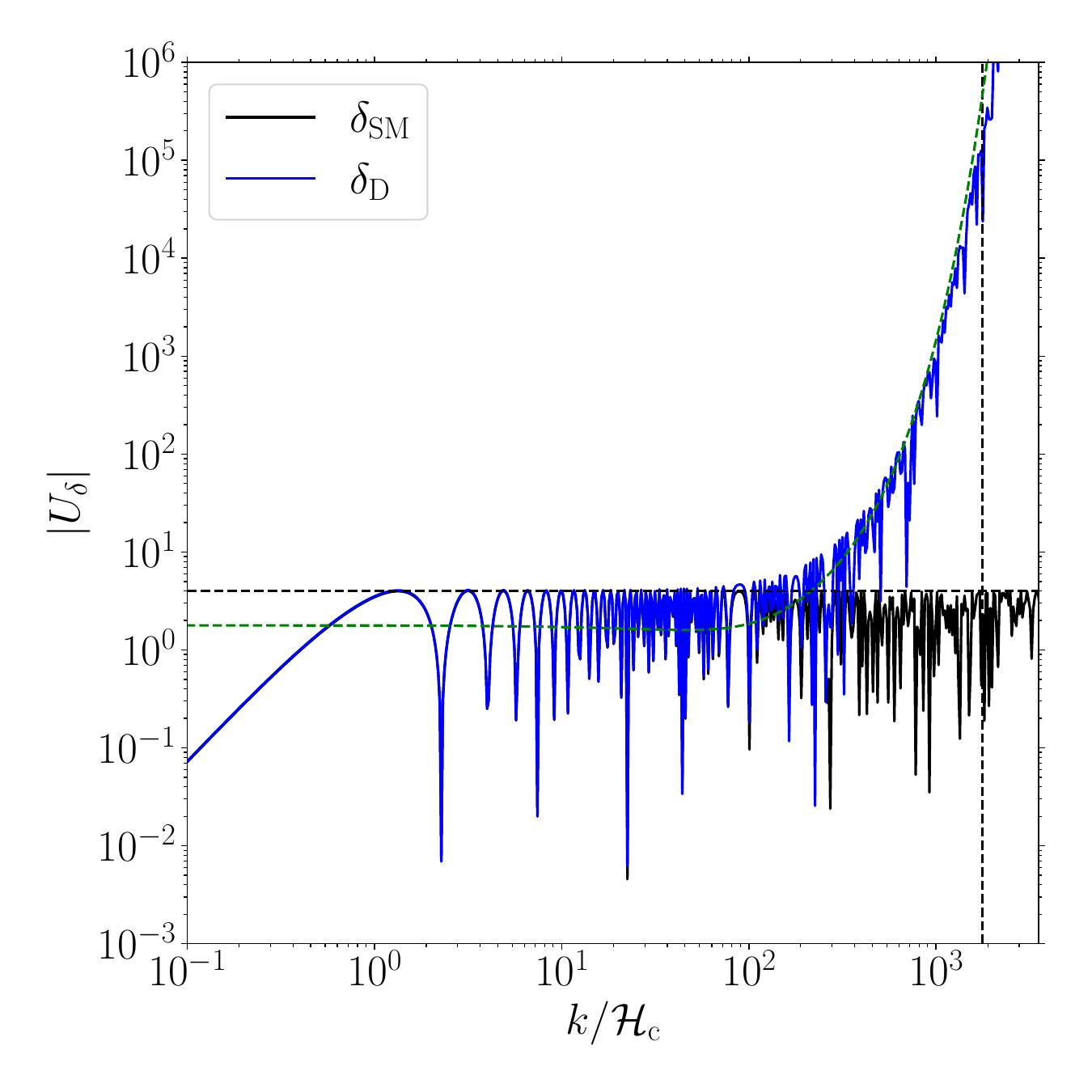}\\
  \includegraphics[scale=0.42]{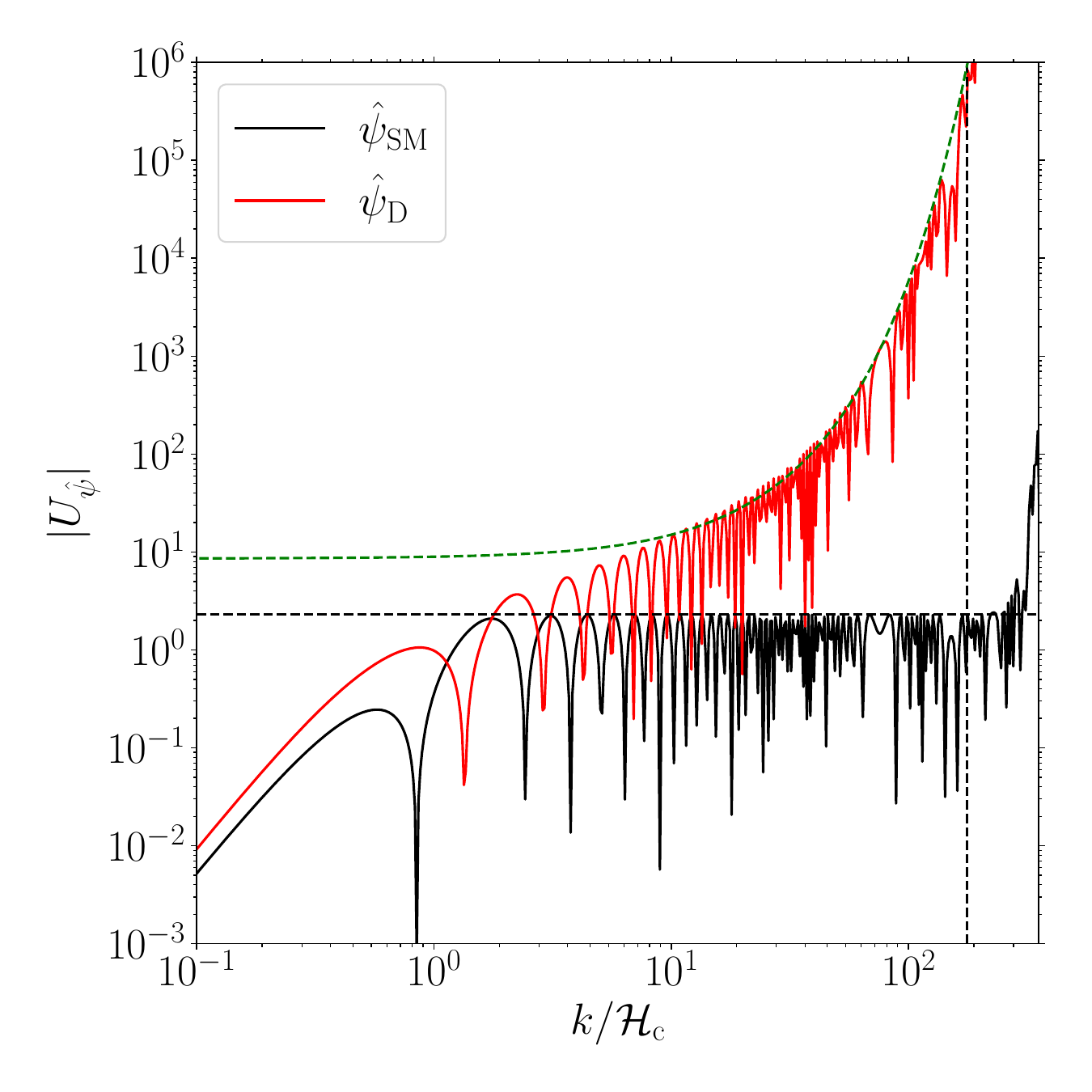}&\includegraphics[scale=0.42]{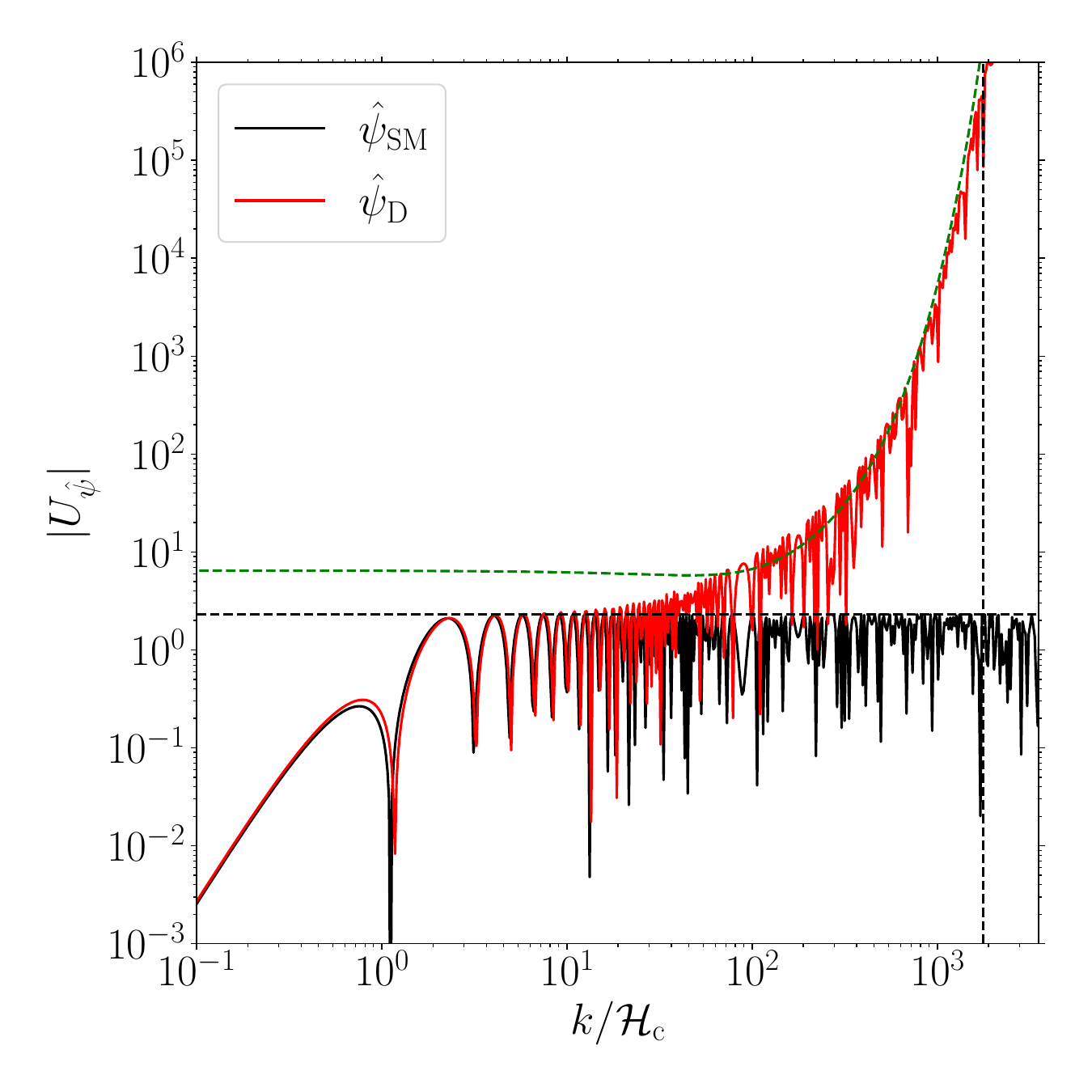}\\
 (a)&(b)
\end{tabular}}
    \caption{The absolute sizes of $U_\delta, U_{\hat{\psi}}$ at the percolation time, for $0.1 < k/\mathcal{H}_{\rm c} < k_{\rm max}/\mathcal{H}_{\rm c}$, and BP-2 (column a) and BP-4 (column b). In each panel we show the dark sector  perturbation (red), SM sector perturbation (blue), and Airy function approximation (green) to the perturbation.}
    \label{fig:Maintransfer} 
\end{figure}

 Ultimately, in calculating the angular momentum of FV remnants, we need the transfer functions at the percolation time. From a practical standpoint, only a finite number of modes can be calculated numerically. We have chosen to solve the perturbation equations up to the percolation time, for $N = 1000$ modes between $10^{-3} < k/\mathcal{H}_{\rm c} < k_{\rm max}/\mathcal{H}_{\rm c}$, where $k_{\rm max} \equiv 5/(H_* R_*)$, $H_*$ is the Hubble parameter at percolation, and $R_*$ is the physical radius of a FV bubble at percolation. In Fig.~\ref{fig:Maintransfer}, we show the absolute values of the transfer functions, evaluated at the percolation time, for modes within $0.1 < k/\mathcal{H}_{\rm c} < k_{\rm max}/\mathcal{H}_{\rm c}$. We take the physical radius $R_*$ of the FV bubble at percolation to be
\begin{eqnarray}
\label{RadiusPrescription}    n_{\rm FV}(t_*)~\frac{4\pi}{3}R_*^3 = F_*\,,
\end{eqnarray}
where $n_{\rm FV}(t_*)$ is the number density of FV bubbles at the percolation time. Equivalently, this critical radius $R_*$ is determined from the condition that no TV bubble nucleates inside, 
\begin{eqnarray}
    \Gamma(t_*)~\frac{R_*}{v_{\rm w,*}}~\frac{4\pi R_*^3}{3} = 1\,,
\end{eqnarray}
{\it i.e.},
\begin{eqnarray}
    R_* = \left(\frac{3}{4\pi}\right)^{1/4}\left(\frac{v_{\rm w, *}}{\Gamma_*}\right)^{1/4}\,.
\end{eqnarray}
 
 From panels (a) and (b) of Fig.~\ref{fig:Maintransfer}, we can see that the superhorizon modes follow the expected scaling behaviors $\delta_{\rm D} \propto (k/\mathcal{H})^2$ and $\hat{\psi}_{\rm D} \propto (k/\mathcal{H})^3$. In the subhorizon regime $k/\mathcal{H}_{\rm c} \gtrsim 1$, the rapid oscillations of $U_\delta$ and $U_{\hat{\psi}}$ are manifestations of the WKB behavior of the solutions in the deep subhorizon regime. For perturbative modes above a certain threshold scale $k/\mathcal{H}_{\rm c}$, we observe an enhancement in the transfer functions for the density perturbations. We can attribute this to the period when $c_{s\rm,D}^2$ exhibits a dip around the percolation time, as seen in the bottom row of Fig.~\ref{fig:SampleEvol}. This behavior is similar to the main results of \cite{Schmid:1998mx}, where the vanishing sound speed of the plasma leads to an enhancement in the density perturbations. For benchmarks BP-2 and BP-4, the width of the dip is $\mathcal{H}_{\rm c} (\eta_f - \eta_i) \sim \mathcal{O}\left(10^{-2}\right)$ and $\mathcal{O}\left(10^{-3}\right)$, respectively. Correspondingly, it is expected that those modes with wavenumbers of around $k/\mathcal{H}_{\rm c} \simeq \frac{1}{\mathcal{H}_{\rm c} (\eta_f-\eta_i)} \sim 100$ and $1000$, should have enough resolution to resolve the dip feature in the temporal evolution of the sound speed.
 
In the limit where $k/\mathcal{H}_{\rm c} \gg 1$, we can understand the growth of the transfer functions by first noting that the density and velocity perturbations at the percolation time can be obtained by using Eq. (\ref{ApproxSubH2}), setting $u = u_*$. We have
\begin{eqnarray}
\label{AiryApprox}    \left(\begin{matrix}
        \delta_*\\
        (k/\mathcal{H}_{\rm c})\hat{\psi}_*
    \end{matrix}\right) = M(k/\mathcal{H}_{\rm c},u_*)M^{-1}(k/\mathcal{H}_{\rm c},u_i)T(k)\left(\begin{matrix}
        \delta_{\rm c}\\
        \sqrt{3}\hat{\psi}_{\rm c}
    \end{matrix}\right)\,.
\end{eqnarray}
Note that
\begin{align}
    u_i &\equiv (-d_1 - d_2\tilde{\eta}_i)(-d_2)^{-2/3} = -\left[\frac{k\left(\tilde{\eta}_*-\tilde{\eta}_i\right)}{\sqrt{3}\mathcal{H}_{\rm c}}\right]^{2/3}\frac{1}{\left(1 + 3\vert c_{s\rm, D*}^2\vert\right)^{2/3}} < 0\,,\\
    u_* &\equiv (-d_1 - d_2\tilde{\eta}_*)(-d_2)^{-2/3} = \left[\frac{k\left(\tilde{\eta}_*-\tilde{\eta}_i\right)}{\sqrt{3}\mathcal{H}_{\rm c}}\right]^{2/3}\frac{3\vert c_{s\rm, D*}^2\vert}{\left(1 + 3\vert c_{s\rm, D*}^2\vert\right)^{2/3}} > 0\,.
\end{align}
In most FOPT scenarios relevant to this work, especially the ones we have featured in Table~\ref{table:FinalBPsOld}, we have $\vert c_{s\rm,D *}^2\vert \gg 1$, which may simplify our asymptotic expressions. We then use the asymptotic expansions for the Airy functions and their derivatives, listed in 
Table~\ref{tab:airy}. 
\begin{table}[t]
    \centering
    \begin{tabular}{ccc}\\\toprule
         & $u > 0, u \gg 1$ & $u < 0, \vert u\vert \gg 1$\\\toprule
        $Ai$ & $\frac{u^{-1/4}}{2\sqrt{\pi}} \exp\left(-\frac{2u^{3/2}}{3}\right)$  & $\frac{\vert u\vert^{-1/4}}{\sqrt{\pi}}\cos\left(\frac{2\vert u\vert^{3/2}}{3}-\frac{\pi}{4}\right)$\\
        $Bi$ & $\frac{u^{-1/4}}{\sqrt{\pi}} \exp\left(\frac{2u^{3/2}}{3}\right)$ & $-\frac{\vert u\vert^{-1/4}}{\sqrt{\pi}}\sin\left(\frac{2\vert u\vert^{3/2}}{3}-\frac{\pi}{4}\right)$\\\midrule
        $Ai'$ & $-\frac{u^{1/4}}{2\sqrt{\pi}} \exp\left(-\frac{2u^{3/2}}{3}\right)$  & $\frac{\vert u\vert^{1/4}}{\sqrt{\pi}}\sin\left(\frac{2\vert u\vert^{3/2}}{3}-\frac{\pi}{4}\right)$\\
        $Bi'$ & $\frac{u^{1/4}}{\sqrt{\pi}} \exp\left(\frac{2u^{3/2}}{3}\right)$ & $\frac{\vert u\vert^{1/4}}{\sqrt{\pi}}\cos\left(\frac{2\vert u\vert^{3/2}}{3}-\frac{\pi}{4}\right)$\\        \bottomrule
    \end{tabular}
    \caption{\label{tab:airy} Asymptotic forms of Airy functions and their derivatives. }
\end{table}
We define 
\begin{eqnarray}
    \Delta \chi_* \equiv \frac{k}{\mathcal{H}_{\rm c}}\vert c_{s\rm,D*}^2\vert^{1/2}\Delta\tilde{\eta}_*\,,\quad \Delta\tilde{\eta}_* \equiv \left(\tilde{\eta}_*-\tilde{\eta}_i\right)\,,
\end{eqnarray}
which is analogous to the argument of the exponential term in the WKB wavefunction, for a particle in the classically forbidden region. One can show that, in the limit $\vert c_{s\rm,D*}^2\vert \gg 1$, we have
\begin{eqnarray}
    u_* \approx (\Delta\chi_*)^{2/3}\,,\quad -u_i \approx \frac{(\Delta\chi_*)^{2/3}}{3\vert c_{s\rm,D*}^2\vert}\,, \quad \left(-d_2\right)^{1/3} \approx \frac{(\Delta\chi_*)^{2/3}}{\Delta\tilde{\eta}_*}\,.
\end{eqnarray}
In terms of the phase acquired by the perturbations up to $\tilde{\eta}_i$, 
\begin{eqnarray}
    \chi_{i} \equiv \frac{k}{\sqrt{3}\mathcal{H}_{\rm c}}\left(1 + \tilde{\eta}_i\right)\,,
\end{eqnarray}
the density and velocity perturbations at percolation are approximately,
\begin{align}
  \nonumber \frac{\delta_*}{\mathcal{R}_k} &\approx \exp\left(\frac{2\Delta\chi_*}{3}\right)\Bigg[-\frac{4}{3^{1/4}\vert c_{s\rm,D*}^2\vert^{1/4}}\left(1+\frac{5}{48\Delta\chi_*}\right)\cos\left(\chi_i + \frac{\pi}{4}+\frac{2\Delta\chi_*}{9\sqrt{3}\vert c_{s\rm,D*}^2\vert^{3/2}}\right)\\
  \nonumber &+\mathcal{O}\left(\vert c_{s\rm,D*}^2\vert^{5/4}/\Delta\chi_*\right)\Bigg]\\
   \nonumber &+\exp\left(-\frac{2\Delta\chi_*}{3}\right)\left[-\frac{2}{3^{1/4}\vert c_{s\rm,D*}^2\vert^{1/4}}\sin\left(\chi_i+\frac{\pi}{4}+\frac{2\Delta\chi_*}{9\sqrt{3}\vert c_{s\rm,D*}^2\vert^{3/2}}\right)+\mathcal{O}\left(\vert c_{s\rm,D*}^2\vert^{5/4}/\Delta\chi_*\right)\right]\,,\\
   \label{ApproxLargePhiDeltaPerc} &
\end{align}
\begin{align}
  \nonumber  \frac{k}{\mathcal{H}_{\rm c}}\frac{\hat{\psi}_*}{\mathcal{R}_k} &\approx \exp\left(\frac{2\Delta\chi_*}{3}\right)\Bigg[-\frac{4\Delta\chi_*}{3^{1/4}\vert c_{s\rm,D*}^2\vert^{1/4}\Delta\tilde{\eta}_*}\left(1-\frac{7}{48\Delta\chi_*}\right)\cos\left(\chi_i + \frac{\pi}{4}+\frac{2\Delta\chi_*}{9\sqrt{3}\vert c_{s\rm,D*}^2\vert^{3/2}}\right)\\
  \nonumber &+\mathcal{O}\left(\vert c_{s\rm,D*}^2\vert^{5/4}/\Delta\chi_*\right)\Bigg]\\
    \nonumber &+\exp\left(-\frac{2\Delta\chi_*}{3}\right)\left[\frac{2\Delta\chi_*}{3^{1/4}\vert c_{s\rm,D*}^2\vert^{1/4}\Delta\tilde{\eta}_*}\sin\left(\chi_i + \frac{\pi}{4}+\frac{2\Delta\chi_*}{9\sqrt{3}\vert c_{s\rm,D*}^2\vert^{3/2}}\right)+\mathcal{O}\left(\vert c_{s\rm,D*}^2\vert^{5/4}/\Delta\chi_*\right)\right]\,.\\
  \label{ApproxLargePhikoHcPsiPerc}&
\end{align}
The leading terms are
\begin{align}
    \frac{\delta_*}{\mathcal{R}_k} &\approx -\frac{4}{(3\vert c_{s\rm,D*}^2\vert)^{1/4}}~\frac{1}{2}\left(2\epsilon_+\cos\chi + \epsilon_- \sin\chi\right)\,,\\
    \frac{k}{\mathcal{H}_{\rm c}} \frac{\hat{\psi}_*}{\mathcal{R}_k} &\approx -\frac{4\Delta\chi_*}{(3\vert c_{s\rm,D*}^2\vert)^{1/4}\Delta\tilde{\eta}_*}~\frac{1}{2}\left(2\epsilon_+ \cos\chi-\epsilon_- \sin\chi\right)\,,\\
 {\rm where}\ \  \epsilon_\pm &\equiv \exp(\pm 2\Delta\chi_*/3)\,,\quad \chi \equiv \chi_i + \frac{\pi}{4} + \frac{2\Delta\chi_*}{9\sqrt{3}\vert c_{s\rm,D*}^2\vert^{3/2}}\,.
\end{align}
Keeping the term of order $(\Delta\chi_*)^0$ for $\delta_*$ (and order $\Delta\chi_*$ for $k/\mathcal{H}_{\rm c} \hat{\psi}_*$) amounts to taking the limit $k/\mathcal{H}_{\rm c} \gg 1$. The respective amplitudes of $\delta_*$ and $\hat{\psi}_*$ are
\begin{eqnarray}
   \label{ApproxDeltaPsiLargekoHc} \frac{\vert \delta_*\vert}{\mathcal{R}_k} \leq \frac{2}{(3\vert c_{s\rm,D*}^2\vert)^{1/4}} \sqrt{4\epsilon_+^2 + \epsilon_-^2}\,,\quad  \frac{\vert \hat{\psi}_*\vert}{\mathcal{R}_k} \leq \frac{2(3\vert c_{s\rm,D*}^2\vert)^{1/4}}{\sqrt{3}} \sqrt{4\epsilon_+^2 + \epsilon_-^2}\,.
\end{eqnarray}
In Fig.~\ref{fig:Maintransfer} we plot the transfer functions for the density and velocity perturbations for BP-2 and BP-4 as solid curves, along with the amplitude of the approximate transfer functions provided in 
Eq.~(\ref{ApproxDeltaPsiLargekoHc}) as dashed green curves. We find that the dashed green curve in both panels captures the exponential growth of the density and velocity perturbations for sufficiently large $k/\mathcal{H}_{\rm c}$. On the other hand, for longer wavelength modes that are still in the subhorizon regime, the amplitudes of the density and velocity transfer functions are close to the amplitudes for the pure radiation background, \textit{i.e.}, 4 and $4/\sqrt{3}$, respectively.

\begin{figure}[t]
\centering
\begin{tabular}{cc}
\includegraphics[scale=0.22]{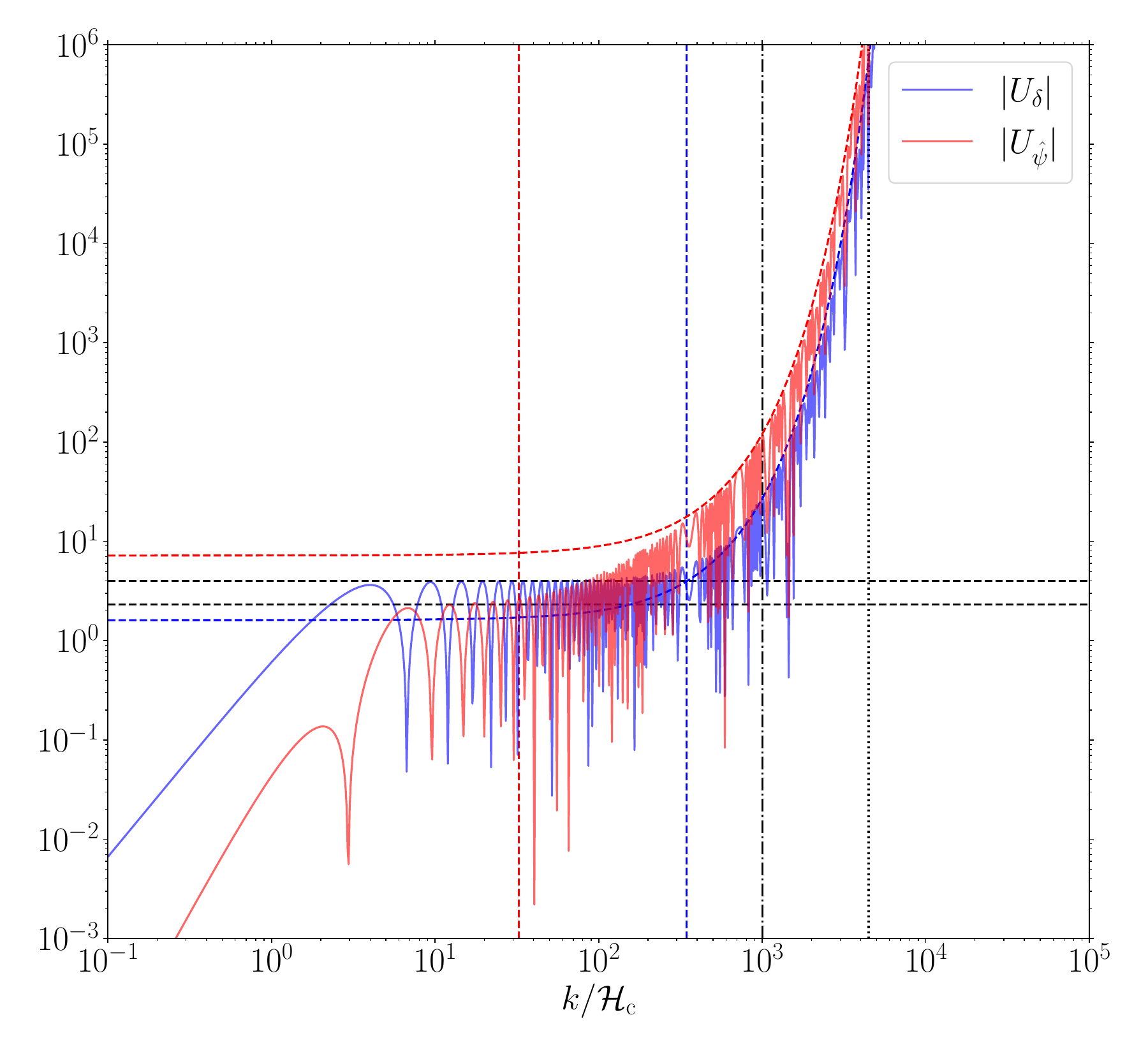}&\includegraphics[scale=0.22]{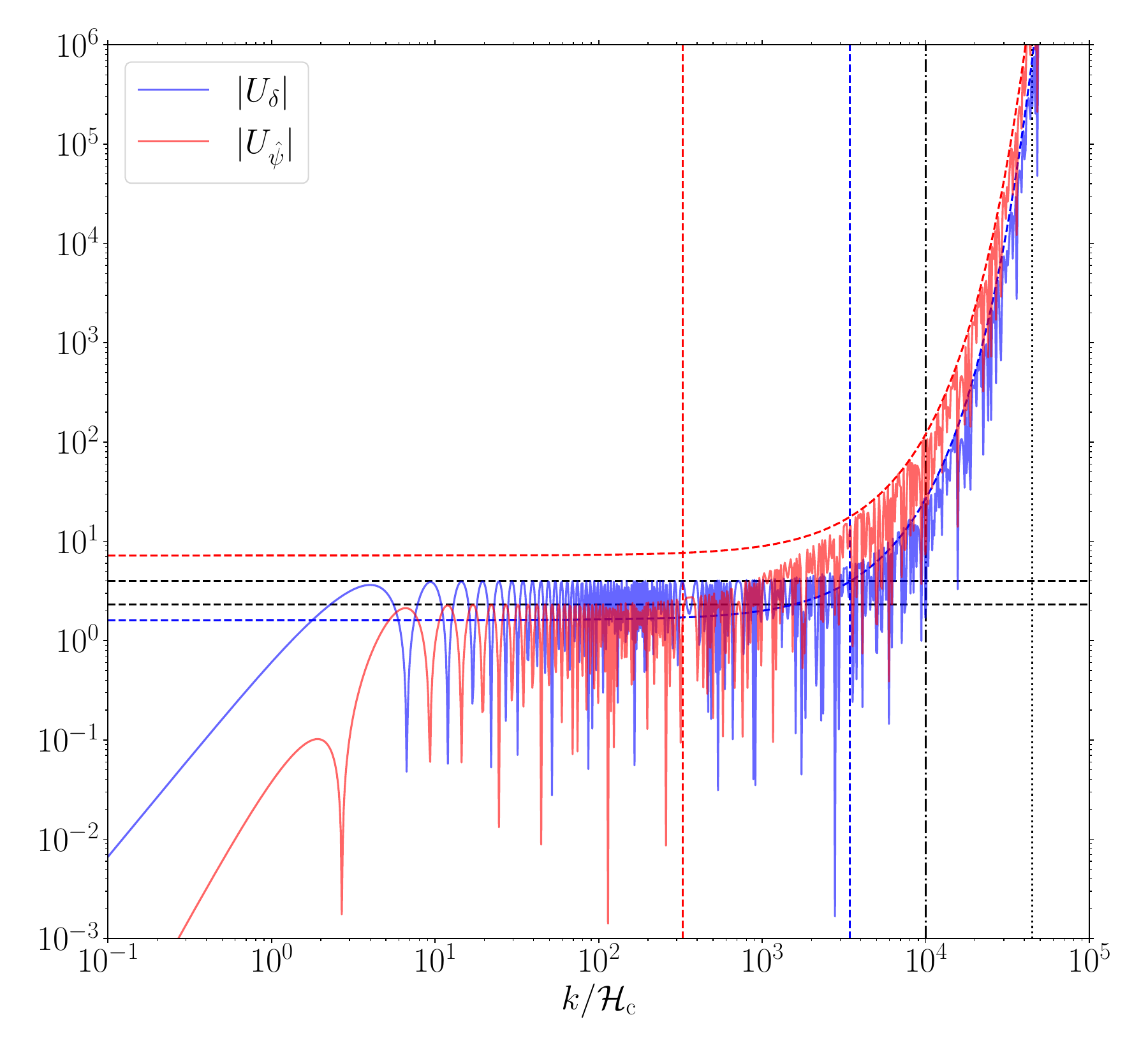}\\
$\Delta\tilde{\eta}_* = 10^{-3}, \tilde{\eta}_* = 0.1, c_{s\rm, D*}^2 = -20$&$\Delta\tilde{\eta}_* = 10^{-4}, \tilde{\eta}_* = 0.1, c_{s\rm, D*}^2 = -20$\\
\includegraphics[scale=0.22]{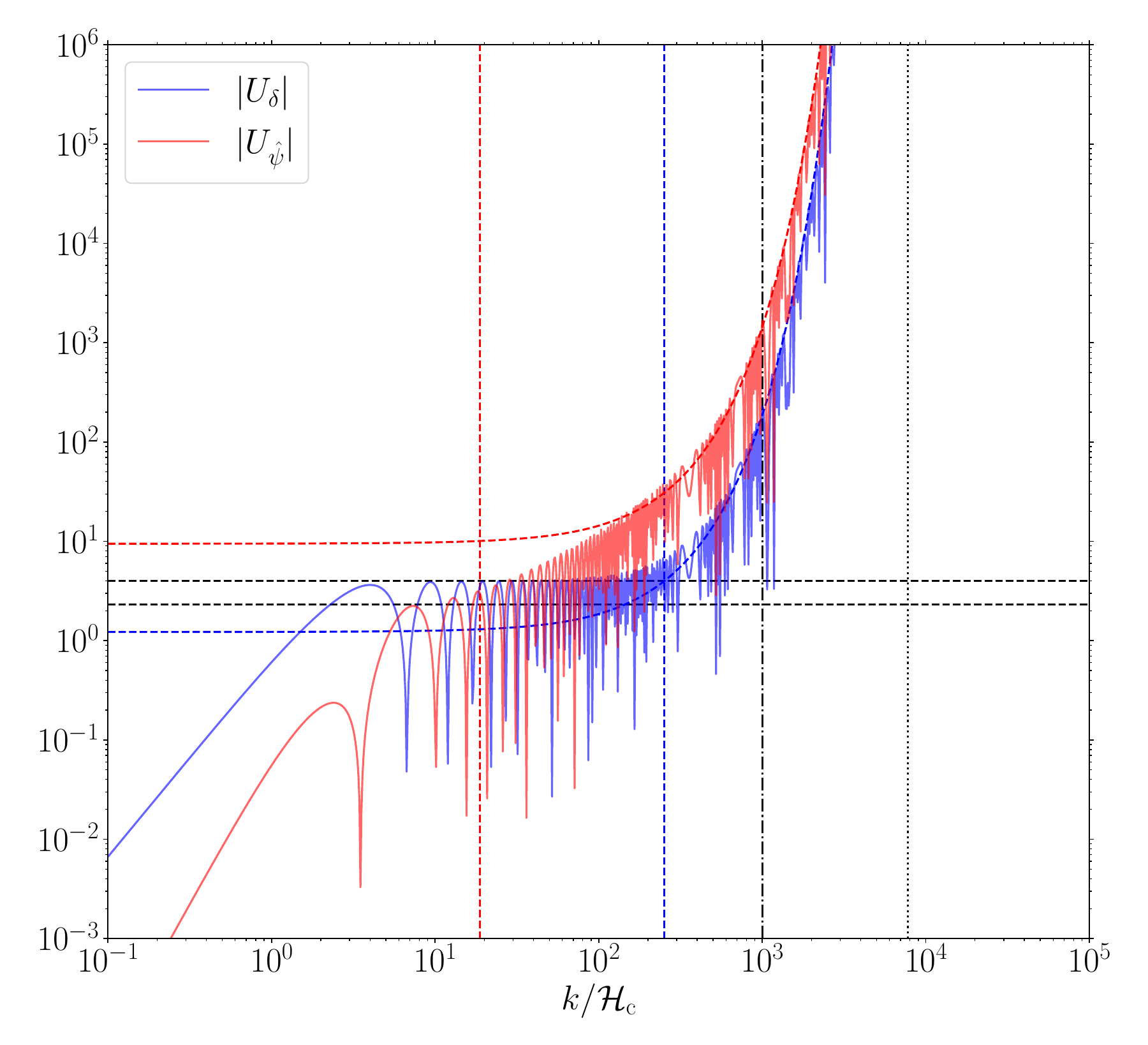}&\includegraphics[scale=0.22]{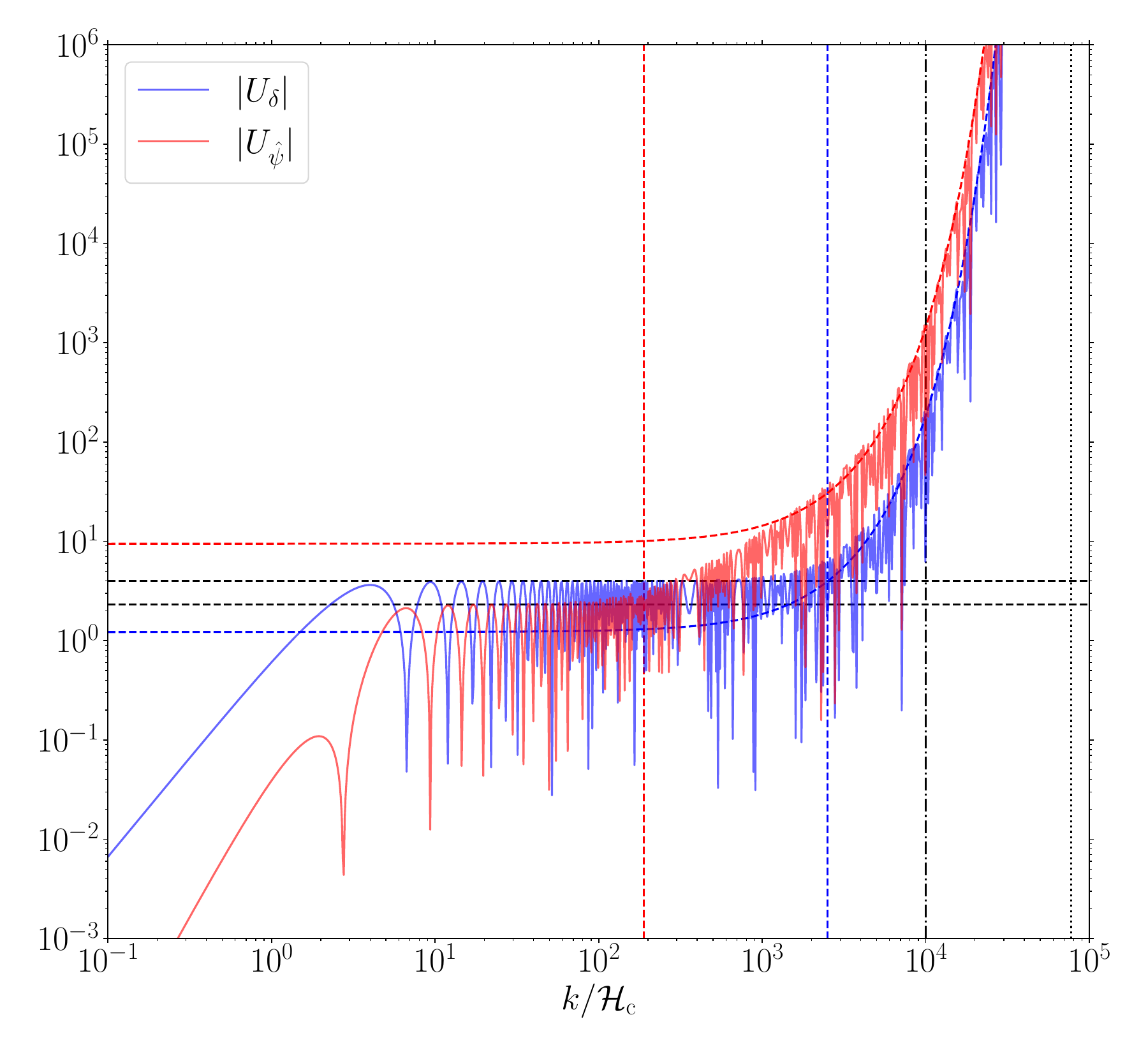}\\
$\Delta\tilde{\eta}_* = 10^{-3}, \tilde{\eta}_* = 0.1, c_{s\rm, D*}^2 = -60$&$\Delta\tilde{\eta}_* = 10^{-4}, \tilde{\eta}_* = 0.1, c_{s\rm, D*}^2 = -60$
\end{tabular}
\caption{\label{fig:Airy} The density and velocity transfer functions, approximating $c_{s\rm,D}^2$ via Eq.~(\ref{cs2DLinApprox}), when the width of the dip in $c_{s\rm,D}^2$ (top) or the sound speed squared at percolation (bottom) is changed. The horizontal dashed lines correspond to the amplitude of the density and velocity transfer functions for perturbations in a pure radiation background. The vertical dashed lines are the critical Fourier scales where the exponential enhancement for the density and velocity perturbations becomes relevant. The dot dashed vertical line is the scale $1/\Delta\tilde{\eta}_*$, and the dotted line marks the Fourier scale beyond which the term of order $\vert c_{s\rm,D*}^2\vert/\Delta\chi_*$ becomes subdominant.}
\end{figure}

We can understand the overall dependence of the transfer functions with the scale of the perturbations by considering the Airy function approximation to the transfer functions, which we plot in Fig.~\ref{fig:Airy}. In this approximation, the shape of the transfer functions are dictated by $\tilde{\eta}_i$, $\tilde{\eta}_*$, and $c_{s\rm,D*}^2$. Furthermore, we assume that we are in the regime where $\vert c_{s\rm,D*}^2 \vert \gg 1$. In the case of the density contrast, the deviation from the expected transfer function in the case of a pure radiation background, according to 
Eq.~(\ref{ApproxLargePhiDeltaPerc}), becomes relevant for scales,
\begin{eqnarray}
  \label{LimScaleDelta}  \exp\left(\frac{2\Delta\chi_*}{3}\right)\frac{1}{\left(3\vert c_{s\rm,D*}^2\vert\right)^{1/4}} \simeq 1 \Rightarrow \frac{k}{\mathcal{H}_{\rm c}} \gtrsim \frac{3}{8}\frac{\ln\left(3\vert c_{s\rm,D*}^2\vert\right)}{\vert c_{s\rm,D*}^2\vert^{1/2}\Delta\tilde{\eta}_*}\,.
\end{eqnarray}
As for the velocity perturbation, 
Eq.~(\ref{ApproxLargePhikoHcPsiPerc}) implies that the deviation from the pure radiation background occurs when
\begin{eqnarray}
  \label{LimScalePsi}  \exp\left(\frac{2\Delta\chi_*}{3}\right) \left(3\vert c_{s\rm,D*}^2\vert\right)^{1/4}\left(1 - \frac{7}{48\Delta\chi_*}\right) \simeq 1 \Rightarrow \frac{k}{\mathcal{H}_{\rm c}} \gtrsim \frac{7}{48\vert c_{s\rm,D*}^2\vert^{1/2}\Delta\tilde{\eta}_*}\,.
\end{eqnarray}
Also, the size of the next to leading order term, relative to the leading term is $\sim \vert c_{s\rm,D*}^2\vert^{3/2}/\Delta\chi_* \sim \vert c_{s\rm,D*}^2\vert/(\Delta\tilde{\eta}_*~k/\mathcal{H}_{\rm c})$. Then $k/\mathcal{H}_{\rm c}$ should be larger than $\sim \vert c_{s\rm,D*}^2\vert/\Delta\tilde{\eta}_*$ to keep the next to leading order smaller than the leading term. In Fig.~\ref{fig:Airy}, vertical blue/red dashed lines, respectively, mark the range of modes identified by Eqs.~(\ref{LimScaleDelta}) and (\ref{LimScalePsi}). For comparison, we mark the scale $k/\mathcal{H}_{\rm c} = 1/(\tilde{\eta}_* - \tilde{\eta}_i)$ by a dashed dot line, and the scale $k/\mathcal{H}_{\rm c} = \vert c_{s\rm,D*}^2\vert/(\tilde{\eta}_* - \tilde{\eta}_i)$ by a dotted line.

The approximate treatment of assuming a linear drop in $c_{s\rm,D}^2$ allows us to identify the factors that affect the transfer functions, namely: the period $\Delta\tilde{\eta}_*$ when the drop in $c_{s\rm,D}^2$ occurs; and the sound speed squared $c_{s\rm, D*}^2$ at the percolation time. We demonstrate the effect of these quantites on the transfer functions in Fig.~\ref{fig:Airy}. In the top row, we changed $\Delta\tilde{\eta}_*$, effectively changing the width of the dip in $c_{s\rm,D}^2$ (across columns). As expected, reducing the width simply moves the vertical lines that mark the modes that suffer enhancement in the density to the right. In the bottom row, we changed $c_{s\rm, D*}^2$ from -20 to -60. Decreasing $c_{s\rm, D*}^2$ has a similar effect of moving the locations of the critical Fourier scales as in the case of changing $\Delta \tilde{\eta}_*$, but they are shifted to the left.
\section{Results and discussion}
\label{sec:MainResultsDisc} 
\subsection{Numerical values of physical quantities}
To calculate the RMS angular momentum and spin of a sphere of dark plasma at the percolation time, we use Eq.~(\ref{JCMrmsmaster}), which requires us to know the enclosed mass and physical radius, as well as the coefficient $\mathcal{C}(\tilde{x}_0, x_0)$.  For an arbitrary conformal time $\eta$ and $\tilde{x}_0$, the RMS values of the angular momentum and spin at any given time can be calculated from
\begin{empheq}[box=\fbox]{align}
   \label{MainJsDefn}    J_{\rm CM,rms}(\eta) = M(\eta) R(\eta) v_{\rm eff}&\,,\quad s_{\rm rms} = \frac{R(\eta)}{G_{\rm N} M(\eta)}v_{\rm eff}\,,\\
    \label{MainMDDefn} M(\eta) = \frac{4\pi}{3}R^3(\eta) \rho_{\rm D}(\eta)&\,,\quad v_{\rm eff} = \frac{3}{4}~A_{\rm s} \tilde{x}_0(\eta) \mathcal{C}^{1/2}(\tilde{x}_0, x_0)\,.
\end{empheq}
From Eq.~(\ref{MainMDDefn}) we see that the spin is a product of $R/(G_{\rm N} M)$, the inverse of the surface potential, and $v_{\rm eff}$, the effective \textit{equatorial velocity} of the FV remnant. The inverse of the surface potential can be calculated by knowing only the background evolution, whereas $v_{\rm eff}$ requires the $\mathcal{C}$ function. At percolation, we simply evaluate $\mathcal{C}(\tilde{x}_0, x_0)$ at $\tilde{x}_0 = \sqrt{3}\mathcal{H}_*x_{0,*} = \sqrt{3} H_* R_*$, which we will then denote by $\mathcal{C}_*$. 

The coefficient $\mathcal{C}_*$, expressed as an integral in Eq.~(\ref{CFunctionMaster}) and evaluated at the percolation time, contains explicit dependence on the primordial power spectrum of curvature perturbations, and the transfer functions for all $k$ modes. We perform the $\mathcal{C}_*$ integration using a general purpose Monte Carlo integrator, where we evaluate the integrand using a lookup table for the transfer functions covering the range $10^{-3} < k/\mathcal{H}_{\rm c} < k_{\rm max}/\mathcal{H}_{\rm c}$, and setting the cutoff scale in the integral at
\begin{eqnarray}
    k_{\rm cut} = \frac{\pi}{R_*}\,,
\end{eqnarray}
which corresponds to a mode whose wavelength matches the FV bubble diameter. We do not consider modes that are shorter than $k_{\rm cut}$, since they can resolve the false and true vacuum regions, and the dark plasma can no longer be considered as a single entity. Furthermore, the cutoff in $k$ can be justified by looking at the volume average, within a sphere of radius $R$, of the mean square density contrast in position space:
\begin{eqnarray}
   \langle \delta^2(\vec{r})\rangle_V &\equiv& \frac{1}{V_R}\int_{R} d^3\vec{r}~\langle \delta^2(\vec{r})\rangle\\
   &=& \int \frac{d^3\vec{k}}{(2\pi)^3}\frac{d^3\vec{k}'}{(2\pi)^3}~\langle \delta_{\vec{k}}\delta_{\vec{k}'}\rangle~\frac{1}{V_R}\int_R d^3\vec{r}~\exp\left[-i\left(\vec{k}-\vec{k}'\right)\cdot\vec{r}\right]\\
   \label{meansqrdelta} &=& \lim_{k_{\rm th} \rightarrow \infty}\int_0^{k_{\rm th}} \frac{dk}{k}~\frac{k^3 P_{\mathcal{R}}(k)}{2\pi^2}\left\vert U_\delta(k)\right\vert^2\,.
\end{eqnarray}
\begin{figure}
    \centering
 \includegraphics[scale=0.26]{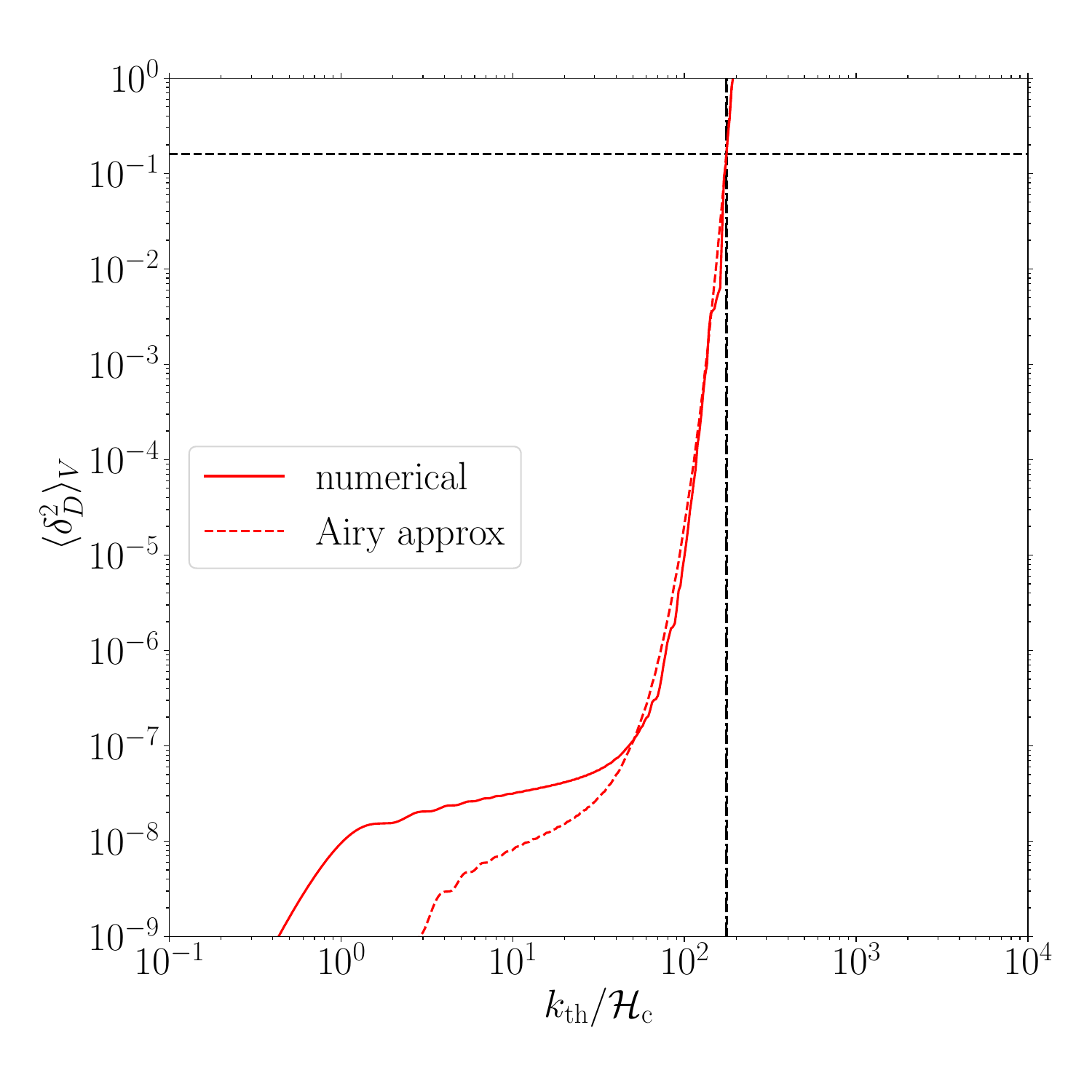}\includegraphics[scale=0.26]{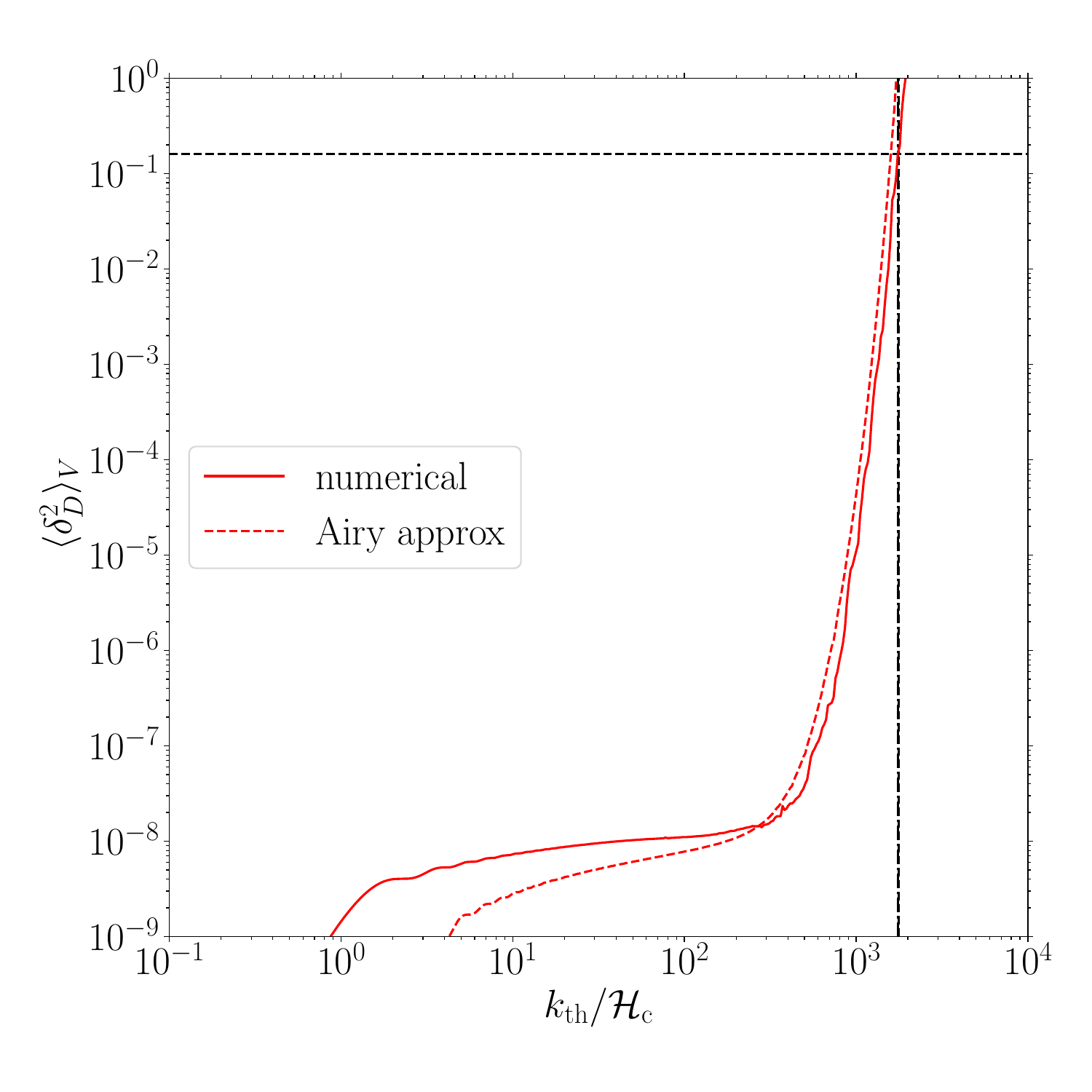}
    \caption{\label{fig:d2} Plot of the mean square density contrast as a function of the integral cutoff $k_{\rm th}/\mathcal{H}_{\rm c}$, for BP-2 (left) and BP-4 (right). The red dashed curves represent the approximate mean squared density contrast, which is obtained by replacing the transfer function in Eq.~(\ref{meansqrdelta}) with Eq.~(\ref{AiryApprox}). In both panels, we indicate, using vertical lines, the value of $k_{\rm th}/\mathcal{H}_c$ at which the RMS density contrast is 0.4 ($k_{\rm BH}/\mathcal{H}_{\rm c}$), and the value of $k_{\rm cut}/\mathcal{H}_{\rm c}$. The lines overlap each other. }
\end{figure}
As Fig. \ref{fig:d2} shows, $\langle \delta^2\rangle_V$ blows up if $k_{\rm th}$ is taken to infinity. A finite cutoff will, of course, lead to a finite value of $\langle \delta^2\rangle_V$, and we are interested in knowing the cutoff, $k_{\rm BH}/\mathcal{H}_{\rm c}$, where the mean square density contrast is at most the square of the threshold density contrast for PBH formation, which we take to be $\delta_{\rm BH}^2 \simeq 0.16$. The scale $k_{\rm BH}$ sets a physical cutoff based on the threshold for PBH formation, so it is desirable to have $k_{\rm cut}$ as close to $k_{\rm BH}$ as possible, as is the case for BP-2 and BP-4.
~\\
\begin{figure}[t]
    \centering
    \resizebox{\columnwidth}{!}{\begin{tabular}{cc}
     \includegraphics[scale=0.26]{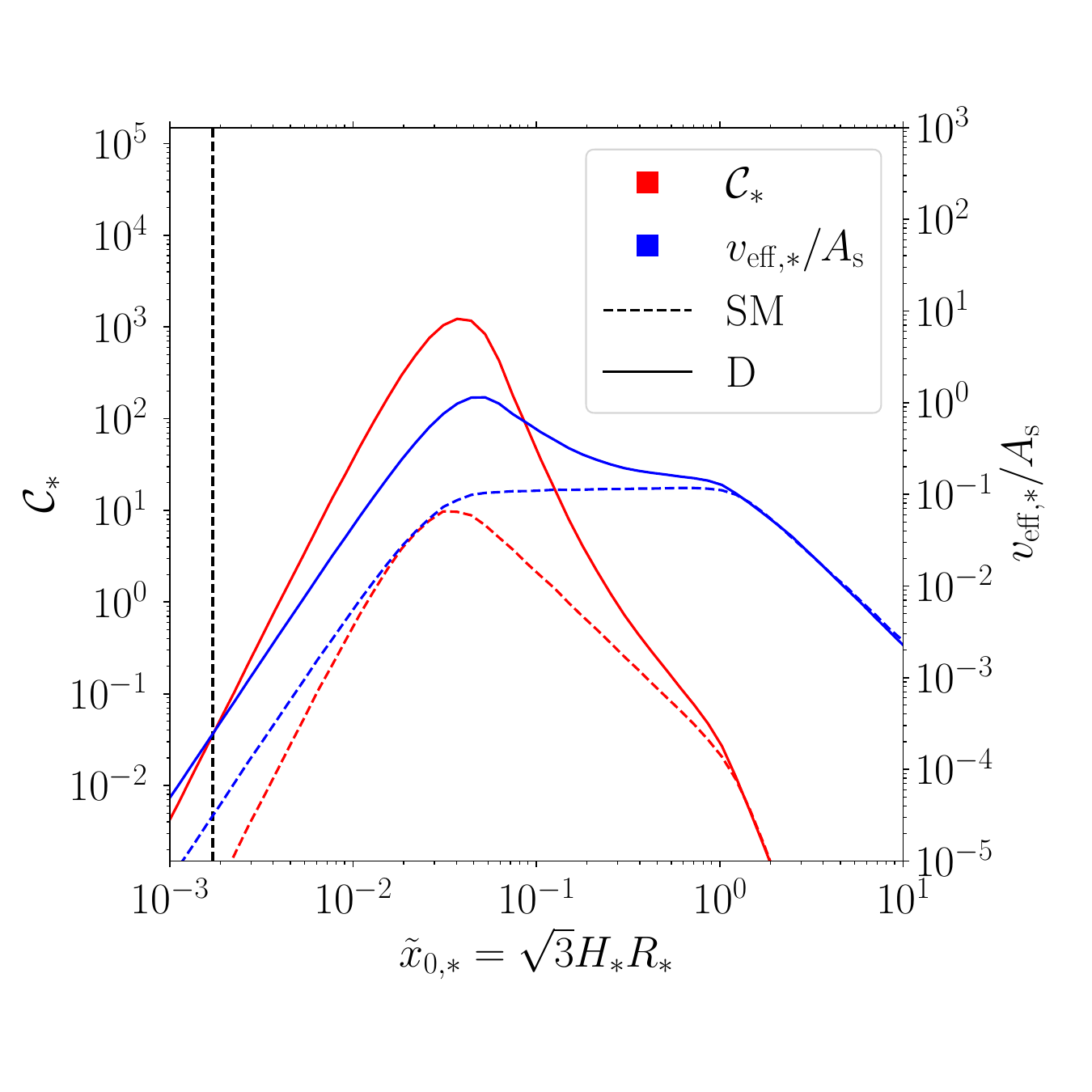}&\includegraphics[scale=0.26]{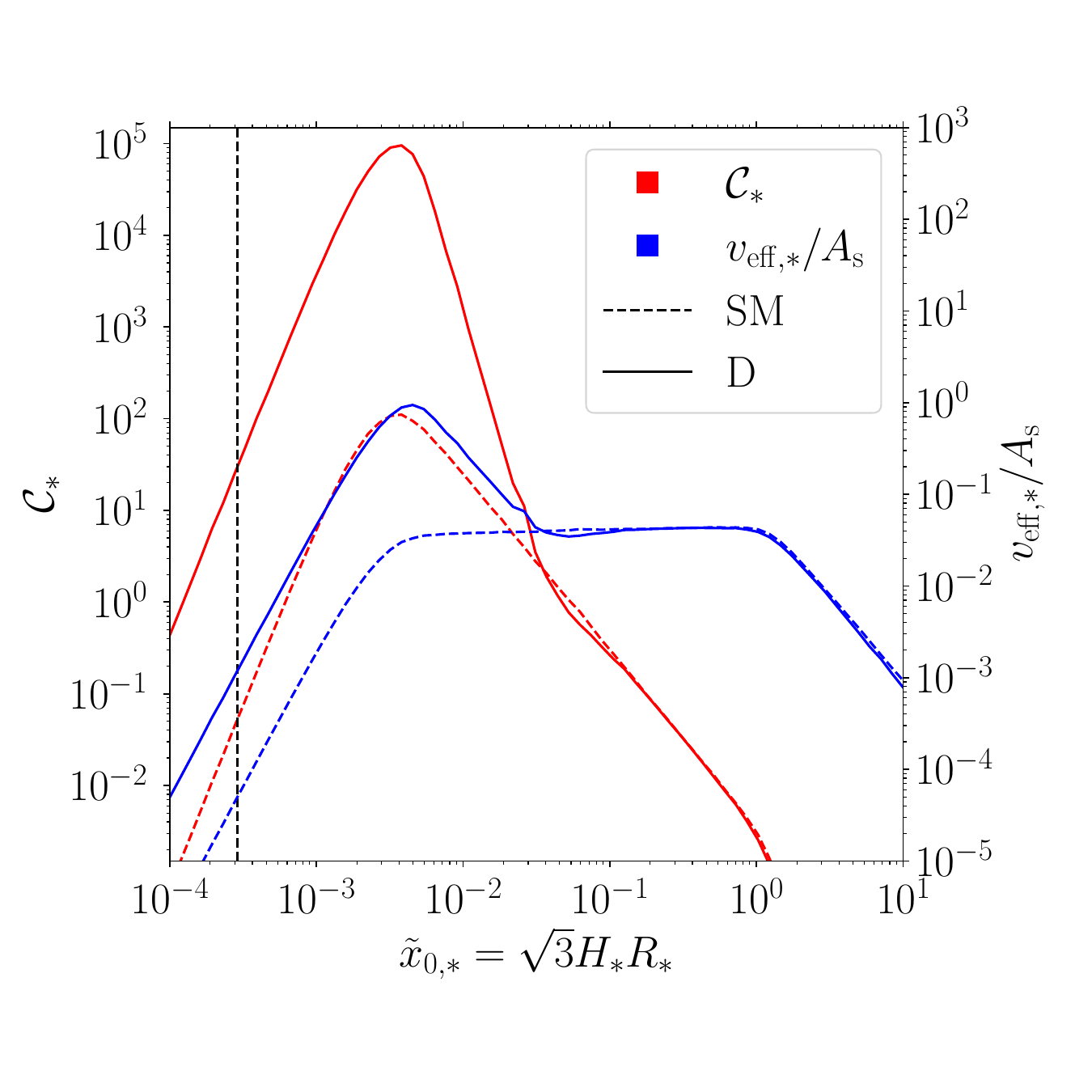}\\
     (a)&(b)
    \end{tabular}}
     \caption{$\mathcal{C}_*$ and $v_{\rm eff,*}$ versus $\tilde{x}_{0,*}$ for the FOPT benchmark cases BP-2 (panel a) and BP-4 (panel b) in Fig.~\ref{fig:SampleEvol}. The vertical line marks the size of the FV bubble $R_{\rm FV,*}$ relative to Hubble constant at percolation.}
    \label{fig:MainCContributions}
\end{figure}
In Fig.~\ref{fig:MainCContributions} we show the dependence of $\mathcal{C}_*$ and $v_{\rm eff,*}$ on the bubble size relative to Hubble in red and blue, respectively. The two panels feature BP-2 and BP-4. The dotted lines, corresponding to the SM approximation, use the exact transfer functions in Eqs.~(\ref{Tdeltapsi}), to calculate $\mathcal{C}_*$ and $v_{\rm eff, *}$, with integration up to $k = \pi/R_*$. The vertical line marks the size of the FV bubble $R_{\rm FV,*}$ relative to Hubble, associated with the prescription in Eq.~(\ref{RadiusPrescription}). As we can see in Fig.~\ref{fig:MainCContributions}, the dark sector $v_{\rm eff,*}$ and the SM approximation can differ by an order of magnitude. The drop in both $\mathcal{C}_*$ and $v_{\rm eff, *}$ for $\tilde{x}_{0,*} < \sqrt{3} H_* R_{\rm FV, *}$ is associated with the fact that $k_{\rm cut}$ is above the largest $k$ mode available in the interpolation table; whenever we encounter this case, we extrapolate the transfer functions to zero. For the benchmark points in Fig.~\ref{fig:MainCContributions}, the relative size of the FV bubble, $\mathcal{C}_*$, $v_{\rm eff,*}$ and $s_{\rm rms,*}$, can be found in Table~\ref{table:FinalBPsOld}. We find that the FV bubbles are much smaller than the Hubble horizon and the effective equatorial velocity is much smaller than the speed of light. 

\subsection{Estimates and scans over the physical region}
\label{sec:MainResEst}

From Eq.~(\ref{MainJsDefn}), the RMS spin of FV bubbles can be expressed in terms of the inverse of the surface potential $R_*/(G_{\rm N} M_*)$ and the equatorial velocity. The former  can be directly estimated from the background evolution as follows. In our prescription, $R_*$ is given by 
Eq.~(\ref{RadiusPrescription}), and the FV fraction is given by Eq.~(\ref{EstimateFVFraction}). We can trade $\Gamma_*$ with $\beta_*$ to find
 \begin{eqnarray}
     \label{RstarApprox}R_* \approx 6^{1/4}\frac{v_{\rm w,*}}{\beta_*}\,.
 \end{eqnarray}
The mass of the FV bubble requires $R_*$ and $\rho_{\rm D,*}$, with the latter given by
\begin{eqnarray}
    \rho_{\rm D,*} = \frac{\pi^2}{30}g_\rho T_*^4 -(1-F_*) \Delta \rho(T_*)\,.
\end{eqnarray}
We can relate $\rho_{\rm D,*}$ with the Hubble scale through the total energy density:
\begin{eqnarray}
\rho_{\rm tot,*} \simeq \frac{\pi^2}{30}g_{\rho, {\rm SM}}(T_*) \frac{T_*^4}{r_{T,*}^4} + \frac{\pi^2}{30}g_\rho T_*^4 -(1-F_*) \Delta\rho(T_*) = \frac{3H_*^2}{8 \pi G_{\rm N}}\,.
\end{eqnarray}
We now formulate the following estimate for the RMS spin, which constitutes one of our main results:
\begin{align}
  \nonumber \boxed{ s_{\rm rms,*} \approx 
      \left(\frac{2}{3}\right)^{1/2}\left(\frac{\beta_*/H_*}{v_{\rm w,*}}\right)^2\left\{1+\frac{g_{\rho,{\rm SM}}(T_*)}{g_{\rho}}r_{T\rm,*}^{-4}\left[1 - (1-F_*) \frac{30}{\pi^2 g_\rho} \frac{\Delta \rho(T_*)}{T_{\rm c}^4}\frac{T_{\rm c}^4}{T_*^4}\right]^{-1}\right\}
      v_{\rm eff,*}\,.}\\
      \label{MainFinalsRMSEstimate}~
\end{align}
Note that the explicit dependence of $s_{\rm rms, *}$ on $v_{\rm w}$ means that the spin depends on the exponent $n_{\rm w}$ in Eq.~(\ref{vwPrescription}). For BP-2 and BP-4, we have checked that for $n_{\rm w}$ values from 2 to 8, the spin gradually decreases by a factor roughly equal to $\left[1 - (T_*/T_{\rm c})^8\right]^2\left[1 - (T_*/T_{\rm c})^2\right]^{-2}$. Note that $\beta_*/H_*$, $v_{\rm eff,*}$, and $T_*$ are marginally sensitive to $n_{\rm w}$. One can show that the term in square brackets is significant when $\Delta \rho(T_*)/T_{\rm c}^4$ hits the maximum possible value allowed by the physical conditions; otherwise, for $\Delta \rho(T_*)/T_{\rm c}^4 \ll g_\rho \pi^2/30$, the term in square brackets is simply unity. On the other hand, we typically find that $v_{\rm eff,*}/A_{\rm s}$ is in the range $\sim 10^{-5}$ to $\sim 10^{-3}$. We have found that the value of $v_{\rm eff,*}/A_{\rm s}$ is sensitive to the spectral index $n_{\rm s}$ of the primordial curvature power spectrum. In the case where $n_{\rm s} = 1$, $v_{\rm eff,*}/A_{\rm s}$, and thus the spin value, is enhanced by roughly an order of magnitude.

\begin{figure}[t]
    \centering
    \begin{tabular}{cc}
          \includegraphics[scale=0.26]{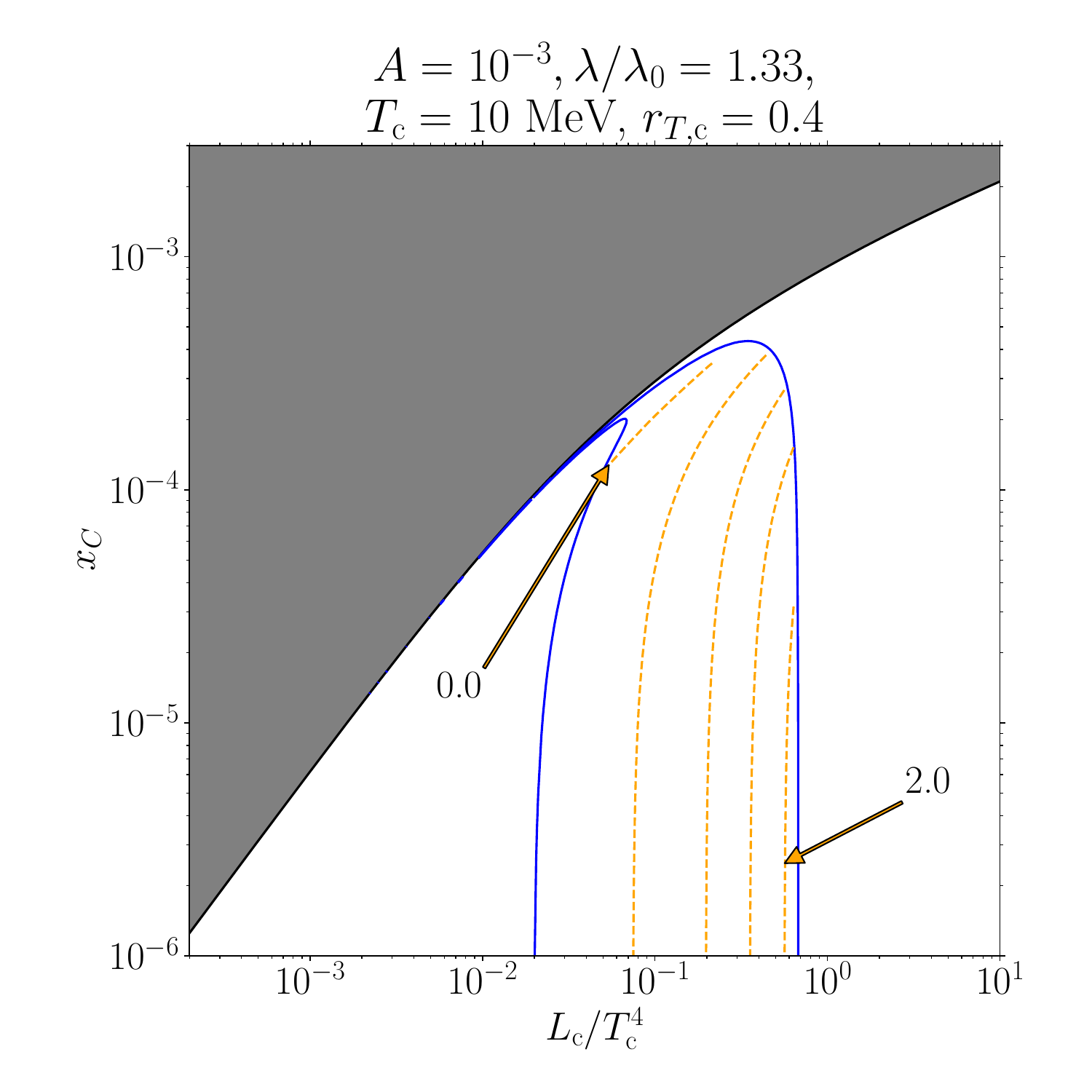} &  \includegraphics[scale=0.26]{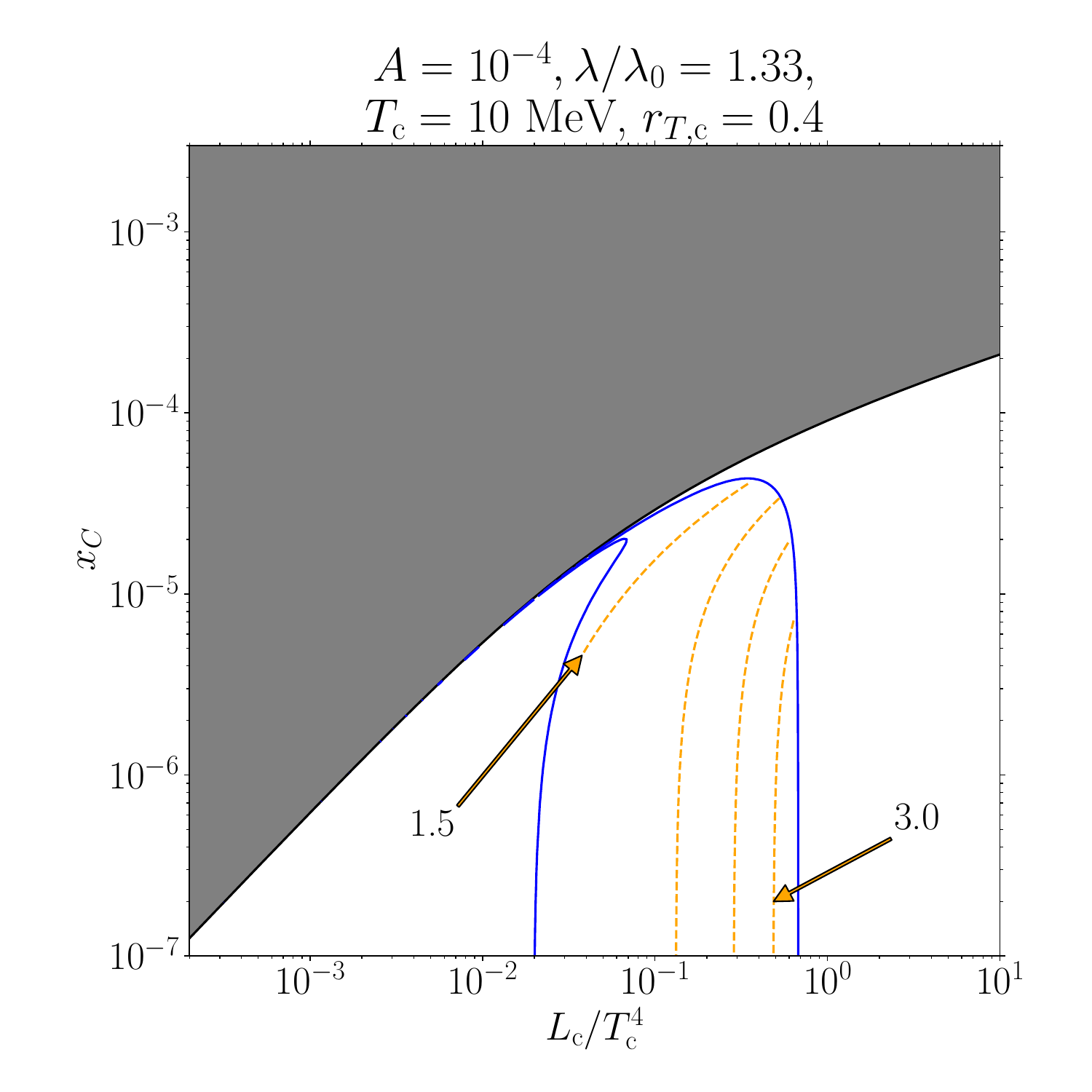}\\
          (a) & (b)\\
          \includegraphics[scale=0.26]{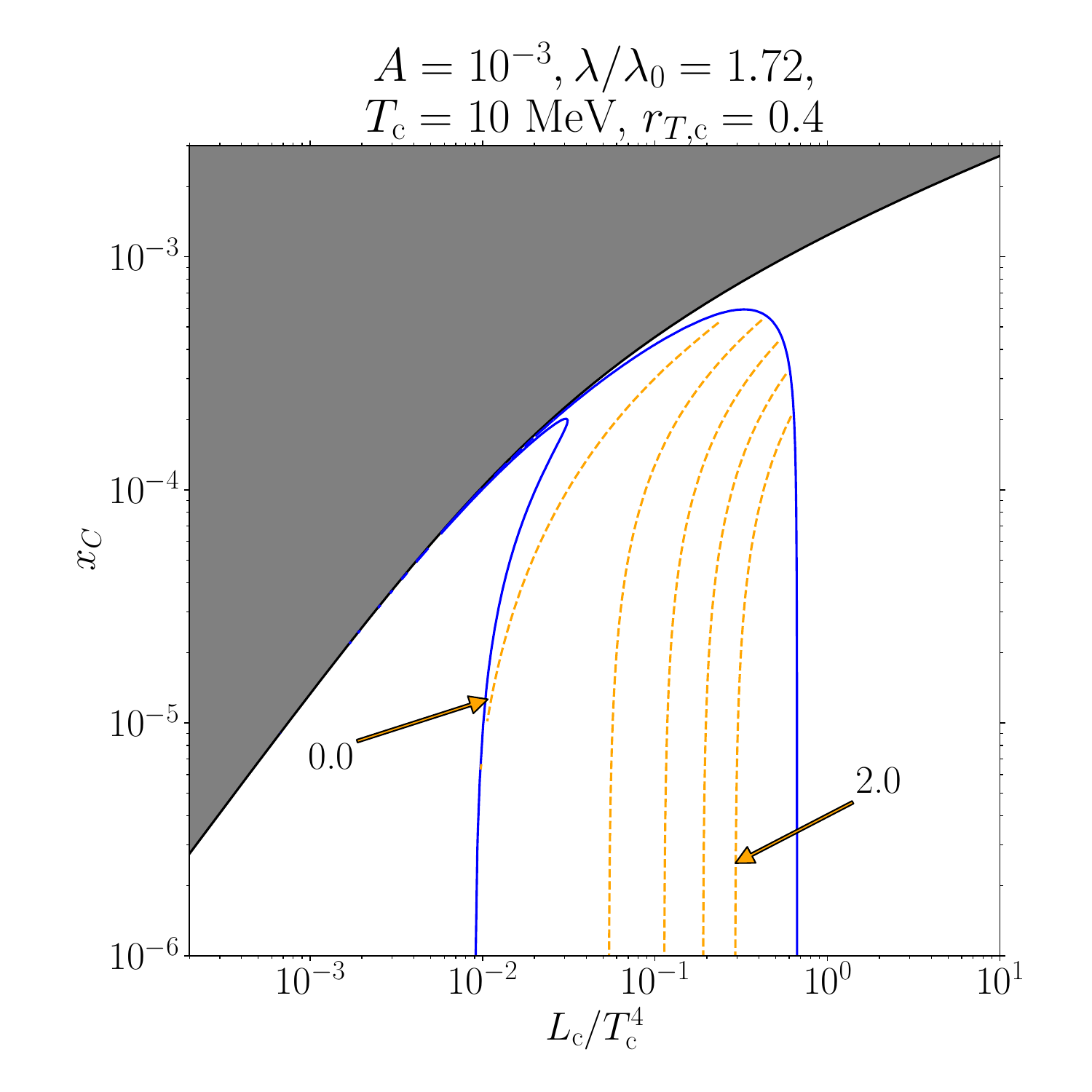}&\includegraphics[scale=0.26]{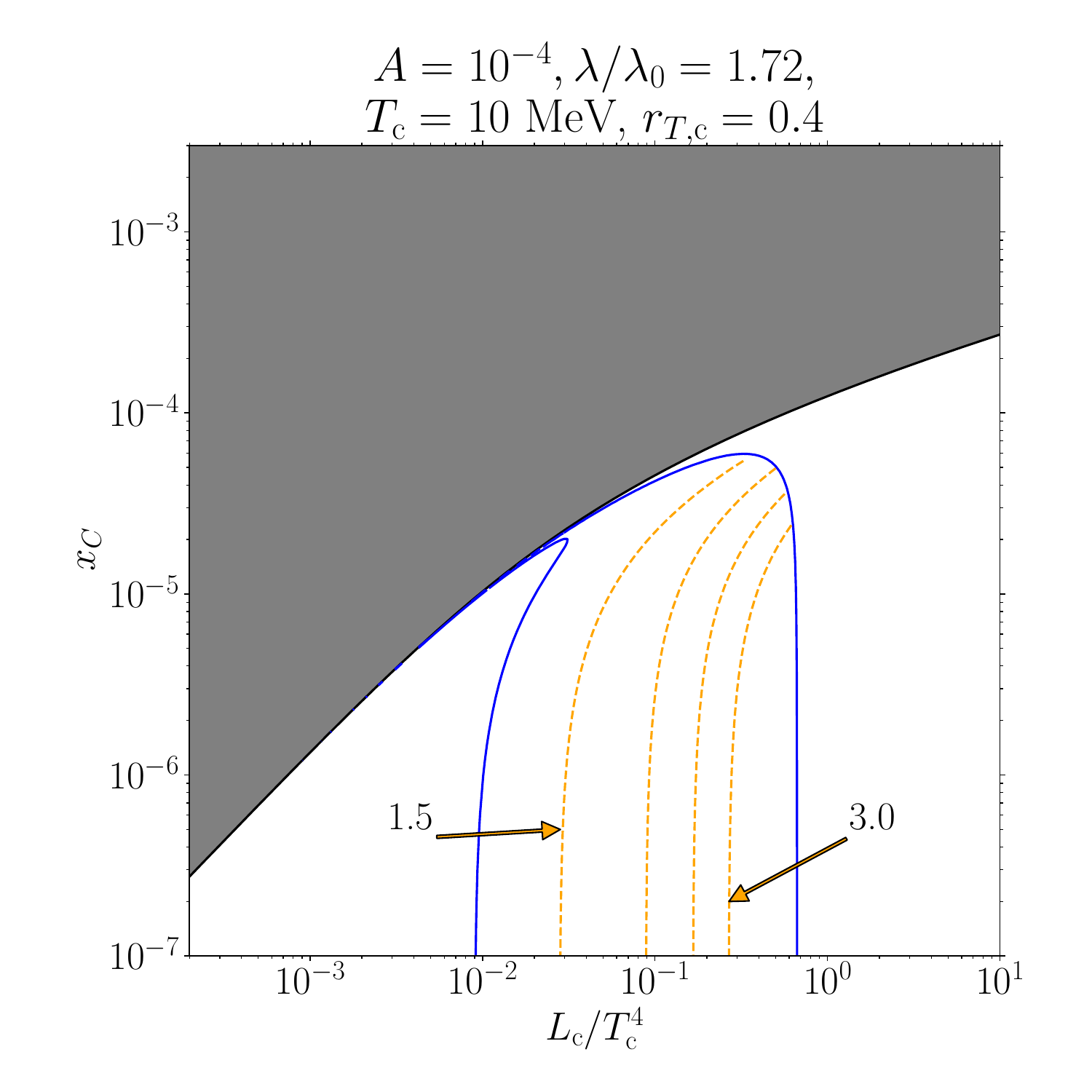}\\
          (c) & (d)
    \end{tabular}
    \caption{\label{fig:ScanPhyssrms}Contours of constant $\log_{10} \left[s_{\rm rms,*}/(v_{\rm eff,*}/A_{\rm s})\right]$ in steps of 0.5 for $A = 10^{-3}, 10^{-4}$, for fixed $\lambda/\lambda_0 \simeq 1.33$ and $\lambda/\lambda_0 \simeq 1.72$, and $T_{\rm c} = \unit[10]{MeV}$, $r_{T\rm,c} = 0.4$. The blue contour marks the boundary of the physical region.}
    \end{figure}

We then combine this result with the scans of $s_{\rm rms,*}/(v_{\rm eff,*}/A_{\rm s})$ in the physical region shown in Fig.~\ref{fig:ScanPhyssrms}. Although it is possible to obtain values of $s_{\rm rms,*}$ that are less than unity, there are cases where the RMS spin can exceed it. Note that a large spin value does not pose a problem because the FV bubble is not a black hole. For comparison, we list some typical dimensionless spin values for a few astrophysical objects in Table~\ref{table:astroSpin}.
\begin{table}[t]
    \centering
    \begin{tabular}{cccc}\toprule
        ~ & Sun & Earth & Neutron star \\\midrule
        Mass ($M_\odot$) & 1.0 & $3.0 \times 10^{-6}$ & 1.5\\
        Radius (km) & $6.963 \times 10^5$ & $6.371 \times 10^3$ & 10 \\
        Orbital period (s) & $2.33 \times 10^6$ & $8.64 \times 10^4$ & $10^{-3}$ \\
        Equatorial velocity & $6.26 \times 10^{-6}$ & $1.55 \times 10^{-6}$ & $0.21$ \\
        Spin & 1.18 & $8.89 \times 10^2$ & 0.379 \\\bottomrule
    \end{tabular}
    \caption{Spin and other physical properties of typical astrophysical objects.}
    \label{table:astroSpin}
\end{table}

\begin{figure}[t]
    \centering
    \begin{tabular}{cc}
        \includegraphics[scale=0.26]{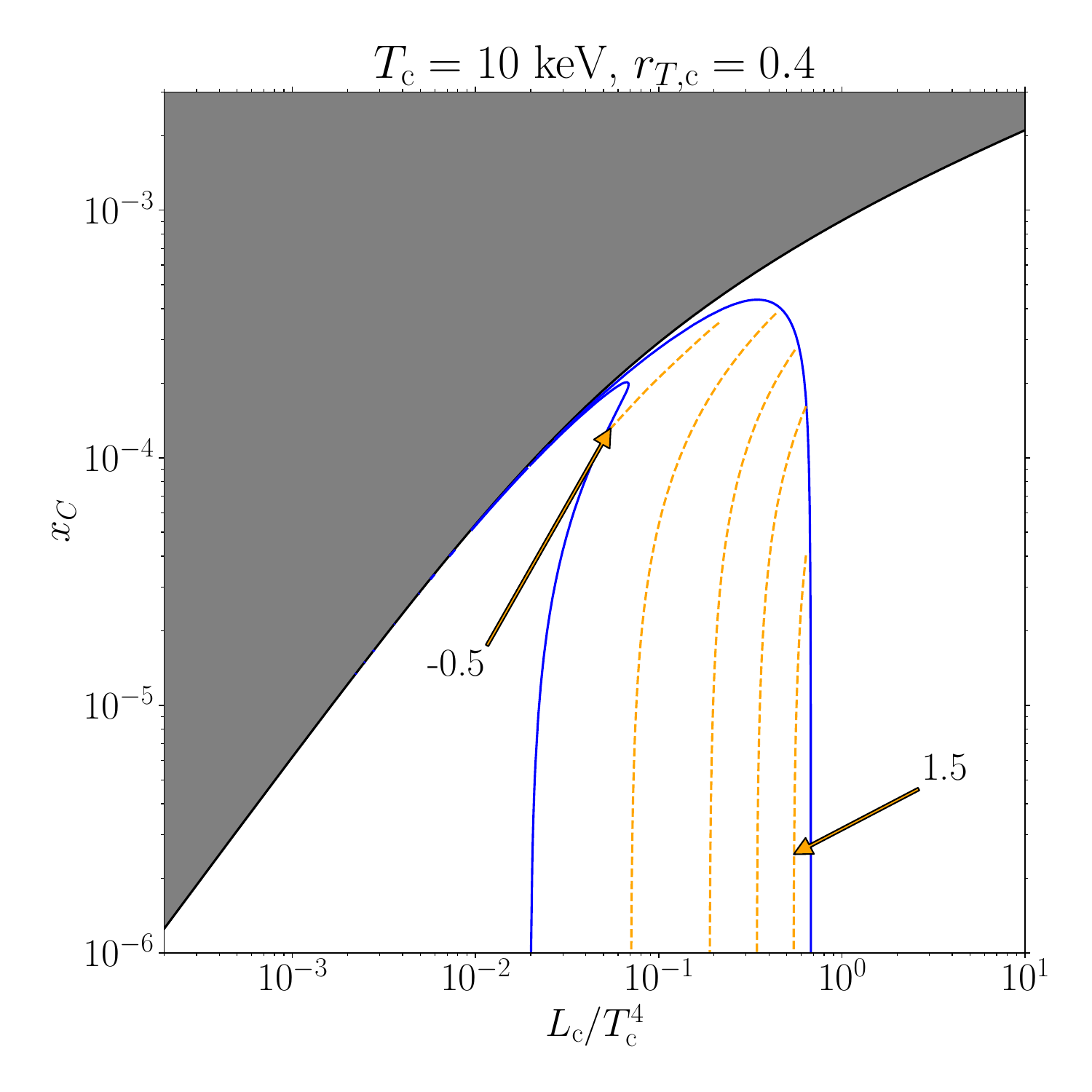} &  \includegraphics[scale=0.26]{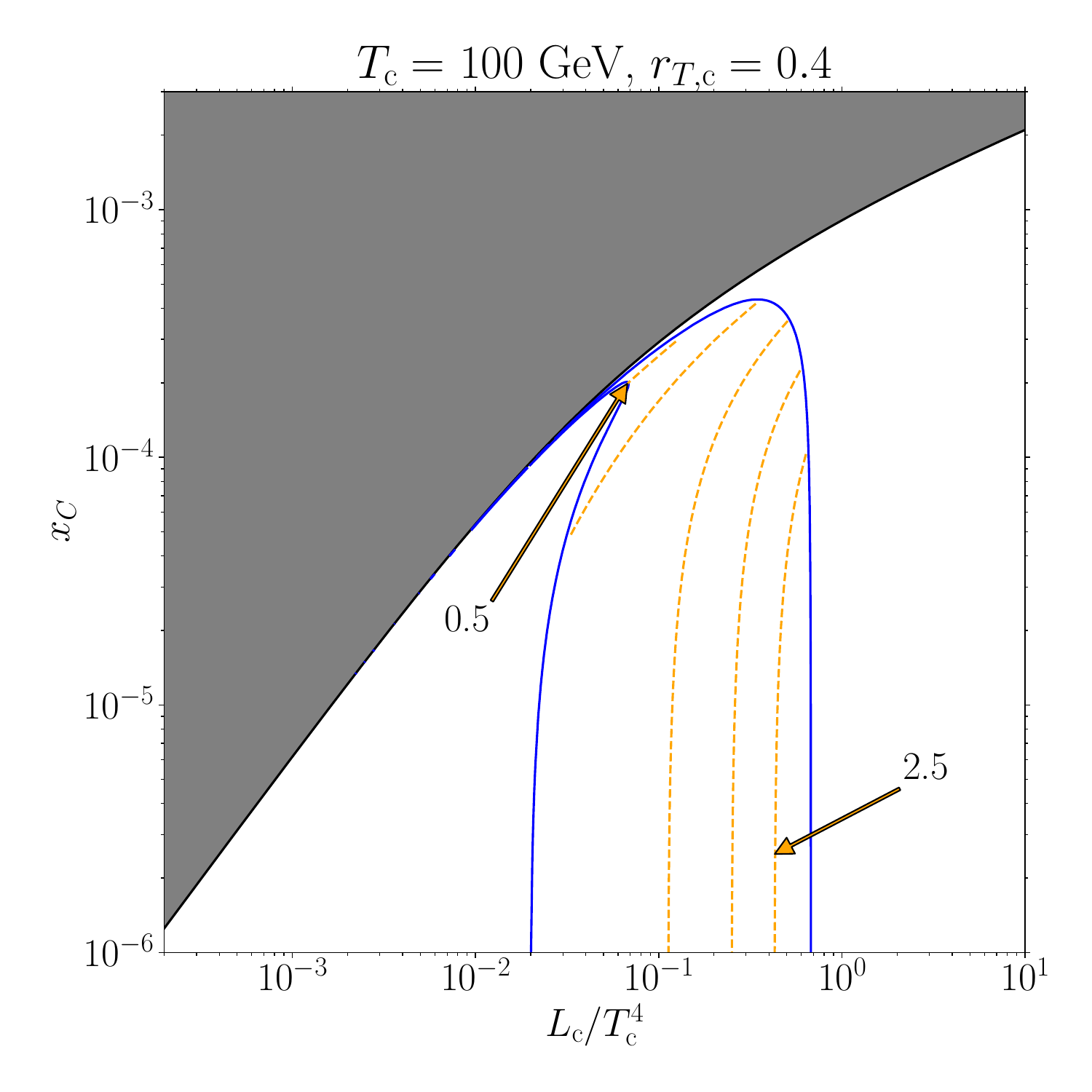}\\
        (a) & (b)\\
          \includegraphics[scale=0.26]{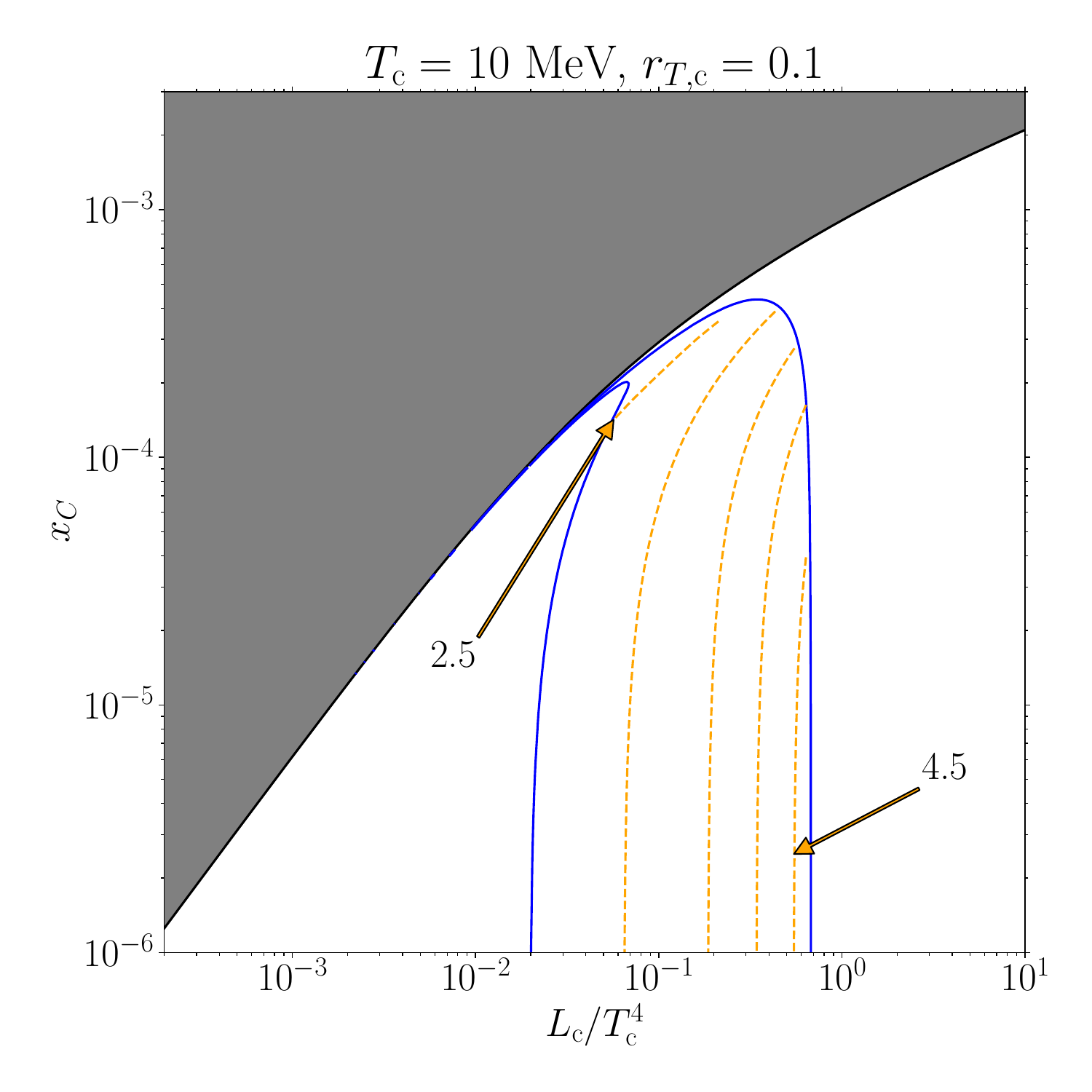} &  \includegraphics[scale=0.26]{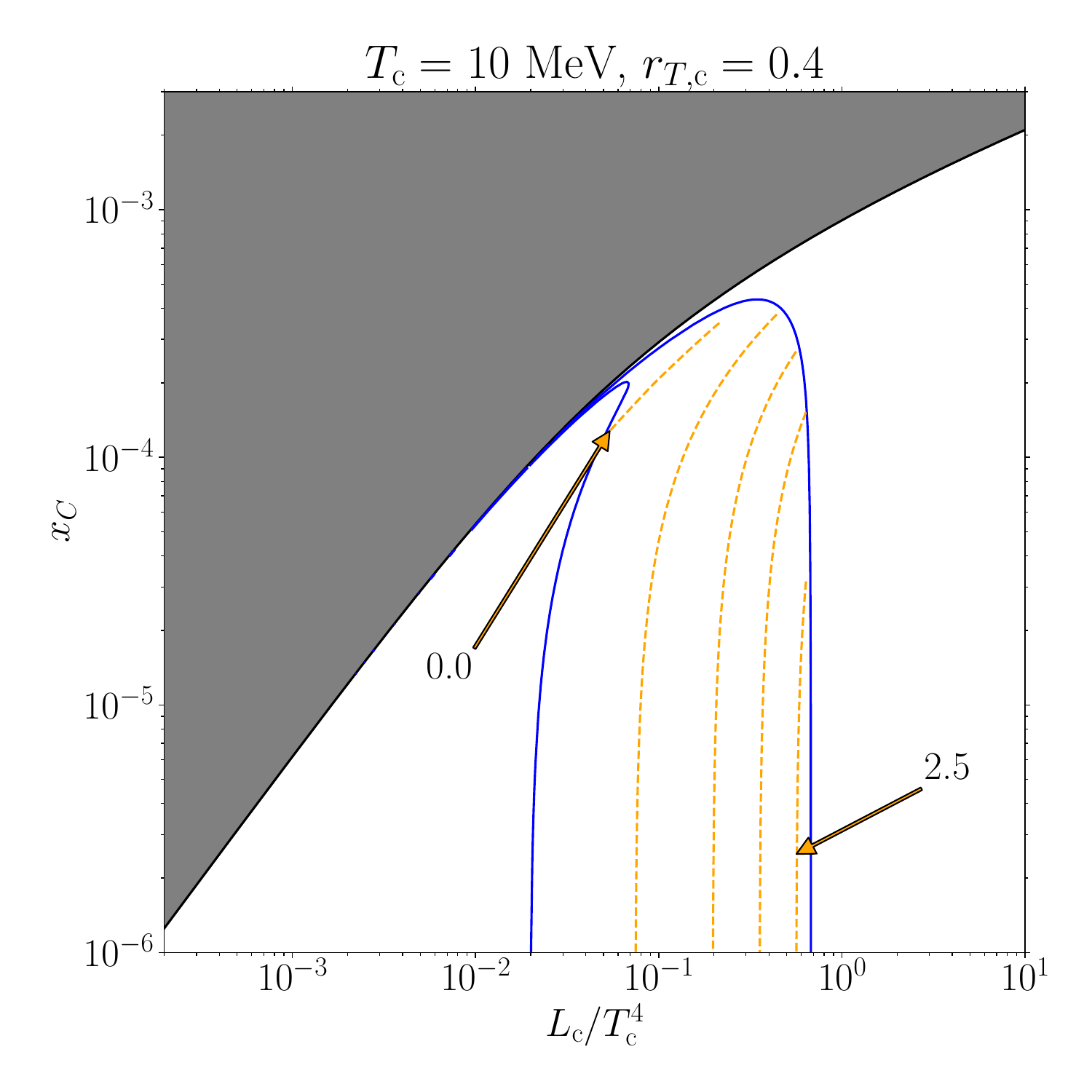}\\
          (c) & (d) 
    \end{tabular}
    \caption{\label{fig:ScanPhyssrmsTcrTc}Contours of constant $\log_{10}\left[s_{\rm rms,*}/(v_{\rm eff,*}/A_{\rm s})\right]$ in steps of 0.5 for $A = 10^{-3}$ and $\lambda/\lambda_0 \simeq 1.33$, changing $T_{\rm c}$ (top) and $r_{T\rm,c}$ (bottom). The blue contour marks the boundary of the physical region.}
    \end{figure}

We can see from Fig.~\ref{fig:ScanPhyssrms} that a wide range of $s_{\rm rms,*}/(v_{\rm eff,*}/A_{\rm s})$ values can be obtained, even for fixed $A$ and $\lambda$, and for fixed $T_{\rm c}$ and $r_{T\rm,c}$. Equation~(\ref{MainFinalsRMSEstimate}) explains this behavior because the percolation temperature and $\beta_*/H_*$ vary widely within the physical region. Furthermore, we see the effect of changing $T_{\rm c}$ and $r_{T\rm,c}$ in the contours of $s_{\rm rms,*}/(v_{\rm eff,*}/A_{\rm s})$ in Fig.~\ref{fig:ScanPhyssrmsTcrTc}. Panel (d) shows the scan for the default choice $T_{\rm c} = \unit[10]{MeV}, r_{T\rm,c} = 0.4$. Panels (a) and (b) show the effect of changing $T_{\rm c}$, which leads to increasing $s_{\rm rms,*}/(v_{\rm eff,*}/A_{\rm s})$ for larger $T_{\rm c}$ due to the increase in $g_{\rho,{\rm SM}}$, consistent with Eq.~(\ref{MainFinalsRMSEstimate}). On the other hand, comparing panels (c) and (d) illustrates the effect of changing $r_{T\rm,c}$, where a smaller $r_{T\rm,c}$ increases the overall $s_{\rm rms,*}/(v_{\rm eff,*}/A_{\rm s})$.
    
\subsection{Benchmarks}
To obtain a general idea of the prediction for the RMS spin value of FV bubbles, we consider a collection of FOPT scenarios that will serve as our benchmark points. We require that these FOPT scenarios are, at least, physically viable. Then we calculate the transfer functions for these benchmarks, in order to eventually obtain $v_{\rm eff,*}$. One criterion we impose on these benchmarks is $T_*/T_{\rm c} \leq 0.95$, which guarantees that the effective potential will have a sufficiently high barrier between the false vacuum and the true vacuum. We also require $k_{\rm cut} \lesssim k_{\rm BH}$.

The above requirements constrain the FOPT scenarios we feature to have $A = 10^{-3}, 10^{-4}$; then for each $A$, we select 
$\lambda/\lambda_0(A) \simeq 1.33, 1.72$. For each pair of $(A,\lambda)$, we scan the physical region in the ($L_{\rm c}/T_{\rm c}^4,x_C$)~plane and obtain physical quantities such as $\alpha_*$, $T_*$, and $s_{\rm rms,*}/(v_{\rm eff,*}/A_{\rm s})$. We choose a random collection of FOPT scenarios and solve the perturbation equations to construct a table of transfer functions versus $k/\mathcal{H}_{\rm c}$. This allows us to calculate $\mathcal{C}_*$ and the effective equatorial velocity $v_{\rm eff,*}/A_{\rm s}$, and finally $s_{\rm rms,*}$. The final benchmark points are presented in Table~\ref{table:FinalBPsOld}, with the rows organized in three groups. The first group (from the top) has parameters that specify the effective potential, critical temperature, and initial temperature ratio. The second group has quantities that can be obtained solely by tracking the evolution of the background cosmology during the phase transition. The third group has quantities that are relevant in inferring the features of the transfer functions, as well as quantities that can be obtained after solving the perturbation equations.

The selected benchmarks have notable features. Most of the benchmark points have $r_{T\rm, c} = 0.4$, except BP-5 which has $r_{T\rm, c} = 0.1$. We provide benchmarks with critical temperatures $T_{\rm c} = \unit[10]{keV}$ (BP-2), \unit[10]{MeV} (BP-3,5), \unit[1]{GeV} (BP-1,4), and \unit[100]{GeV} (BP-6). The values of $\lambda$ are such that $\lambda/\lambda_0(A) \simeq 1.33, 1.72$. We restrict $k_{\rm BH}/k_{\rm cut}$ between 1 and 2, with BP-2 and BP-4 representing very special cases where the cutoff scale is very close to the threshold scale for PBH formation. The constraint on $k_{\rm BH}/k_{\rm cut}$ pushes the parameter space to the left edge of the physical region in Fig.~\ref{fig:PhysRegion} so that the benchmark points have values of $L_{\rm c}/T_{\rm c}^4$ close to ${\cal O}(10^{-2})$. The percolation temperatures of the benchmarks range from $\sim 0.2$ to 0.3 times the critical temperature, $T_*/T_{\rm c}$ being the smallest for BP-2. While points on the left edge of the physical region correspond to $T_0/T_{\rm c}$ slightly above 0.3 for $x_C \ll A$, $T_0/T_{\rm c}$ can reach values between 0.2 and 0.3 when $x_C \lesssim A$. BP-2 and BP-3 correspond to $T_*/T_{\rm c}$ less than 0.3, where $x_C \simeq 0.24 A$. For BP-5 and BP-6, we have $x_C \sim \mathcal{O}\left(10^{-3}, 10^{-2}\right)A$, so $0.3 < T_0/T_{\rm c} < T_*/T_{\rm c}$. Regarding the duration of the phase transition, we find that the rescaled elapsed conformal time since the critical point $\tilde{\eta}_*$ is at least 1.5, with the longest duration corresponding to BP-2 at $\tilde{\eta}_* \simeq 3.22$. Such phase transitions correspond to  $(1/2)\ln(1+2\tilde{\eta}_*)$ $e$-folds ranging from 0.6 to 1.0. The period for which the dark sector sound speed squared deviates significantly from $1/3$ lasts for $\sim 10^{-3}$ to $10^{-2}$. Across the benchmarks, the value of $c_{s\rm,D*}^2$ can be as small as $-7$ and can reach values as low as around $-40$. 

We now discuss the RMS spins for the benchmark points. We use Eq.~(\ref{MainFinalsRMSEstimate}) as our guide to understand the dependence of $s_{\rm rms,*}/(v_{\rm eff,*}/A_{\rm s})$ --- which we will call the RMS spin prefactor, since it excludes $v_{\rm eff,*}$ --- on $T_{\rm c}$, $r_{T,\rm *}$, and $(\beta/H)_*/v_{\rm w,*}$. Firstly, the dependence on $T_{\rm c}$ enters in $g_{\rho,\rm SM}$, so we expect the benchmark with the lowest $T_{\rm c}$ ({\it i.e.}, BP-2) to give the smallest RMS spin prefactor. Since all of the benchmarks have $v_{\rm w,*}$ close to unity, the RMS spin prefactor effectively scales with $(\beta/H)_*$. The RMS spin prefactor for BP-3 is close to that of BP-2 which has a lower $T_{\rm c}$, because the inverse FOPT time scale $(\beta/H)_*$ for BP-3 is lower than that for BP-2. Similarly, while BP-1 and BP-4 have the same $T_{\rm c}$, the inverse FOPT time scale for BP-4 is larger by about a factor of 5, so the RMS spin prefactor of BP-4 is larger by about a factor of 25. Comparing BP-3 and BP-5, which have the same $T_{\rm c}$ and roughly the same $(\beta/H)_*$, BP-5 has a $r_{T,\rm c} \approx r_{T\rm,*} \approx 0.1$. Since the RMS spin prefactor scales as $r_{T\rm, c}^{-4}$, it will be larger by $4^4 = 256$ for BP-5. Ultimately, the trend in $s_{\rm rms,*}$ is determined both by the RMS spin prefactor and $v_{\rm eff,*}$. The dependence of the latter on the FOPT scenario can only be determined once the transfer functions at percolation are known. With our analysis, we can only conclude that $v_{\rm eff,*}/A_{\rm s}$ can range from $\mathcal{O}(10^{-5})$ to $\mathcal{O}(10^{-3})$. For our benchmark points, the RMS spin ranges from $\mathcal{O}(10^{-5})$ to $\mathcal{O}(10)$.

\section{Conclusions}
\label{sec:Conclusion}
Questions of how the angular momentum of primordial black holes is generated and how to characterize their spin distribution have been the subject of a lot of work. Most approaches assume an enhancement in the primordial curvature power spectrum in order to increase the probability of obtaining high peaks that can collapse to form primordial black holes. In contrast, the PBH formation mechanism we have in mind relies on dynamics during a first-order phase transition, which does not necessarily require an enhancement in the primordial curvature power spectrum. We restricted our calculation of the angular momentum and spin to the case of false vacuum bubbles. They can either collapse directly to PBHs or form Fermi balls that eventually collapse to PBHs. A minimal realization of the dark sector model consists of a singlet real scalar field, which triggers the FOPT by acquiring a nonzero vacuum expectation value through a temperature-dependent effective potential. To accommodate the Fermi ball formation scenario, we also introduce dark fermions which couple to the scalar field through a Yukawa interaction.

We adopt a picture in which the angular momentum is induced by cosmological perturbations, which ultimately come from initial Gaussian and nearly scale-invariant curvature perturbations. Taking the false vacuum bubbles to be spherical, we confirmed that the leading-order contribution arises at second order in perturbation theory. During the FOPT we treat the dark sector plasma as a single fluid consisting of two phases and track the evolution of cosmological perturbations up to the percolation time, at which epoch we evaluate the angular momentum and spin. We ensure that the FOPT scenarios considered are physical, such that the energy density in the true vacuum is positive and that there is a minimum temperature at which there are two local minima in the effective potential.  

 We worked in the uniform Hubble gauge to track the evolution of cosmological perturbations during the FOPT, and numerically obtained the density, velocity, and metric perturbations for each Fourier mode within a chosen range of comoving wavenumbers. Carrying out this task requires knowledge of the evolution of the sound speed and equation of state of the tightly coupled dark fluid. In general, it is possible that the perturbations in the dark sector fluid may deviate from those of a pure radiation fluid, in cases where the equation of state and sound speed squared differ from $1/3$. In particular, we find that the sound speed squared decreases around the time when the false vacuum fraction has a noticeable drop, takes a value smaller than $1/3$ at percolation, and then approaches the sound speed squared in the true vacuum. In particular, there is a short period in which the sound speed squared becomes negative. Thus, subhorizon modes that have sufficiently short wavelengths will be exponentially enhanced. Because of this enhancement, the mean squared value of the density contrast diverges. This signifies a natural threshold scale, above which the density contrast is sufficiently large to form black holes.  Once the perturbations at the percolation time are known, the RMS values of the angular momentum and spin are calculable. We perform this through phase-space integration, over pairs of Fourier modes, of a product of density and velocity perturbations weighted by a form factor associated with the size of the spherical false vacuum bubble. The phase-space integration over Fourier modes is limited to a scale corresponding to the radius of the FV bubbles, and we consider scenarios where this critical scale is below the threshold scale for black hole production.
 
 From a classical mechanics standpoint, the RMS spin of the false vacuum bubble is dependent on the equatorial velocity, radius, and mass of the sphere. We chose the radius of the FV bubble to be the maximal volume at the percolation time, within which no true vacuum bubble can nucleate. We have found a simple scaling relation, which relates the RMS spin with the inverse FOPT time scale $\beta_*/H_*$, the wall velocity, the degrees of freedom in the dark and SM sectors, and the initial dark-to-SM temperature ratio $r_{T\rm,c}$. In particular, the RMS spin is mainly determined by $(\beta_*/H_*)/v_{\rm w,*}$, $r_{T\rm,c}$, and $T_*$. The RMS spin scales as $(\beta_*/H_*)^2/v_{\rm w,*}^2$ and is inversely proportional to $r_{T\rm,c}^4$. The quantity $(\beta_*/H_*)/v_{\rm w,*}$ is mainly determined by the particular FOPT scenario. The critical temperature $T_{\rm c}$ also plays a role in determining the RMS spin through the direct proportionality between $s_{\rm rms,*}$ and $g_{\rho,{\rm SM}}$. We have found that the RMS spin value can take a wide range of values. For our sample scans with $T_{\rm c}$ between $\unit[10]{keV}$ and $\unit[100]{GeV}$, and $r_{T\rm,c}$ ranging from 0.1 to 0.4, it can be as low as $\mathcal{O}(10^{-5})$ to a few $\mathcal{O}(10)$. The upper limit on $r_{T\rm,c}$ arises from the Planck constraint on the effective number of relativistic degrees of freedom, $N_{\rm eff}$, and the lower limit on $r_{T\rm,c}$ ensures that an FOPT occurs. A spin larger than unity does not pose an issue, since the false vacuum bubbles are not black holes.

\newpage
\section*{Acknowledgments}
 We thank Pin-Jung Chen for collaboration during the early stages of this work. JTA thanks Reginald Bernardo,  Kuan-Yen Chou, Daniel Harsono, Mehrdad Mirbabayi, Michiyasu Nagasawa, Martin Spinrath, Yuhsin Tsai, and Ian Vega for fruitful discussions and insights.  JTA and PYT acknowledge the kind support of the National Science and Technology Council of the Republic of China (formerly the Ministry of Science and Technology), with grant number
NSTC 111-2811-M-007-018-MY2. DM acknowledges financial support from the U.S. Department of Energy under Grant No. DE-SC0010504. This work used high-performance computing facilities operated by the Center for Informatics and Computation in Astronomy (CICA) at the National Tsing Hua University. This equipment was funded by the Ministry of Education of Taiwan, the National Science and Technology Council of Taiwan, and National Tsing Hua University. We also thank the National Institute of Physics, UP Diliman for allowing us to use the in-house high performance computing cluster.

%\newpage
 \bibliographystyle{JHEP}
 \bibliography{JCAP_AMT}

%% or
%% [B] Manual formatting (see below)
%% (i) We suggest to always provide author, title and journal data or doi:
%% in short all the informations that clearly identify a document.
%% (ii) please avoid comments such as "For a review'', "For some examples",
%% "and references therein" or move them in the text. In general, please leave only references in the bibliography and move all
%% accessory text in footnotes.
%% (iii) Also, please have only one work for each \bibitem.

\end{document}